\documentclass[a4paper,12pt]{report}
\pdfoutput=1
\usepackage{iiscthes}
\usepackage{graphicx}
\usepackage{ dsfont }
\usepackage{amsmath,amssymb,cite}

\def \beq{\begin{equation}}
\def \eeq{\end{equation}}
\def \beqa{\begin{eqnarray}}
\def \eeqa{\end{eqnarray}}

\def \e{\varepsilon}
\def \l{\left(}
\def \r{\right)}

\begin{document}

\begin{frontmatter}
%
%


\title{TOPICS IN NONCOMMUTATIVE GAUGE THEORIES AND DEFORMED RELATIVISTIC THEORIES}
\author{Nitin Chandra}

\submitdate{JULY 2012}
\dept{Centre for High Energy Physics}
\sciencefaculty

\iisclogotrue 


\maketitle
\begin{dedication}
\begin{center}
TO \\[2em]
\large\it Bhaiji (Shri Dhiraj Kumar Mishra)\\
and\\
\large\it Pappaji (Shri Karunakar Dutt)
\end{center}
\end{dedication}


\declaration

I hereby declare that the work presented in this thesis entitled
``TOPICS IN NONCOMMUTATIVE GAUGE THEORIES AND DEFORMED RELATIVISTIC THEORIES''
is the result of the investigations carried out by me under the 
supervision of Prof. Sachindeo Vaidya at the Centre for High Energy Physics,
Indian Institute of Science, Bangalore, India, and that it has not been
submitted elsewhere for the conferment of any degree or diploma
of any Institute or University. Keeping with the
general practice, due acknowledgements have been made wherever the work
described is based on other investigations. \\

\bigskip

Dated: \quad \quad \quad \quad \quad \quad  \quad \quad \quad \quad 
\quad \quad \quad \quad \quad \quad \quad \quad  \quad \quad \quad \quad Nitin Chandra

\bigskip
\bigskip
\bigskip
\bigskip
\bigskip
\bigskip
{\bf Certified} \\
Sachindeo Vaidya\\
\quad Associate Professor \\
\quad Centre for High Energy Physics,\\
\quad Indian Institute of Science,\\
\quad Bangalore - 560012,\\
\quad Karnataka, India\\

\acknowledgements

First and foremost I thank my PhD adviser Sachindeo Vaidya, whose guidance made my Ph.D. possible. Despite of my slow progress in research, he has been very patient, cooperative and helpful.  Timely completion of this thesis is a result of his support and encouragement. I also acknowledge my sincere thanks to Manu Paranjape of Universit\'e de Montreal. I simply can't explain in words how wonderful person he is. His friendly behaviour is the prime reason for my stay in Montreal being a wonderful and memorable one. I thank IISc and UdeM for giving me an opportunity to pursue my research and learn a lot in the process. I also thank Canadian Commonwealth Scholarship Program and IISc for providing the financial assistance. 

I thank all CHEP members, especially the student friends for all the fun including BABA and chicken Parties. Thanks to all my friends who were also my collaborators. I learnt a lot about independent research while working with Sandeep on the problem of DSR. Thanks to all the friends from IISc including those in Physics, Bihari Samiti, my RKM friends, Integrated PhD batch-mates, B-Mess friends for making this 7 year-long stay enjoyable. I also want to thank all my friends outside IISc in Bangalore for the memorable moments I spent with them. Presence of my friend SP makes me forget all my worries. I am thankful to all my childhood friends who kept contact with me throughout these years. 

I thank Pappu Bhaiya and his family. They supported me as local guardians throughout my stay in Bangalore. I dedicate my thesis to Bhaiji and Pappaji. Bhaiji is the person with whom I discussed all my curiosities in science when I was a child. Pappaji taught me English when I needed it most. Finally and most importantly I am grateful to my parents and family. I am forever indebted to their love and support which gives me greatest satisfaction and happiness.

\publications

\begin{enumerate}
\item 
  {\it Time dependent transitions with time-space noncommutativity and its implications in Quantum Optics} \\
  N.~Chandra \\
  J.\ Phys.\ A A {\bf 45}, 015307 (2012) 
  [arXiv:1104.1059 [hep-th]].
  
  \item 
  {\it Thermodynamics of Ideal Gas in Doubly Special Relativity} \\
  N.~Chandra and S.~Chatterjee \\
  Phys.\ Rev.\ D {\bf 85}, 045012 (2012) 
  [arXiv:1108.0896 [gr-qc]].

\item 
  {\it Noncommutative Vortices and Instantons from Generalized Bose Operators} \\
  N.~Acharyya, N.~Chandra and S.~Vaidya \\
  JHEP {\bf 1112}, 110 (2011)
  [arXiv:1109.3703 [hep-th]].
\end{enumerate}

\begin{abstract}
\sl
There is a growing consensus among physicists that the classical notion of spacetime has to be drastically revised in order to find a consistent formulation of quantum mechanics and gravity.
One such nontrivial attempt comprises of replacing functions of continuous spacetime coordinates with functions over noncommutative algebra.
Dynamics on such noncommutative spacetimes (noncommutative theories) are of great interest for a variety of reasons among the physicists.
Additionally arguments combining quantum uncertainties  with classical gravity provide an alternative motivation for their study, and it is hoped that these theories can provide a self-consistent deformation of ordinary quantum field theories at small distances, yielding non-locality, or create a framework for finite truncation of quantum field theories while preserving symmetries.

In this thesis we study the gauge theories on noncommutative Moyal space.
We find new static solitons and instantons in terms of the so-called generalized Bose operators (GBO).
GBOs are constructed to describe reducible representation of the oscillator algebra.
They create/annihilate $k$-quanta, $k$ being a positive integer.
We start with giving an alternative description to the already found static magnetic flux tube solutions of the noncommutative gauge theories in terms of GBOs. The Nielsen-Olesen vortex solutions found in terms of these operators also reduce to the ones known in the literature. On the other hand,
we find a class of new instanton solutions which are unitarily inequivalent to the ones found from ADHM construction on noncommutative space. The charge of the instanton has a description in terms of the index representing the reducibility of the Fock space representation, i.e., $k$.
After studying the static soliton solutions in noncommutative Minkowski space and the instanton solutions in noncommutative Euclidean space we go on to study the implications of the time-space noncommutativity in Minkowski space.  To understand it properly we study the time-dependent transitions of a forced harmonic oscillator in noncommutative 1+1 dimensional spacetime. We also provide an interpretation of our results in the context of non-linear quantum optics.
We then shift to the so-called DSR theories which are related to a different kind of noncommutative ($\kappa$-Minkowski) space. DSR (Doubly/Deformed Special Relativity) aims to  search for an alternate relativistic theory which keeps a length/energy scale (the Planck scale) and a velocity scale (the speed of light scale) invariant. We study thermodynamics of an ideal gas in such a scenario.

In first chapter we introduce the subjects of the noncommutative quantum theories and the DSR.
Chapter \ref{instanton} starts with describing the GBOs.
They correspond to reducible representations of the harmonic oscillator algebra. 
We demonstrate their relevance in the construction of topologically non-trivial solutions in 
noncommutative gauge theories, focusing our attention to flux tubes, vortices, and instantons.
Our method provides a simple new relation between the topological charge and the number of times the basic irreducible representation occurs in the reducible representation underlying the GBO.
When used in conjunction with the noncommutative ADHM construction, we find that these 
new instantons are in general not unitarily equivalent to the ones currently known in literature.

Chapter \ref{qo} studies the time dependent transitions of quantum forced harmonic oscillator (QFHO) in noncommutative $\mathds{R}^{1,1}$
perturbatively to linear order in the noncommutativity $\theta$.
We show that the Poisson distribution gets modified, and that the vacuum state evolves into a ``squeezed'' state rather than a coherent state.
The time evolutions of uncertainties in position and momentum in vacuum are also studied 
and imply interesting consequences for modelling nonlinear phenomena in quantum optics.

In chapter \ref{DSR} we study thermodynamics of an ideal gas in Doubly Special Relativity.
We obtain a series solution for the partition function and derive thermodynamic quantities.
We observe that DSR thermodynamics is non-perturbative in the SR and massless limits.
A stiffer equation of state is found.
We conclude our results in the last chapter. 

\end{abstract}


\makecontents

%
%

\notations

The metric for $1+d$ Minkowski space has been taken to be 
\begin{equation}
\eta_{\mu\nu} = {\rm dia} (1,-1,-1,...)
\end{equation} 
where the first element comes for time coordinate.

The Abbreviations used in the thesis are
\begin{itemize}
\item GBO: {\it Generalized Bose Operator}
\item UIR: {\it Unitary Irreducible Representation}
\item ADHM: {\it Atiyah, Drinfeld, Hitchin and Manin}
\item SD: {\it Self Dual}
\item ASD: {\it Anti--self Dual}
\item SHO: {\it Simple Harmonic Oscillator}
\item FHO: {\it Forced Harmonic Oscillator}
\item QFHO: {\it Quantum Forced Harmonic Oscillator}
\item SR: {\it Special Relativity}
\item DSR: {\it Doubly/Deformed Special Relativity}
\end{itemize}

\end{frontmatter}
\chapter{Introduction}

There is a growing consensus among physicists that the classical notion of spacetime has to be drastically revised in order to find a consistent formulation combining quantum mechanics and gravity \cite{Doplicher:1994tu, Seiberg:2006wf, Rovelli:2003cu}.
One such nontrivial attempt comprises of replacing functions of continuous spacetime coordinates with functions over noncommutative algebra \cite{Douglas:2001ba, Balachandran:2010wc}.
Dynamics on such noncommutative spacetime (noncommutative theories) are of interest for a variety of reasons among the physicists for a long time.
Let us start with introducing such noncommutative theories.

\section{Noncommutative theory}

The idea of extension of noncommutativity to the coordinates was first suggested by Heisenberg as a possible solution for removing the infinite quantities of field theories before the renormalization procedure was developed and had gained acceptance. They seem to have first appeared in a letter from Heisenberg to Peierls in 1930~\cite{letter}. The first paper on the subject was published in 1947 by Hartland Snyder who sought to use it to regularize Quantum Field Theories~\cite{Snyder}. The success of renormalization theory however drained interest from the subject for some time.\\

The renewed interest by particle physics community started to grow after the paper by Seiberg and Witten \cite{Seiberg:1999vs}.
Arguments combining quantum uncertainties  with classical gravity also provide an alternative motivation for their study \cite{Doplicher:1994tu}, and these theories can provide a self-consistent deformation of ordinary quantum field theories at small distances, yielding non-locality \cite{Bahns:2002vm,Bahns:2003vb,Bahns:2004fc}, or create a framework for finite truncation of quantum field theories while preserving symmetries  \cite{Grosse:1995ar,Grosse:1995pr,Grosse:1995jt,Grosse:1996mz,Baez:1998he,Balachandran:1999hx}.
Also, the emergence of noncommutative field theories in string theory  \cite{Dhar:1994ib,Douglas:1997fm,Seiberg:1999vs} has provided a considerable impetus to their investigation.

The noncommutativity of the coordinates can be described by the following commutation relation
\begin{equation} \label{basic_comm}
\left[x_\mu,x_\nu\right]=i\theta_{\mu\nu}
\end{equation}   
where $\theta_{\mu\nu}$ is an anti-symmetric tensor. The simplest case corresponds to $\theta_{\mu\nu}$ being constant. The field theories on such spacetimes are described by elements of unital algebra (associative, but not commutative) generated by $x_\mu$'s. 
There is a sense in which the algebra over the noncommutative space $\mathds{R}^d_\theta$ and the algebra over commutative space  $\mathds{R}^d_0$ ($d$ being the dimension of the space) have the same topology. This algebra can be considered as a deformation of the algebra over $\mathds{R}^d_0$ such that the elements of the algebra have the same addition law, but a different (noncommutative) multiplication law which reduces to the commutative point-wise multiplication in the limit $\theta_{\mu\nu}\rightarrow 0$. This noncommutative multiplication law is often denoted by $f\star g$ or {\it the star product} to distinguish it from the ordinary pointwise multiplication of functions.
Let us give an example of such star product. If we denote by $m_0$ the pointwise multiplication map
\begin{equation}
m_0 \left(f\otimes g\right) (x)= f(x) g(x),
\end{equation} 
then the star product is generally given by
\begin{equation}
\left(f\star g\right) (x) = m_\theta  \left(f\otimes g\right) (x) = m_0 \circ \mathcal{F}_\theta  \left(f\otimes g\right) (x)
\end{equation}  
For instance,  $ \mathcal{F}_\theta$ corresponding to the Moyal product is given by
\begin{equation}
 \mathcal{F}_\theta = e^{\frac{i}{2}\theta_{\mu\nu}\left(\partial_{x_\mu} \otimes \partial_{x_\nu} -\partial_{x_\nu} \otimes \partial_{x_\mu} \right)}.
\end{equation}
One can easily check that the following is satisfied:
\begin{equation} \label{star_comm}
\left[x_\mu,x_\nu\right]_\star = x_\mu\star x_\nu - x_\nu\star x_\mu = i\theta_{\mu\nu}.
\end{equation}

\subsection{The Poincar\'e invariance}

Noncommutativity of the spacetime coordinates apparently conflicts with Poincar\'e invariance, as one can see in (\ref{basic_comm}). We circumvent this problem by ``twisting" the Poincar\'e symmetry \cite{Chaichian:2004yh}.
It goes as follows.
First of all one needs to define the action of a Poincar\'e group element on a tensor product space, which is generally given by the {\it coproduct}. The usual coproduct used is as follows:
\begin{equation}
\Delta_0 (\Lambda) = \Lambda\otimes \Lambda,
\end{equation}
$\Lambda$ being an element of the Poincar\'e group. We twist the coproduct to
\begin{equation}
\Delta_\theta (\Lambda)= \mathcal{F}_\theta^{-1} \Delta_0 (\Lambda) \mathcal{F}_\theta = \mathcal{F}_\theta^{-1} (\Lambda \otimes\Lambda) \mathcal{F}_\theta
\end{equation}
Action of a Poincar\'e element on star product of two functions is given by
\begin{equation}
\Lambda \triangleright \left(f\star g\right) = \Lambda \triangleright m_\theta \left(f\otimes g\right) = m_\theta \,\, \Delta_\theta (\Lambda) \left(f\otimes g\right)
\end{equation}
Putting $f,g = x_\mu, x_\nu$ one can check that the commutation relation (\ref{star_comm}) remains invariant under the action of Poincar\'e group. This modification in coproduct changes the standard Hopf algebra structure associated with the Poincar\'e group (the Poincar\'e-Hopf algebra) to the twisted Poincar\'e-Hopf algebra. 
This twist in turn give rise to twisted statistics leading to variety of interesting consequences \cite{Balachandran:2006pi, Oeckl:2000eg, Balachandran:2005eb, Balachandran:2005pn, Balachandran:2007vx, Akofor:2007fv, Chakraborty:2006hv, Balachandran:2010xk, Balachandran:2010wq, Srivastava:2012av, Basu:2010qm}. 
We will not be discussing them here to avoid any deviation from the desired topic.

In this thesis we study the gauge theories on noncommutative Moyal space \linebreak $\left(\theta_{\mu\nu} = constant\right)$. We start by studying the implications of the construction of so-called generalized Bose operators (GBO) \cite{bg} in noncommutative gauge theories. GBOs correspond to the reducible representation of the oscillator algebra
\begin{equation}
\left[a,a^\dagger\right] = 1.
\end{equation}  
We discuss them in detail in chapter \ref{instanton}. 
The noncommutative theory discussed here contains commutative time. 
Having studied gauge theories with space-space noncommutativity through the construction of GBOs we proceed to understand the implications of noncommutative time coordinate. 
There have been even claims that quantum field theories based on noncommutative time coordinate are nonunitary \cite{Chaichian:2000ia}. 
In contrast, in a series of fundamental papers, Doplicher et al. \cite{Doplicher:1994zv, Doplicher:1994tu} have studied (\ref{basic_comm}) in complete generality, without assuming that $\theta_{0i}= 0$ and developed unitary quantum field theories which are ultraviolet finite to all orders. 
In this thesis we study a simple quantum mechanical model of forced harmonic oscillator in $1+1$ dimensional noncommutative spacetime to get the essence of the effect of noncommutative time coordinate. 
For this we follow the unitary formulation of quantum mechanics with time-space noncommutativity discussed in \cite{Balachandran:2004rq}.

There are kinds of noncommutative spaces other than the Moyal space. In particular, we study the thermodynamics in the so-called Doubly/Deformed special relativistic (DSR) theories which are related to a noncommutative space known as $\kappa$-Minkowski space \cite{Ghosh:2007ai, Daszkiewicz:2004xy, KowalskiGlikman:2002jr, KowalskiGlikman:2002we, KowalskiGlikman:2001ct, Heuson:2003zt}. 
DSR (Doubly/Deformed Special Relativity) aims to search for an alternate relativistic theory which keeps a length/energy scale (the Planck scale) and a velocity scale (the speed of light scale) invariant.

\section{Doubly/Deformed Special Relativity}
Attempts to combine gravity with quantum mechanics in search of the theory 
for quantum gravity always seem to give rise to the Planck length 
$\left(l_P =\sqrt{\frac{G c}{h^3}}\right)$ 
that provides the scale at which the quantum effects of gravity will show up \cite{Mead:1964zz, Padmanabhan:1986ny, Padmanabhan:1985jq, Veneziano:1986zf, Amati:1988tn, Konishi:1989wk, Greensite:1990jm, Maggiore:1993rv, Garay:1994en, Rovelli:1994ge, AmelinoCamelia:2002vw}.
The existence of such a length scale is in conflict with the equivalence principle
because observers in different inertial frames will not agree on $l_P$ due to the 
Lorentz-Fitzgerald contraction. To understand in a better way consider the following:
 If the Planck scale is to play the role of the threshold for the discrete spacetime, it may come in the quantities involving the metric (for example the action).
 For equivalence principle to hold, i.e., for physics to be independent of the inertial frames the action and hence the Planck scale has to remain invariant under the relativistic transformation.

It has been shown that it is possible to still have equivalence 
principle by deforming Special Relativity (SR). These classes of theories fall under 
the name Doubly/Deformed Special Relativity (DSR) \cite{AmelinoCamelia:2000ge, AmelinoCamelia:2000mn, AmelinoCamelia:2002vy, AmelinoCamelia:2002wr}. In DSR, apart from the constancy of speed-of-light scale, the Planck length $l_P$ or equivalently Planck energy $\kappa$ is also 
constant under coordinate transformation from one inertial frame to another. It was first proposed by G. Amelino-Camelia \cite{AmelinoCamelia:2000mn}. To proceed one can take lessons from the transition from Galilean to Lorentz transformations. The Galilean transformation is linear in velocity and hence it does not keep any velocity scale invariant. To have an invariant velocity scale we needed to accept the Lorentz transformation which is nonlinear in velocity. Thus if one wants to have Planck length/energy scale invariant the relativistic transformation has to be nonlinear in position/momentum space. One also keep in mind the possibility of the symmetry group being different from the usual Lorentz group. This symmetry group must act nonlinearly in the position/momentum space representation to keep Planck length/energy invariant. In the special relativistic limit, Planck length/energy $\rightarrow 0/\infty$ the symmetry group has to go to the Lorentz group.

J. Magueijo and L. Smolin have argued that the symmetry group for the DSR transformations remains to be the Lorentz group \cite{Magueijo:2001cr, Magueijo:2002am}. But in DSR prescription they act nonlinearly on the position/momentum space. They have also suggested the following nonlinear representation of the Lorentz transformation on momentum space
\begin{eqnarray}
\Lambda(v):\varepsilon\rightarrow \varepsilon^\prime &=& \frac{\gamma (\varepsilon - vp_z)}
{1 + (\gamma -1)\frac{\varepsilon}{\kappa} - \gamma v \frac{p_z}{\kappa}} \label{p0_trans} \\
\Lambda(v):p_z\rightarrow p_z^\prime &=& \frac{\gamma (p_z - v\varepsilon)}{1 + (\gamma -1)\frac{\varepsilon}{\kappa} - \gamma v \frac{p_z}{\kappa}} \label{p3_trans} \\
\Lambda(v):p_x \rightarrow p_x^\prime &=& \frac{p_x}{1 + (\gamma -1)\frac{\varepsilon}{\kappa} - \gamma v \frac{p_z}{\kappa}} \label{p1_trans} \\
\Lambda(v):p_y \rightarrow p_y^\prime &=& \frac{p_y}{1 + (\gamma -1)\frac{\varepsilon}{\kappa} - \gamma v \frac{p_z}{\kappa}} \label{p2_trans}
\end{eqnarray}
with $\gamma = \frac{1}{\sqrt{1-v^2}}, c=1$. The above transformation corresponds to the boost along $z$-axis.
Note that they also satisfy the group multiplication law of the usual Lorentz group, i.e.,
\begin{equation}
\Lambda (u) \Lambda (v) = \Lambda \left(\frac{u+v}{1+uv}\right)
\end{equation}
 The transformation keeps the following mass invariant
 \begin{equation}
 m = \frac{1-\frac{\varepsilon}{\kappa}}{\sqrt{ \varepsilon^2-p^2}}
 \end{equation}
We call it the {\it invariant mass}. Note that unlike the special relativistic (SR) case, this is not same as the rest mass energy.

There have been attempts to study the representation on the position space too \cite{Gao:2003ex, Kimberly:2003hp, Hossenfelder:2006rr}. 
But there is no proper understanding of its relation to the momentum space representation. In this thesis we have used the momentum space representation given by  (\ref{p0_trans}) -- (\ref{p2_trans}) which in turn implies the modification in the dispersion relation
\begin{equation}
 \varepsilon^2-p^2=m^2\left(1-\frac{\varepsilon}{\kappa}\right)^2.
\end{equation}  
Consequences of the modified dispersion relations on the thermodynamics are being studied extensively to infer the effect of Planck scale physics \cite{AmelinoCamelia:2004xx, Camacho:2006qg, Gregg:2008jb, AmelinoCamelia:2009tv, Camacho:2007qy, Alexander:2001ck, Bertolami:2009wa}. The effect of modified dispersion relations in loop-quantum-gravity on black hole thermodynamics was studied in \cite{AmelinoCamelia:2004xx}. The same as a Lorentz violating phenomena on the thermodynamics of macroscopic systems (like white dwarfs) \cite {Camacho:2006qg, Gregg:2008jb, AmelinoCamelia:2009tv} and as a noncommutative phenomena on cosmology and astrophysical systems \cite{Alexander:2001ck, Bertolami:2009wa} have also been studied. Moreover, photon gas thermodynamics in the context of modified dispersion relations \cite{Camacho:2007qy} and DSR \cite{Das:2010gk} are being investigated. In \cite{Das:2010gk} the effect comes solely because of the presence of a maximum energy scale as the photon dispersion relation remains unmodified.
In this thesis we study the thermodynamics of an ideal gas in the above described DSR scenario.

The organization of the thesis is as follows:
Chapter \ref{instanton} starts with describing the GBOs.
They correspond to reducible representations of the harmonic oscillator algebra. 
We demonstrate their relevance in the construction of topologically non-trivial solutions in 
noncommutative gauge theories, focusing our attention to flux tubes, vortices, and instantons.
Our method provides a simple new relation between the topological charge and the number of times the basic irreducible representation occurs in the reducible representation underlying the GBO.
When used in conjunction with the noncommutative ADHM construction, we find that these 
new instantons are in general not unitarily equivalent to the ones currently known in literature.

Chapter \ref{qo} studies the time dependent transitions of quantum forced harmonic oscillator (QFHO) in noncommutative $\mathds{R}^{1,1}$
perturbatively to linear order in the noncommutativity $\theta$.
We show that the Poisson distribution gets modified, and that the vacuum state evolves into a ``squeezed'' state rather than a coherent state.
The time evolutions of uncertainties in position and momentum in vacuum are also studied 
and imply interesting consequences for modelling nonlinear phenomena in quantum optics.

In chapter \ref{DSR} we study thermodynamics of an ideal gas in Doubly Special Relativity.
New type of special functions (which we call Incomplete Modified Bessel functions) emerge.
We obtain a series solution for the partition function and derive thermodynamic quantities.
We observe that DSR thermodynamics is non-perturbative in the SR and massless limits.
The equation of state found makes the {\it pressure vs energy density} graph stiffer.

We conclude our results in the last chapter. We take this opportunity to mention that we have corrected some typos in the notations and some results in the published version of our papers  \cite{Acharyya:2011bx} and \cite{Chandra:2011tf}. While the other corrections are trivial (or they come in the middle of calculations) we must mention the corrected results in (\ref{uncertainties_X1})-(\ref{uncertainties_X1X2}).

\chapter{Noncommutative Vortices and Instantons from GBOs} \label{instanton}

\section{Introduction}


Detailed investigations of noncommutative gauge theories have led to the discovery of localized static 
classical solutions in noncommutative spaces \cite{Gopakumar:2000zd,Aganagic:2000mh,Gross:2000ss,Lechtenfeld:2001aw, Grisaru:2003gt, Hamanaka:2005cd,Harvey:2000jt,Harvey:2000jb,Nekrasov:2000ih}. Among the models of gauge theories in noncommutative spaces, one of the simplest is the 
Abelian-Higgs model possessing vortex--like solutions \cite{Jatkar:2000ei, Bak:2000ac, Bak:2000im, Lozano:2000qf}.
Another interesting class of solutions in noncommutative euclidean space are instantons in $U(N)$ Yang-Mills theories. 
Nekrasov and Schwarz developed a generalization of ADHM construction as given in \cite{Atiyah:1978ss} to find these noncommutative instantons \cite{Nekrasov:1998ss}. 
The $U(N)$ Yang-Mills instantons in $\mathds{R}_{NC}^4 $ and  $\mathds{R}_{NC}^2 \times\mathds{R}_{C}^2 $ were further studied in 
\cite{Furuuchi:1999kv,Kim:2000msa,Chu:2001cx,Kim:2001ai, Lechtenfeld:2001ie, FrancoSollova:2002nn,Furuuchi:2000vc}. 
Pedagogical reviews can be found in \cite{Nekrasov:2000zz, Sako:2010zza}. Apart 
from these, there are multitudes of other solutions in noncommutative gauge fields like merons \cite{FrancoSollova:2002nn}, flux tubes 
\cite{Polychronakos:2000zm}, monopoles \cite{Gross:2000wc}, dyons \cite{Hashimoto:1999zw}, 
skyrmions \cite{Ezawa:2005cj}, false vacuum bubbles \cite{Bak:2000ac}, to name just a few. 

In this chapter we present a new construction of such topological objects, based on an analysis of the 
{\it reducible} representations of the standard harmonic oscillator algebra. Our method gives rise to 
new instanton solutions (i.e. not gauge equivalent to the known ADHM instantons) on a noncommutative space, and in the process provides a 
simple interpretation for the instanton number: it simply ``counts" the number of copies of the basic 
irreducible representation. 
 
Our construction relies on operators called Generalized Bose Operators (GBO) \cite{bg,katriel} which provide
 an explicit realization of the reducible representations of the oscillator algebra, and are well-known 
in the quantum optics literature (see for examples \cite{katriel2, buzek}). As a warm-up, 
we first study the significance of GBOs in constructing fluxes and vortices with higher 
winding numbers, and then discuss instanton solutions in noncommutative Yang--Mills theories.
 
The chapter is organized as follows. We start with a brief review of GBOs and their representations in section \ref{bg_review}. In section \ref{static_soln}, 
we discuss the flux tube solutions of \cite{Polychronakos:2000zm} in the language of 
GBOs and then we go on to show the relevance of these operators in 
noncommutative Nielsen-Olesen vortices. Section \ref{instantons} discusses the noncommutative 
instantons. Using the GBO in conjunction with the ADHM construction, we construct 
a class of new instantons and compute 
their topological charges.
  Notations and formalism of the theories discussed in sections \ref{sec:flux_sec}, \ref{nov} and \ref{instantons} are all independent to each other and they should not be mixed.

\section{Generalized Bose Operators -- A Brief Review}\label{bg_review}

Brandt and Greenberg \cite{bg} give a construction of Generalized Bose Operators that 
change the number of quanta of the standard Bose operator $a$ by 2 (or more generally 
by a positive integer $k$). We briefly recall their construction in this section.

Consider the infinite-dimensional Hilbert space $\mathcal{H}$ spanned by 
a complete orthonormal basis $\{|n \rangle, n=0,1,\cdots, \infty \}$ labeled by a 
non-negative integer $n$. Vectors in $\mathcal{H}$ are of the form $|\psi \rangle = \sum_n c_n |n \rangle, c_n \in {\mathds C} \quad \forall n$ such that $\sum_n |c_n|^2 < \infty$. We formally write
\begin{equation}
\mathcal{H} = \left\{\sum_n c_n |n \rangle \, : \,\,  c_n \in {\mathds C} \, \forall n \,\, {\rm and} \, \sum_n |c_n|^2 < \infty \right\}.
\end{equation}
The standard bosonic annihilation operator $a$ acts on this basis as
\begin{equation}
 a |n\rangle = n^{\frac{1}{2}} |n-1\rangle, \quad \forall n\geq 1 \quad \mathrm{and} 
 \quad a|0\rangle=0.
\end{equation}
The annihilation operator is unbounded, and hence comes with a domain of definition:
\begin{equation}
\mathcal{D}_a = \left\{ \sum_n c_n |n \rangle \, : \,\, \sum_n n |c_n|^2 < \infty \right\}.
\end{equation}
Its adjoint $a^\dagger$ satisfies
\begin{equation}
a^\dagger |n\rangle = (n+1)^{\frac{1}{2}} |n+1\rangle, \quad \forall n \geq 0
\end{equation}
and has the same domain $\mathcal{D}_a$. The number operator 
\begin{equation} \label{number_operator}
N \equiv a^\dagger a
\end{equation} 
is defined with a domain
\begin{equation}
\mathcal{D}_N = \left\{ \sum_n c_n |n \rangle \, : \,\, \sum_n n^2 |c_n|^2 <\infty \right\}.
\end{equation}
The basis vectors $\{|n\rangle\}$ are eigenstates of $N$:
\begin{equation}
N|n\rangle = n |n\rangle .
\end{equation}
On $\mathcal{D}_N$, the operators $a$ and $a^\dagger$ satisfy
\begin{equation}
[a,a^\dagger]=1.
\label{basiccomm}
\end{equation}
While $N$ counts the number of quanta in a state, the $a$ and $a^\dagger$ 
destroy and create respectively a single quantum. They satisfy the following commutation relations
\begin{equation}
\left[N,a\right] = -a, \quad \left[N,a^\dagger\right] = a^\dagger
\end{equation}
We denote this irreducible representation of the oscillator algebra by $(a,\mathcal{H})$. This, up to a unitary equivalence, is the unique irreducible representation \cite{kirillov}.
The Hilbert space $\mathcal{H}$ can be split into two disjoint subspaces $\mathcal{H}_+$ and $\mathcal{H}_-$ given by
\begin{equation} \label{H+}
\mathcal{H}_+=\left\{\sum c_{2n}|2n\rangle  \in \mathcal{H}\right\}
\end{equation}
\begin{equation}
\mathcal{H}_- 
=\left\{\sum c_{2n+1}|2n+1\rangle  \in \mathcal{H}\right\}
\end{equation}
such that 
\begin{equation}
\mathcal{H}=\mathcal{H}_+\oplus\mathcal{H}_-
\end{equation}
The projection operators
\begin{equation}
\Lambda_+ = \sum_{n=0}^\infty |2n\rangle\langle2n| 
=\cos ^2 \left(\frac{\pi N}{2}\right),
\quad {\rm and} \quad \Lambda_- = \sum_{n=0}^\infty |2n+1 \rangle \langle 2n+1|
=\sin ^2 \left(\frac{\pi N}{2}\right)
\end{equation}
project onto the subspaces $\mathcal{H}_+$ and $\mathcal{H}_-$ respectively. One can define the operators $b_\pm$ and its adjoint $b^\dagger_\pm$ on the subspaces $\mathcal{H}_{\pm}$ by\begin{equation}
 b_+ |2n\rangle = n^{\frac{1}{2}} |2n-2\rangle, \quad   b^\dagger_+ |2n\rangle = (n+1)^{\frac{1}{2}} |2n+2\rangle,   \quad  b_+ |0\rangle =0
 \end{equation}
 \begin{equation}
b_-|2n+1\rangle = n^{\frac{1}{2}} |2n-1\rangle,  \quad b^\dagger_-|2n+1\rangle = (n+1)^{\frac{1}{2}} |2n+3\rangle, \quad b_- |1\rangle =0
\end{equation}
with domain of closure $\mathcal{D}_a \cap \mathcal{H}_{\pm}$. On the domain $\mathcal{D}_N \cap \mathcal{H}_{\pm}$ we have 
\begin{equation} \label{comm_b+-}
[b_\pm,b_\pm^\dagger ]=1.
\end{equation}
Thus $(b_-,\mathcal{H}_-)$, $(b_+,\mathcal{H}_+)$ and $(a,\mathcal{H})$ are isomorphic to 
each other. In other words, there exist unitary operators $U_\pm$ such that 
\begin{equation}
U_\pm b_\pm U_\pm^\dagger = a.
\end{equation}

Using the projection operators $\Lambda_\pm$, one can define an operator $b$ as
\begin{equation}
 b=b_+\Lambda_+ +b_-\Lambda_- 
\label{irreducible_decomp}
\end{equation}
on $\mathcal{H}$ whose action  on the basis vectors $|n\rangle$ is
\begin{equation}
 b|2n\rangle = n^\frac{1}{2} |2n-2\rangle, \quad b|2n+1\rangle = n^\frac{1}{2} |2n-1\rangle
\label{action_b}.
\end{equation}
Notice that both $|0\rangle$ and $|1\rangle$ are annihilated by $b$.
The operator $b$ satisfies the commutation relation
\begin{equation}
[N, b] =-2b.
\end{equation}
The adjoint of $b$ satisfies
\begin{equation}
 b^\dagger |2n\rangle = (n+1)^\frac{1}{2} |2n-2\rangle, \quad b^\dagger |2n+1\rangle = (n+1)^\frac{1}{2} |2n-1\rangle
\label{action_b_dagger}.
\end{equation}
and
\begin{equation}
[N, b^\dagger] =2b^\dagger.
\end{equation}
A new number operator can be defined as 
\begin{equation} \label{new_number_operator}
M=b^\dagger b= \frac{1}{2}\left(N-\Lambda_-\right)
\end{equation}
which has the same eigenstates $|n\rangle$, but each eigenvalue 
is two-fold degenerate! We denote these eigenvalues  by 
\begin{equation}
m_n=\frac{1}{2} (n-\lambda_{n-})
\end{equation}
where we define $\lambda_{n\pm}$ as
\begin{equation} \label{lambda_n+}
 \lambda_{n+} = \langle 
n|\Lambda_+|n\rangle = \cos^2\left(\frac{n\pi}{2}\right) = \left\{
\begin{array}{cl}
1 & \mbox{{\rm if n is even}} \\
0 & \mbox{{\rm if n is odd}}
\end{array} \right.
\end{equation}
\begin{equation} \label{lambda_n-}
 \lambda_{n-} = \langle 
n|\Lambda_-|n\rangle = \sin^2\left(\frac{n\pi}{2}\right) = \left\{
\begin{array}{cl}
0 & \mbox{{\rm if n is even}} \\
1 & \mbox{{\rm if n is odd}}
\end{array} \right.
\end{equation}
We rewrite the action of $b$ and $b^\dagger$ in a compact form
\begin{equation} \label{action_b_1}
 b|n\rangle=m_n^\frac{1}{2}|n-2\rangle \quad \mathrm{and} \quad  b^\dagger|n\rangle=(m_n+1)^\frac{1}{2}|n+2\rangle.
\end{equation}
The operators $b$ and $b^\dagger$ have $\mathcal{D}_a$ as their domain of closure and they satisfy
\begin{equation} \label{basic_comm_k=2}
[b,b^\dagger] =1
\end{equation} 
in the domain $\mathcal{D}_N$. Hence ($b,\mathcal{H}$) forms a reducible representation of the oscillator algebra characterized by (\ref{basiccomm}) having (\ref{irreducible_decomp}) as its irreducible 
decomposition.

One can generalize the above construction and formulate an operator $b^{(k)}$ which lowers a state 
$|n\rangle$ by $k-$steps. First of all we note that $\mathcal{H}$ can be split as
\begin{equation}
\mathcal{H} = \mathcal{H}_0^{(k)} \oplus \mathcal{H}_1^{(k)} \oplus ... \oplus \mathcal{H}_{k-1}^{(k)}
\end{equation}
for an integer $k$ where $\mathcal{H}_i^{(k)}$'s are defined as
\begin{equation}
\mathcal{H}_i^{(k)} =\left\{\sum_n c_{kn+i} |k n +i\rangle \in \mathcal{H}\right\}.
\end{equation}
We define projection operators $\Lambda_i^{(k)}$ as
\begin{equation}
\Lambda_i^{(k)} = \sum_{n=0}^\infty |k n +i\rangle \langle k n + i |, \quad i = 0,1, \cdots k-1.
\end{equation} 
which project onto the subspace $\mathcal{H}_i^{(k)}$.
In each subspace $\mathcal{H}_i^{(k)}$, one can define operators $b_i^{(k)}$ and their adjoints 
$b_i^{(k)\dagger}$ that satisfy
\begin{equation}
\left[b_i^{(k)}, b_i^{(k)\dagger}\right]=1
\end{equation}
and hence correspond to the UIR of the oscillator algebra. They act on the states as
\begin{equation}
b_i^{(k)} |kn+i\rangle = \sqrt{n} |kn+i-k\rangle, \quad b_i^{(k)\dagger} |kn+i\rangle = \sqrt{n+1} |kn+i+k\rangle
\end{equation}
We can easily check that
\begin{equation}
\left[b_i^{(k)},\Lambda_i^{(k)}\right]= \left[b_i^{(k)\dagger},\Lambda_i^{(k)}\right]=0
\end{equation}
A reducible representation is given by
\begin{equation}
 b^{(k)} = \sum_{i=0}^{k-1} b_i^{(k)} \Lambda_i^{(k)}, \quad b^{(k)\dagger} = \sum_{i=0}^{k-1} b_i^{(k)\dagger} \Lambda_i^{(k)}.
\label{b_red}
\end{equation}
They satisfy the oscillator algebra commutation relation
\begin{equation}
[b^{(k)},b^{(k)\dagger}]=1
\end{equation}
on the domain $\mathcal{D}_N$. Thus $(b_i^{(k)},\mathcal{H}_i^{(k)})$ for $i=0,1,...,k-1$ are isomorphic to $(a,\mathcal{H})$ and hence $(b^{(k)},\mathcal{H})$ 
forms a reducible representation of the oscillator algebra.
Discussions from (\ref{H+}) to (\ref{basic_comm_k=2}) represent the case $k=2$, the simplest non-trivial example of this construction.
Henceforth we will use $b$ for $b^{(2)}$. An explicit expression for $b$ is \cite{katriel} 
\begin{equation}
 b= \frac{1}{\sqrt{2}} \left( a\frac{1}{\sqrt{N}}a\Lambda_+ +a\frac{1}{\sqrt{N+1}}a\Lambda_-\right)
\label{expression_b}
\end{equation}

Before we end this section, let us point out a minor generalization of the 
Brandt-Greenberg construction. Under any unitary transformation $U_\pm$ defined on $\mathcal{H}_\pm$ that transforms $b_\pm$ as
$b_\pm \rightarrow U_\pm b_\pm U_\pm^\dagger$, the fundamental commutation relation (\ref{comm_b+-}) remains unchanged. In particular, if we choose 
\begin{equation}
U_\pm(z_\pm) = e^{z_\pm b_\pm^\dagger - \bar{z}_\pm b_\pm},
\end{equation} 
then we find that 
\begin{equation}
b_\pm(z_\pm) \equiv U_\pm(z_\pm) b_\pm U_\pm^\dagger(z_\pm) = b_\pm - z_\pm,
\end{equation}
i.e., we get the ``translated" annihilation operator. One can construct a reducible representation using 
$b_+(z_+)$ and $b_-(z_-)$ as
\begin{equation}
b(z_+,z_-) = b_+(z_+) \Lambda_+ + b_-(z_-) \Lambda_- = b_+ \Lambda_+ + b_-\Lambda_- -z_+\Lambda_+ -z_-\Lambda_-.
\end{equation}
Here $b$  gets translated by different amounts in different subspaces $\mathcal{H}_{\pm}$ and the ``translated'' operator 
$b(z_+,z_-)$ is unitarily related to $b$ as
\begin{equation}
b(z_+,z_-) =U(z_+,z_-)b U^\dagger(z_+,z_-),  \quad U (z_+,z_-) = U_{+}(z_+)\Lambda_{+}+U_{-}(z_-)\Lambda_{-}.
\label{translation_operator}
\end{equation}
More generally, using (\ref{b_red}) we can write
\begin{equation}
b^{(k)}(z_0,z_1,\cdots,z_{k-1}) = b^{(k)} - \sum_{i=0}^{k-1} z_i \Lambda_i^{(k)}
\end{equation}
and the unitary operator is 
\begin{equation}
U(z_0,z_1,\cdots,z_{k-1}) = \sum_{i=0}^{k-1} U_{i}(z_i) \Lambda_i^{(k)}.
\end{equation}
Though minor, this generalization will play a role in the construction of noncommutative 
multi-instantons.

There exist other possibilities as well. For example, choosing 
\begin{equation}
U_\pm(\xi_\pm) = e^{\frac{\xi_\pm}{2}(b_\pm^2 -b_\pm^{\dagger 2})}
\end{equation}
gives
\begin{equation}
b_\pm(\xi_\pm) = b_\pm \cosh \xi_\pm + b_\pm^\dagger \sinh \xi_\pm.
\end{equation}
The above is the well known squeezed annihilation operator. A reducible representation may be constructed:
\begin{equation}
b(\xi_+,\xi_-) = b_+(\xi_+)\Lambda_+ + b_-(\xi_-) \Lambda_-
\end{equation}
and more generally,
\begin{equation} \label{reducible_squeeze}
b(\xi_0,\xi_1, \cdots,\xi_{k-1}) = \sum_{i=0}^{k-1} b_i^{(k)}(\xi_i)\Lambda_i^{(k)}
\end{equation}
We discussed this ``squeezing" operators just to give an example and will not study it in detail as they are not directly used further.

Fields in noncommutative Moyal space  are generally interpreted in terms of the simple harmonic oscillator algebra elements. Having found the reducible representation of this algebra in terms of the so-called GBOs, we try to seek nontrivial solutions of field theories on noncommutative spaces in terms of these operators. 

\section{Static Solutions In Noncommutative Gauge Theories }\label{static_soln}

As we are interested in exploring the relevance of the GBOs $b^{(k)}$ 
in noncommutative field theories, we  start with the following two simple situations where this relevance is most visible:
\begin{itemize}
\item the flux tube solution in $(3+1)-$dimensional pure gauge theory \cite{Polychronakos:2000zm}
\item the vortex solution in $(2+1)-$dimensional abelian Higgs model \cite{Jatkar:2000ei}.
\end{itemize}
We find a new interpretation for the already known solutions of the above theories in terms of the GBOs.

\subsection{Flux Tube Solution In Noncommutative Gauge Theories}{\label{sec:flux_sec}}
Consider pure $U(1)$ gauge theory in $(3+1)$-dimensional spacetime with only 
spatial noncommutativity. This theory incorporates magnetic flux tube solutions \cite{Polychronakos:2000zm} which  are important in the context of monopoles and strings discussed in \cite{Gross:2000wc}. 
We will search for non-trivial solutions of the static equation 
of motion. These solutions do not  possess a smooth 
$\theta\rightarrow 0$ limit \cite{Polychronakos:2000zm}, implying that they have 
no commutative counterpart, i.e, the origin of this effect is entirely due  the noncommutativity of 
the underlying space.   
The noncommutativity is only among the space coordiantes  (time is commutative)
\begin{equation}
\left[\hat{x}^i,\hat{x}^j\right] = i\theta^{ij}; \quad \quad i,j=1,2,3.
\end{equation}
The anti-symmetry of the real $\theta$-matrix, i.e., 
\begin{equation}
\theta^{ij} = -\theta^{ji}
\end{equation}
guarantees a choice of axes in which the noncommutativity becomes 
\begin{equation}
 [\hat{x}^1,\hat{x}^2]=i\theta, \quad [\hat{x}^1,\hat{x}^3]=0, \quad [\hat{x}^2,\hat{x}^3]=0
\end{equation}
so that only the $\hat{x}^1-\hat{x}^2$ plane is noncommutative. 
On a noncommutative space, ``functions" are elements of the noncommutative algebra generated by the operators $\hat{x}^i$. 
Derivatives in the $\hat{x}^1$ and $\hat{x}^2$ 
directions are defined via the adjoint action
\begin{equation}
 \partial_{x^1}f=\frac{i}{\theta}\left[\hat{x}^2,f\right], \quad \partial_{x^2}f=-\frac{i}{\theta}\left[\hat{x}^1,f\right]
\label{derivative}
\end{equation}
while the derivatives in the $\hat{x}^3$ and $t$ directions are the same as in the commutative case.
We can  define  a set of complex (noncommuting) variables $z$ and $\bar{z}$ and a set of creation-annihilation operators as
\begin{equation} \label{z_zbar}
 z= \frac{1}{\sqrt{2}}(\hat{x}^1+i\hat{x}^2),  \quad  \bar{z}= \frac{1}{\sqrt{2}}(\hat{x}^1-i\hat{x}^2), \quad a=\frac{1}{\sqrt{\theta}}z,
\quad  a^\dagger=\frac{1}{\sqrt{\theta}}\bar{z}
\end{equation}
Here $a$ and $a^\dagger$ satisfy (\ref{basiccomm}). With this convention, the derivatives with respect to the complex coordinates are 
given as
\begin{equation} \label{derivative_z}
 \partial_{z}f=-\frac{1}{\sqrt{\theta}}\left[a^\dagger,f\right], \quad \partial_{\bar{z}}f=\frac{1}{\sqrt{\theta}}
\left[a,f\right].
\end{equation}
Integration on $x^1-x^2$ plane is replaced by trace over Fock space $\mathcal{H}$:
\begin{equation} \label{integration_trace}
 \int dx^1 dx^2 f(x^1,x^2) \rightarrow 2\pi\theta \, \mathrm{Tr}_{\mathcal{H}}\hat{f}(\hat{x}^1,\hat{x}^2) = 2\pi\theta \sum_{n=0}^\infty \langle n | \hat{f}(\hat{x}^1,\hat{x}^2) | n \rangle.
\end{equation}
where $|n\rangle$'s are the number eigenstates and the factor $2\pi\theta$ ensures the proper commutative limit ($\theta\rightarrow0$). 

We will follow the construction of gauge theories on a noncommutative space as given in \cite{Polychronakos:2007df}.
In accordance with the commutative theory the field strength is defined as
\begin{equation}
\hat{F}_{\mu\nu}=[\mathcal{D}_{\mu},\mathcal{D}_{\nu}]
\end{equation}
where the noncommutative covariant derivative $\mathcal{D}_\mu$ is defined as 
 \begin{equation}
  \mathcal{D}_{0} = \partial_t + A_{0}, \quad   \mathcal{D}_{3} = \partial_{x^3} + A_{3}, \quad  \mathcal{D}_{1} = \frac{i}{\theta} \hat{x}^2 + A_{1}, \quad  \mathcal{D}_{2} = -\frac{i}{\theta} \hat{x}^1 + A_{2}.
 \end{equation}
 Here $A_\mu$ is an anti-hermitian operator field
 \begin{equation}
 A_{\mu}^\dagger = -A_{\mu}
 \end{equation}
The gauge transformation is such that it transforms $\mathcal{D}_\mu$ covariantly
\begin{equation}
\mathcal{D}_\mu \rightarrow U\mathcal{D}_\mu U^{-1}
\end{equation}
Noncommutative gauge theory has an alternative (and possibly more natural) formulation in terms 
of $\mathcal{D}_{\mu}$ rather than $A_{\mu}$. In terms of $\mathcal{D}_{\mu}$, for a pure gauge field, one works with 
the action 
\begin{equation}
 \hat{S} = -\frac{\pi\theta}{2g^2}\int dx^3 dt \, \mathrm{Tr} \lbrace\hat{F}^2_{\mu\nu}\rbrace\end{equation}
This formulation of noncommutative gauge theory is classically equivalent to the standard formulation where action is given by
\begin{equation}
S=-\frac{\pi\theta}{2g^2}\int dx^3 dt \, \mathrm{Tr} \lbrace F^2_{\mu\nu}\rbrace
\end{equation}
with standard field strength given by
\begin{equation}
F_{\mu\nu} = \partial_\mu A_\nu - \partial_\nu A_\mu + [A_\mu,A_\nu].
\end{equation}
The derivatives for $\mu,\nu = 1,2$ are defined in (\ref{derivative}). One can verify
\begin{equation}
\hat{F}_{\mu\nu} = F_{\mu\nu} + \omega_{\mu\nu}, \quad \quad \omega_{\mu\nu} = \frac{i}{\theta} \epsilon_{0\mu\nu 3}.
\end{equation}
The anti-symmetric Levi-civita tensor is defined as
\begin{equation}
\epsilon^{\mu\nu\rho\sigma} = \left\{
\begin{array}{cl}
1 & \mbox{{\rm if $\mu\nu\rho\sigma$ is in even permutation of 0123}} \\
-1 & \mbox{{\rm if $\mu\nu\rho\sigma$ is in odd permutation of 0123}} \\
0 & \mbox{{\rm otherwise}}
\end{array}
\right.; \quad \quad \epsilon_{\mu\nu\rho\sigma} = -\epsilon^{\mu\nu\rho\sigma}.
\end{equation}
This also implies
\begin{equation}
\omega_{\mu\nu} = -\omega^{\mu\nu}
\end{equation}
Now it is just a matter of simple algebra to see that
\begin{equation}
\hat{S}=S +\frac{\pi}{g^2 \theta}\int dt dx^3\, \mathrm{Tr} \left(1+ 2i\theta F_{12}\right).
\end{equation}
While each term of $F_{12}$ is a commutator or a total derivative and hence will contribute only to boundary terms, the term with $\mathrm{Tr}\,1$ is only an infinite irrelevant term. As the variations in $\mathcal{D}_\mu$ and $A_\mu$ are same, both $\hat{S}$ 
and $S$ give the same equations of motion and hence are classically equivalent! 

For static, magnetic configurations $(\partial_t=0 = A_{0})$ and with the choice 
$\partial_{3}A_{i}=0, A_{3}=0$, the equations of motion in terms of $\mathcal{D}$'s reduce to
\begin{equation}
 [\mathcal{D},[\mathcal{\bar{D}},\mathcal{D}]]=0, \quad \quad \mathrm{with} \quad \mathcal{D}=\frac{1}{\sqrt{2}}( \mathcal{D}_{1}+i\mathcal{D}_{2}), 
\quad \mathcal{\bar{D}}=\frac{1}{\sqrt{2}}( \mathcal{D}_{1}-i\mathcal{D}_{2}) = -\mathcal{D}^\dagger.
\label{eom_flux}
\end{equation}
It is easy to check that the standard ``vacuum" configuration corresponding to $A_1=A_2=0$ (and hence $F_{\mu\nu} = 0$) will satisfy the equation of motion (\ref{eom_flux}) and corresponds to
\begin{equation}
\mathcal{D}=\frac{a}{\sqrt{\theta}}.
\end{equation} 
We can construct solutions about this vacuum by taking a rotationally invariant ansatz  $\mathcal{D}=af(N)$ and it 
can be shown that there exists a  solution of the form (see Appendix \ref{flux_tube_calculation} for a detailed calculation)
\begin{equation} \label{polychronakos_sol}
 \mathcal{D}= \frac{a}{\sqrt{\theta}}\sqrt{\frac{N-n_0}{N}}\sum_{n=n_0}^\infty |n\rangle \langle n|, \quad N=a^\dagger a, \quad n_0 =0,1,2.... 
\end{equation}
with $N$ being the number operator as given in (\ref{number_operator}). This solution corresponds to
\begin{equation}
\hat{F}_{\mu\nu} = \omega_{\mu\nu} \sum_{n=n_0}^{\infty} |n\rangle \langle n|.
\end{equation}
The standard field strength becomes
\begin{equation} \label{F_mu_nu}
F_{\mu\nu} = \hat{F}_{\mu\nu} - \omega_{\mu\nu} = - \omega_{\mu\nu} \sum_{n=0}^{n_0-1} | n \rangle \langle n |.
\end{equation}
Thus it represents a  classical localized static circular magnetic flux tube in $x^3$ direction centred about origin of the $(x^1,x^2)$ 
plane with $n_0$ related to its radius. The total magnetic flux $\Phi$ gets quantized
\begin{equation}
\Phi = 2 \pi \theta\, \mathrm{Tr} (i F_{12}) = -2 \pi n_0.
\end{equation}
The choice $n_0 =0$ corresponds to the vacuum configuration and has zero magnetic flux.

By virtue of the relation (\ref{basic_comm_k=2}),
\begin{equation} \label{D_b}
\mathcal{D}=\frac{b}{\sqrt{\theta}}
\end{equation} 
is also a vacuum solution (though reducible) of (\ref{eom_flux}).  
Again, we can start with the ansatz $\mathcal{D}= b \, G(N)$ to construct the following solution (see Appendix \ref{flux_tube_calculation} for a detailed calculation)
\begin{equation} \label{flux_tube_sol_b}
 \mathcal{D}= \frac{b}{\sqrt{\theta}}\sqrt{\frac{N-n_{0+}\Lambda_+ -n_{0-} \Lambda_-}{2M}}
\left(\sum_{n=n_{0+}}^\infty |n\rangle \langle n|\Lambda_+ +\sum_{n=n_{0-}}^\infty |n\rangle \langle n|\Lambda_-\right).
\end{equation}
Here $n_{0+} = 0,2,4,...$ and $n_{0-} = 1,3,5,...$.
This solution can be re-written in the form 
\begin{equation}
\mathcal{D}= \sum_{n=0}^\infty g(n)|n\rangle \langle n+2|
\end{equation}
with
\begin{equation}
g(2n) = \sqrt{n-n_{0+}^{\prime} +1} \quad \quad \mbox{{\rm for}}\,\, n\geq n_{0+}^{\prime}, \quad \quad
n_{0+}^{\prime}= \frac{n_{0+}}{2} 
\end{equation} 
\begin{equation}
g(2n+1) =\sqrt{n-n_{0-}^{\prime} +1} \quad \quad \mbox{{\rm for}} \,\,   n\geq n_{0-}^{\prime}, \quad \quad n_{0-}^{\prime}= \frac{n_{0-}-1}{2}.
\end{equation}
This solution is same as the higher moment solution obtained in \cite{Polychronakos:2000zm} starting with the ansatz $\mathcal{D}=F(N)a^2$. 
Again, for the choice $n_{0+}=0$ and $ n_{0-}=1$, this solution reduces to (\ref{D_b}).

 Furthermore,  the rotationally invariant ansatz of $\mathcal{D}= f(n)b^{(k)} $ about the reducible ``vacuum"
 \begin{equation}
 \mathcal{D}=\frac{b^{(k)}}{\sqrt{\theta}}
 \end{equation}
 for $k\geq 2$ gives the following solution of the equation of motion
 \begin{equation} \label{flux_tube_sol_bk}
 \mathcal{D}= \frac{b^{(k)}}{\sqrt{\theta}}\sqrt{\frac{N-\displaystyle{\sum_{i=0}^{k-1}} n_{0i}\Lambda_i^{(k)}}{2M^{(k)}}}
\sum_{i=0}^{k-1}\sum_{n=n_{0i}}^\infty |n\rangle \langle n|\Lambda_i^{(k)},\quad M^{(k)}= b^{(k)\dagger}b^{(k)}
\end{equation}
with $n_{0i} = kn_0+i, n_0=0,1,2,...$. These solutions are also same as the higher moment solutions obtained from the ansatz $\mathcal{D}=g(n)a^k$ in \cite{Polychronakos:2000zm}. Further computation of the magnetic field and other things for the above solution can be found in \cite{Polychronakos:2000zm}.
 These solutions with the GBO represent static magnetic flux tubes with localized flux, with size of the tube in the $i$-th irreducible part of the Fock space related to $n_{0i}$. Thus the radial profile of the magnetic field is determined by the set $\lbrace n_{0i}\rbrace$. 

Let us try to understand the above construction in the following manner: 
The flux tube solution using irreducible representation $(a,\mathcal{H})$ can be described by a single integral index $n_0$ which is related to the extension of the magnetic field in the Fock space (see (\ref{F_mu_nu})).
The use of reducible representation $\left(b^{(k)}, \mathcal{H}\right)$ separates the Fock space in $k$-parts, resulting in $k$ numbers of indices ${n_{0i}}$ each of which is related to the extension of magnetic field in the corresponding subspace of the Fock space.
Increase in the number of indices (charges) clearly expand the moduli space of the static magnetic flux tube solutions!
These solutions are nonperturbative in $\theta$ as the reducibility of the Fock space and hence the notion of GBOs are typically attributed to noncommutative spaces!

In this section we saw that there already exist  certain solutions of noncommutative gauge theory which can be re-written in terms of the GBOs. 
This fact shows the importance of GBOs and motivates us to seek such solutions in other noncommutative gauge theories.

\subsection{Noncommuative Abelian Higgs Model - Nielsen-Olesen Vortex Solution \label{nov}}

The abelian Higgs model in noncommutative spaces is of some interest because of its simplicity 
as a noncommutative gauge theory and the existence of vortex solutions. 
Various topologically non-trivial vortex solutions in this context have been studied in detail. 

An interesting class of vortex solutions in this 
theory is studied in  \cite{Jatkar:2000ei}, which are analogous to the Nielsen-Olesen Vortices in the commutative space \cite{Nielsen:1973cs}.
The model is in $(2+1)-$dimensions, and consists of a complex Higgs field $\Phi$ which is a left gauge module (the gauge fields multiply the 
complex Higgs field $\Phi$ from left and $\bar{\Phi}$ from right). Minimizing the static noncommutative energy functional,
 Bogomolnyi equations are generalized to the noncommutative space and  $1/\theta$-expansion is done in large $\theta$ limit. 
The equations are then solved order by order and the corrections to the leading order equation converge rapidly. In the large distance limit 
(which is the commutative limit in this case), the solution reduces to the Nielsen-Olesen vortex solution in ordinary (commutative) 
abelian Higgs model. Here we will give a brief formalism of the standard noncommutative theory. For extensive calculation one can refer to \cite{Jatkar:2000ei}.

The noncommutativity is same as that in section (\ref{sec:flux_sec}), with the only difference that now the space is 
$2-$dimensional, so the direction $x^3$ is absent:
\begin{equation}
 [\hat{x}^1,\hat{x}^2]=i\theta.
\end{equation}
The notations defined in (\ref{derivative}), (\ref{z_zbar}), (\ref{derivative_z}) and (\ref{integration_trace}) remain intact.
The energy functional in the  static 
configuration is given by \cite{Jatkar:2000ei}
\begin{equation}
 \mathcal{E} = \mathrm{Tr} \left[\frac{1}{2} (B+ \Phi\bar{\Phi}-1)^2+\mathcal{D}_{\bar{z}}\Phi\mathcal{D}_z\bar{\Phi} +
\mathcal{T}\right]
\label{energy_func}
\end{equation}
where $\mathcal{D}$ is a covariant derivative with the gauge field $A$ and is given by
\begin{eqnarray}
\mathcal{D}_z \Phi = \partial_z \Phi - i A_z \Phi, && \mathcal{D}_{\bar{z}} \Phi = \partial_{\bar{z}} \Phi - i A_{\bar{z}} \Phi \nonumber \\
\mathcal{D}_z \bar{\Phi} = \partial_z \bar{\Phi} + i \bar{\Phi} A_z, && \mathcal{D}_{\bar{z}} \bar{\Phi} = \partial_{\bar{z}} \bar{\Phi} + i \bar{\Phi} A_{\bar{z}}.
\end{eqnarray}
Note the left and right actions of the gauge fields on $\Phi$ and $\bar{\Phi}$ respectively. The magnetic field $B$ is defined as 
\begin{equation}
B=-i(\partial_z 
A_{\bar{z}}-\partial_{\bar{z}}A_z)-[A_z,A_{\bar{z}}]
\end{equation}
With the abuse of notation a dimensionful quantity $\beta = \frac{2\lambda}{e^2}$ ($e$ and $\lambda$ are the coefficients of gauge coupling and the self-coupling of the Higgs fields respectively) has been set to $1$. For more details see \cite{Jatkar:2000ei}.
$\mathcal{T}$ is the topological term defined as
\begin{equation}
 \mathcal{T}= \partial_m S^m-i[A_m,S^m] + B.
\end{equation}
where $m$ takes the values $z$ and $\bar{z}$. Here 
\begin{equation}
S^m= \frac{i}{2} \epsilon^{mn} (\Phi \mathcal{D}_n\bar{\Phi}- \mathcal{D}_n \Phi\bar{\Phi}).
\end{equation}
with the convention $\epsilon^{z\bar{z}} = 1$. It can  be 
shown that $\mathrm{Tr}\mathcal{T}$ corresponds to the topological charge.
Our prime interest is to study the Bogomolnyi equations. Minimizing (\ref{energy_func}) we get the following operator 
equations:
 \begin{equation}
 \mathcal{D}_{\bar{z}}\Phi=0, \quad \mathcal{D}_z\bar{\Phi}=0, \quad B=1-\Phi\bar{\Phi}.
 \end{equation}
which are the noncommutative Bogomolnyi equations.
Now one can do a $1/\theta$ expansion of the Higgs and the gauge fields, in the large $\theta$ limit 
\begin{eqnarray}
 \begin{array}{lll}
  \Phi &=& \Phi_\infty +\frac{1}{\theta}\Phi_{-1}+..... \\
   \bar{\Phi} &=& \bar{\Phi}_\infty +\frac{1}{\theta}\bar{\Phi}_{-1}+..... \\
     A_z &=& \frac{1}{\sqrt{\theta}} (A_\infty+\frac{1}{\theta}A_{-1}+.....) \\
     A_{\bar{z}} &=& \frac{1}{\sqrt{\theta}} (\bar{A}_\infty+\frac{1}{\theta}\bar{A}_{-1}+.....)
 \end{array}
\end{eqnarray}
The factor of $\frac{1}{\sqrt{\theta}}$ is used for scaling the variables $A$ as it is
a $1$-form. This makes sure that the fields $\Phi_\infty$ and $A_\infty$ are dimensionless. From the definition of the magnetic field we see
\begin{equation}
B =\frac{1}{\theta} (B_\infty+\frac{1}{\theta}B_{-1}+.....)
\end{equation}
with
\begin{eqnarray}
B_\infty & = & i\left(\left[a,A_\infty\right]+\left[a^\dagger,\bar{A}_\infty\right]\right) - \left[A_\infty, \bar{A}_\infty\right] \label{B_infinity} \\
B_{-1} &=& i\left(\left[a,A_{-1}\right]+\left[a^\dagger,\bar{A}_{-1}\right]\right) - \left(\left[A_\infty,\bar{A}_{-1}\right]+\left[A_{-1},\bar{A}_\infty\right]\right).
\end{eqnarray}
With this expansion, we can get the leading order $O(0)$ Bogomolnyi equation as
\begin{equation}
 \Phi_\infty\bar{\Phi}_\infty=1.
\label{eom_lead}
\end{equation}
This equation admits a solution \cite{Witten:2000nz} 
\begin{equation}
 \Phi_\infty= \frac{1}{\sqrt{a^n a^{\dagger n}}}a^n, \quad \bar{\Phi}_\infty=a^{\dagger n} \frac{1}{\sqrt{a^n a^{\dagger n}}}
\label{witten_vortex}
\end{equation}
which represents an $n$-vortex at origin. A more general solution is discussed in \cite{Jatkar:2000ei} which represents $n$ single 
vortices at $n$ different points in the noncommutative plane and (\ref{witten_vortex}) is a special case of that general solution. But for our discussion, (\ref{witten_vortex}) 
is sufficient and due to its simple form, computation and understanding becomes easier. For the solution (\ref{witten_vortex}) we can check
\begin{equation} \label{Phibar_Phi}
\bar{\Phi}_\infty \Phi_\infty = \sum_{m=n}^{\infty} |m\rangle\langle m| \neq 1.
\end{equation}
The next order Bogomolnyi equations become
\begin{equation}
 [a,\Phi_\infty]=i\bar{A}_\infty \Phi_\infty, \quad \left[a^\dagger,\bar{\Phi}_\infty\right] = i\bar{\Phi}_\infty A_\infty
\label{eom_sublead}
\end{equation}
which can be solved to get (for details see  appendix \ref{det_gauge_field})
\begin{equation} \label{NO_gauge_fields}
 \bar{A}_\infty = -i \frac{1}{\sqrt{N+1}} a\left(\sqrt{N}-\sqrt{N+n}\right), \quad {A}_\infty= i \left(\sqrt{N}-\sqrt{N+n}\right) a^\dagger  \frac{1}{\sqrt{N+1}} 
\end{equation}
$N$ being the number operator (\ref{number_operator}).

In the coherent state $|\omega\rangle $ ($a|\omega\rangle=\omega|\omega\rangle$), the expectation of the field $\Phi_\infty$ is 
\begin{equation}
 \langle\omega|\Phi_{\infty}| \omega \rangle= \omega^n \langle\omega|\frac{1}{\sqrt{a^n a^{\dagger n}}}| \omega \rangle
\quad \mathrm{with} \quad \omega=|\omega| e^{i \varphi}.
\end{equation}
The phase dependence is $ e^{i n\varphi}$, which comes solely from  $\omega^n$ as the other factor
$\langle\omega|\frac{1}{\sqrt{a^n a^{\dagger n}}}$ $| \omega \rangle$ is purely real, signifying a vortex in the noncommutative plane.
The large distance behavior is given by the  large 
$\omega $ limit or equivalently  large $\langle N\rangle $ limit \cite{Jatkar:2000ei}. The coherent state expectations in this limit becomes (for details see appendix \ref{l_d_b})
\begin{eqnarray}
 \langle\omega|\Phi_{\infty}| \omega \rangle \approx e^{i n\varphi},\quad \langle\omega|\bar{A}_\infty| \omega \rangle
\approx  i\frac{n}{2\bar{\omega}},\quad \langle\omega|A_\infty| \omega \rangle\approx - i\frac{n}{2\omega},
\label{large_distance1}
\end{eqnarray}
which  is exactly like the commutative $n$-Nielsen-Olesen Vortex.

It is interesting to note that the leading order magnetic field is (for details see \ref{appendix_B})
\begin{equation} \label{B_witten}
B_\infty= n|0\rangle\langle 0|
\end{equation}
which means that the magnetic 
field of the solution is localized and the magnetic fluxes are confined. The flux (trace of the field) is quantized and is characterized by the integer $n$. 

We try to seek the vortex solutions in terms of the GBOs. One can easily check that
\begin{equation}
 \Phi^{new}_\infty= \frac{1}{\sqrt{b^n b^{\dagger n}}}b^n, \quad \bar{\Phi}_\infty^{new}= b^{\dagger n} \frac{1}{\sqrt{b^n b^{\dagger n}}}
\label{new_vortex}
\end{equation}
satisfies (\ref{eom_lead}). This gives (details in appendix \ref{det_gauge_field})
 \begin{equation}
 \bar{A}^{new}_\infty=-i\left(a- \frac{1}{\sqrt{b^n b^{\dagger n}}}b^nab^{\dagger n}\frac{1}{\sqrt{b^n b^{\dagger n}}}\right), 
\quad {A}^{new}_\infty= i\left(a^\dagger- \frac{1}{\sqrt{b^n b^{\dagger n}}}b^na^\dagger b^{\dagger n}\frac{1}{\sqrt{b^n b^{\dagger n}}}\right).
\label{new_gauge_field}
\end{equation}
The new solutions satisfy
\begin{equation} \label{Phibar_Phi_new}
\bar{\Phi}_\infty \Phi_\infty = \sum_{m=2n}^{\infty} |m\rangle\langle m| \neq 1.
\end{equation}
The expectation value in the coherent state $|\omega\rangle $ (eigenstate of $a$) gives a phase dependence of $e^{i 2n\varphi}$ 
(as $\Phi^{new}_\infty$ can always be reduced to the form $F(N)a^{2n}$ ), a characteristic feature of $2n$ vortex in noncommutative
plane. In the  large $\omega $ limit it gives the large distance behavior:
\begin{eqnarray}
 \langle\omega|\Phi^{new}_{\infty}| \omega \rangle \approx e^{i 2n\varphi},\quad \langle\omega|\bar{A}^{new}_\infty| \omega \rangle
\approx  i\frac{n}{\bar{\omega}},\quad \langle\omega|A^{new}_\infty| \omega \rangle\approx - i\frac{n}{\omega}
\label{large_distance2}
\end{eqnarray}
which is exactly the commutative $2n$ Nielsen-Olesen vortex. One can also calculate the magnetic field for the new solution to be (see Appendix \ref{appendix_B}).
\begin{equation}
B_\infty^{new} = 2n |0\rangle \langle 0|.
\end{equation}

Till now all the calculated properties of the new solutions matched exactly with those of the Witten's solution (\ref{witten_vortex}) of vortex number $2n$. This stimulates us to compare the expression of the new solution (\ref{new_vortex}) with that of the Witten's solution. For the simplicity of 
expression and better understanding of the underlying algebra, we take $n=1$ in (\ref{new_vortex}). Using the explicit expression of 
the GBO $b$, the new vortex solution can be
written as  
\begin{eqnarray}
  \Phi^{new}_\infty =& \frac{1}{\sqrt{bb^\dagger}} b 
&= \frac{1}{\sqrt{M+1}} \frac{1}{\sqrt{2}} \left(a\frac{1}{\sqrt{N}}a \Lambda_+ + a\frac{1}{\sqrt{N+1}}a\Lambda_-\right)
\label{reduction1} 
\end{eqnarray}
$M$ being the reducible number operator (\ref{new_number_operator}). Further simplification can be done and the expression (\ref{reduction1}) for the new vortex reduces to 
\begin{equation}
 \Phi^{new}_\infty=\left(\frac{1}{\sqrt{N-\Lambda_-+2}}\frac{1}{\sqrt{N+1}}\Lambda_++\frac{1}{\sqrt{N-\Lambda_-+2}}\frac{1}{\sqrt{N+2}}
\Lambda_-\right) a^2.
\label{reduction2} 
\end{equation}
The eigenvalues of the projection operators $\Lambda_{\pm}$ are $0$ and $1$ and they never contribute simultaneously.
Owing to this fact (also keep in mind that $\Lambda_\pm$ commute with $N$) the expression 
(\ref{reduction2}) simplifies to 
\begin{equation}
 \Phi^{new}_\infty=\frac{1}{\sqrt{N+2}}\frac{1}{\sqrt{N+1}}(\Lambda_++\Lambda_-) a^2
=\frac{1}{\sqrt{(N+1)(N+2)}} a^2
=\frac{1}{\sqrt{a^2 a^{\dagger 2}}}a^2
\end{equation}
which is same as the $n=2$  Witten's vortex. 
 This calculation can be generalized for any $n$ and it can be always shown that 
$n$-new vortex solution is same as the $2n$-Witten vortex for all $n$.
It is also easy to show that $\Phi_\infty=\frac{1}{\sqrt{(b^{(k)})^n(b^{(k)\dagger})^n}} (b^{(k)})^n $ is also a solution of (\ref{eom_lead}) 
and this solution is same as the $kn$-Witten vortex. Note that the leading order gauge field and hence the magnetic field are determined uniquely by (\ref{expression_gauge_field}) and (\ref{B_infinity}) for a given Higgs field. Hence equality of $(\Phi_\infty^{new}, \bar{\Phi}_\infty^{new})$ to $(\Phi_\infty, \bar{\Phi}_\infty)$ with vortex number $kn$ ensures that the new gauge fields and the magnetic field are also equal to the old ones.


\section{Instantons}\label{instantons}

\subsection{Instantons In Commutative Gauge Theories}
Instantons are localized finite action solutions of the classical Euclidean field equations of a theory (for a review see \cite{Rajaraman:1982is}).
The finite action condition is satisfied only if the Lagrangian density of the theory vanishes at boundary.
This in turn can lead to different topological configurations of the field characterized by its 
``topological charge''.
For  Yang-Mills theories,  the instantons  are further classified as 
Self-Dual (SD) or Anti-Self-Dual (ASD) with their topological charges having opposite signs.
A  simple prescription to construct (anti-) self-dual instantons in the Yang-Mills theory is given in \cite{Atiyah:1978ss}.
 Let us first review this construction. We will not distinguish between lower and upper indices in this section as the space is Euclidean.

In order to describe charge $k$ instantons with gauge group $U(N)$ on $\mathds{R}^4$ one starts with the following data:
\begin{enumerate}
\item A pair of complex hermitian vector spaces $V=\mathds{C}^k$ and $W=\mathds{C}^N$.
\item The operators $B_1,B_2 \in \rm{Hom}(V,V)$, $I \in \rm{Hom}(W,V)$, $J \in \rm{Hom}(V,W)$, which must obey the equations 
\begin{equation} \label{adhm_condition}
[B_1,B_1^\dagger]+[B_2,B_2^\dagger]+II^\dagger - J^\dagger J = 0, \quad
[B_1,B_2]+IJ = 0.
\end{equation}
\end{enumerate}
For $z=(z_1,z_2) \in \mathds{C}^2 \approx \mathds{R}^4$, define an operator 
$\mathcal{D}:V\oplus V \oplus W \rightarrow V \oplus V$ as
\begin{equation} \label{d_asd}
\mathcal{D}^\dagger = \left(
\begin{array}{c}
 \tau \\
\sigma^\dagger
\end{array}
\right), \quad
\tau = \left(
\begin{array}{ccc}
 B_2 - z_2 & B_1-z_1 & I
\end{array}
\right), \quad
\sigma = \left(
\begin{array}{c}
 -B_1+z_1 \\
B_2-z_2 \\
J
\end{array}
\right)
\end{equation}
for anti-self-dual instantons and by
\begin{equation} \label{d_sd}
\mathcal{D}^\dagger = \left(
\begin{array}{c}
 \tau \\
\sigma^\dagger
\end{array}
\right), \quad
\tau = \left(
\begin{array}{ccc}
 B_2 - \bar{z}_2 & B_1+z_1 & I
\end{array}
\right), \quad
\sigma = \left(
\begin{array}{c}
 -B_1-z_1 \\
B_2-\bar{z}_2 \\
J
\end{array}
\right)
\end{equation}
for self-dual instantons. 
Given the matrices $B_1, B_2,I$ and $J$ obeying all the conditions above, 
the actual instanton solution, $A_\alpha: W \rightarrow W,$ is determined by the following rather explicit formulae:
\begin{equation} \label{gauge_field}
A_\alpha = \Psi^\dagger\partial_\alpha\Psi,
\end{equation}
($\alpha=1,2,3,4$) where $\Psi: W \rightarrow V \oplus V \oplus W$ is the normalized mode of the operator $\mathcal{D}^\dagger$
\begin{equation} \label{zero_mode}
 \mathcal{D}^\dagger\Psi = 0, \quad  \Psi^\dagger\Psi = 1.
\end{equation}
 Here $\partial_\alpha$ is derivative with respect to the spacetime coordinates $x_\alpha$
which are related to the $z$-coordinates as
\begin{equation}
z_1 = \frac{x_1 + ix_2}{\sqrt{2}}, \quad  \bar{z}_1 = \frac{x_1 - ix_2}{\sqrt{2}}, \quad 
z_2 = \frac{x_3 + ix_4}{\sqrt{2}}, \quad \bar{z}_2 = \frac{x_3 - ix_4}{\sqrt{2}}.
\end{equation}
For given ADHM data and the zero mode condition (\ref{zero_mode}), the following completeness
relation has to be satisfied 
\begin{equation}
\label{completeness}
\mathcal{D}\frac{1}{\mathcal{D}^\dagger \mathcal{D}}\mathcal{D}^\dagger + \Psi\Psi^\dagger = 1.
\end{equation}
 It has been shown in \cite{Kim:2000msa} that (\ref{completeness}) can be satisfied even for noncommutative spaces. 
Note that the fields $A_\alpha$ are anti-hermitian, consistent with $\Psi^\dagger\Psi=1$.
The field strength $F_{\alpha\beta}$ and its dual $\tilde{F}_{\alpha\beta}$ are given as
\begin{equation}
F_{\alpha\beta} = \partial_\alpha A_\beta - \partial_\beta A_\alpha + [A_\alpha,A_\beta],
\quad \tilde{F}_{\alpha\beta} = \frac{1}{2}\epsilon_{\alpha\beta\gamma\delta}F_{\gamma\delta}.
\end{equation}
The instantons found by the ADHM construction satisfy both the Yang-Mills equation of motion and the (anti-) self-duality condition
\begin{equation}
  D_\alpha F_{\alpha\beta} = 0, \quad \tilde{F}_{\alpha\beta} = \pm F_{\alpha\beta} 
\end{equation}
 and has topological charge  given by
\begin{equation}
 Q=-\frac{1}{16\pi^2}\int d^4x \, Tr(\tilde{F}_{\alpha\beta}F^{\alpha\beta}).
\end{equation}
The trace is over the space $W=\mathds{C}^N$.

\subsection{Noncommutative Euclidean Space}
To study instantons on a noncommutative $\mathds{R}^4$, we will use the notation outlined below.
The 4-dimensional noncommutative euclidean space is defined by the following noncommutative coordinates:
\begin{equation}\label{commutation_x}
 [\hat{x}^\alpha , \hat{x}^\beta ] = i\theta^{\alpha\beta} ,\quad \alpha,\beta=1,2,3,4,
\end{equation}
and $\theta^{\alpha\beta}$ is a constant anti-symmetric $4\times4$ matrix.
We denote the algebra generated by these $\hat{x}^\alpha$'s  by $\mathcal{A}_\theta$.
There are three distinct cases one may consider:
\begin{enumerate}
\item $\theta$ has rank 0 ($\theta^{\alpha\beta}=0 $ $\forall$ $ \alpha, \beta$).
In this case $\mathcal{A}_\theta$ is isomorphic to the algebra of functions on the ordinary $\mathds{R}^4$.
This space may be denoted by $\mathds{R}_C^4$.
\item $\theta$ has rank 2.
In this case $\mathcal{A}_\theta$ is the algebra of functions on the ordinary $\mathds{R}^2$ 
times the noncommutative $\mathds{R}^2$, which may be denoted by $\mathds{R}_{NC}^2 \times \mathds{R}_{C}^2$. Without 
 loss of generality, we can choose
$$\theta^{\alpha\beta} =
\left[ \begin{array}{cccc}
0 & -\theta & 0 & 0 \\
\theta &0 & 0 &0 \\
0& 0 & 0& 0\\
0&0 & 0 & 0\end{array} \right].$$
Let us define a system of complex coordinates and a set of operators as
\begin{eqnarray}\hspace{-1 cm} \left.
\begin{array}{cccc}
 z_1=\frac{1}{\sqrt{2}}(x^1+i x^2), & \bar{z}_1=\frac{1}{\sqrt{2}}(x^1-i x^2), & z_2=\frac{1}{\sqrt{2}}(x^3+i x^4), &\bar{z}_2=\frac{1}{\sqrt{2}}(x^3-i x^4)\\
a_1=\frac{\bar{z}_{1}}{\sqrt{\theta}}, 
& a_1^{\dagger}=\frac{z_{1}}{\sqrt{\theta}}, 
& a_2=\frac{\bar{z}_{2}}{\sqrt{\theta}},
& a_2^{\dagger}=\frac{z_{2}}{\sqrt{\theta}}
\end{array}
\right.
\label{complex_coordinates}
\end{eqnarray}
which reduces the algebra (\ref{commutation_x}) to
\begin{equation}
[\bar{z}_{1},z_{1}]=\theta, \quad [\bar{z}_{2},z_{2}]=0, \quad [a_1,a_1^{\dagger}]=1 , \quad  [a_2,a_2^{\dagger}]=0. 
\label{commutation_z1}
\end{equation}
Here $a_1$ and $a_1^\dagger$ are like annihilation and creation operators respectively while $a_2^\dagger,a_2$ are ordinary complex numbers.
We can define a number operator by $N_1 = a_1^\dagger a_1$. 
The Fock space on which the elements of $\mathcal{A}_\theta$ act, consists of states denoted by $|n_1,z_2\rangle$.
Here $n_1$ denotes the eigenvalues of the number operator $N_1$ and can take only non-negative integral values, 
while $z_2$ can be any complex number and denotes the eigenvalues of $z_2$.

\item $\theta$ has rank 4.
In this case $\mathcal{A}_\theta$ is the noncommutative $\mathds{R}^4$.
We choose $\theta$ to be of the form given by 
$$\theta^{\alpha\beta} =
\left[ \begin{array}{cccc}
0 & \theta^{12} & 0 & 0 \\
-\theta^{12} &0 & 0 &0 \\
0& 0 & 0 & \theta^{34}\\
0&0 & -\theta^{34} & 0\end{array} \right]=
\left[ \begin{array}{cccc}
0 & -\theta & 0 & 0 \\
\theta &0 & 0 &0 \\
0& 0 & 0& -\theta\\
0&0 & \theta & 0
\end{array} \right],$$ 
where we have assumed $\theta^{12}=\theta^{34}=-\theta$. Again, we can define a system 
of complex coordinates and a set of operators as in (\ref{complex_coordinates}) but now the algebra (\ref{commutation_x}) becomes 
\begin{equation}
[\bar{z}_{1},z_{1}]=[\bar{z}_{2},z_{2}]=\theta, \quad [a_1,a_1^{\dagger}]=[a_2,a_2^{\dagger}]=1.
\label{commutation_z2}
\end{equation}
The Fock space on which the elements of $\mathcal{A}_\theta$ act consists of states denoted by $|n_1,n_2\rangle$.
Here $n_1$ and $n_2$ denote the eigenvalues of the number operators $N_1 = a_1^\dagger a_1$ and $N_2 = a_2^\dagger a_2$ respectively 
which can take only non-negative integral values.

\end{enumerate}
As already mentioned in section \ref{sec:flux_sec}, differentiation in the noncommutative space  is implimented as 
an adjoint as in (\ref{derivative}),i.e.,
\begin{equation}
\partial_\alpha f = -i\tilde{\theta}_{\alpha\beta} \left[x^\beta,f\right]
\end{equation}
where $\tilde{\theta}_{\alpha\beta}$ is the inverse of the matrix $\theta_{\alpha\beta}$, i.e.,
\begin{equation}
\theta_{\alpha\gamma}\tilde{\theta}_{\gamma\beta}=\delta_{\alpha\beta}.
\end{equation}
and is given by
\begin{equation}
\tilde{\theta}_{\alpha\beta} = -\frac{1}{\theta^2}\theta_{\alpha\beta}
\end{equation}
Differentiation with respect to the complex coordinates are
\begin{equation}
\partial_{z_a} f = \frac{1}{\sqrt{\theta}} \left[a_a,f\right], \quad \partial_{\bar{z}_a} f = -\frac{1}{\sqrt{\theta}} \left[a_a^\dagger, f\right]; \quad \quad a=1,2.
\end{equation}
Again, the integration is implemented by a suitable trace.

\subsection{Noncommutative ADHM Construction}
ADHM construction for instantons has been generalized to a noncommutative space in \cite{Nekrasov:1998ss}.
The construction effectively remains same as in the commutative case, only change being the replacement of 0 in the right hand side
of the first equation of (\ref{adhm_condition}) by the noncommutative parameter $\theta$ 
for the case of $\mathds{R}^2_{NC}\times \mathds{R}^2_C$ and by $2\theta$ for the case of 
$\mathds{R}^2_{NC}\times \mathds{R}^2_{NC}$ respectively:
\begin{equation} \label{adhm_condition_r2}
[B_1,B_1^\dagger]+[B_2,B_2^\dagger]+II^\dagger-J^\dagger J = \theta, \quad
[B_1,B_2]+IJ = 0
\end{equation}
for $\mathds{R}^2_{NC}\times \mathds{R}^2_C$  as in \cite{Kim:2001ai} and
\begin{equation} \label{adhm_condition_r4}
[B_1,B_1^\dagger]+[B_2,B_2^\dagger]+II^\dagger-J^\dagger J = 2\theta, \quad
[B_1,B_2]+IJ = 0
\end{equation}
for $\mathds{R}^2_{NC}\times \mathds{R}^2_{NC}$. 
The Yang-Mills gauge connection $A_{x^\alpha}$ is given by
\begin{equation}
A_{x^\alpha} = -i\tilde{\theta}_{\alpha\beta} \Psi^\dagger \left[x^\beta, \Psi\right]
\end{equation}
The gauge field, as in section \ref{sec:flux_sec}, is related to the gauge connection by
\begin{equation} \label{instanton_field_connection}
\hat{D}_{x^\alpha} = -i \tilde{\theta}_{\alpha\beta} x^\beta +  A_{x^\alpha}
\end{equation}
and is given by
\begin{equation} \label{instanton_field}
\hat{D}_{x^\alpha}=-i \tilde{\theta}_{\alpha\beta}\Psi^{\dagger}x^\beta\Psi
\end{equation}
Both $A_{x^\alpha}$ and $\hat{D}_{x^\alpha}$ are again anti-hermitian.
In $\mathds{R}^2_{NC}\times \mathds{R}^2_C$, the components of the gauge field along the commutative directions will be given by (\ref{gauge_field}), while
those along the noncommutative axes by $\hat{D}_{x^\alpha}$.

Let us first discuss the usual single anti-self-dual $U(1)$ instanton solutions ($k=1, N=1$) 
in  $\mathds{R}^2_{NC}\times \mathds{R}^2_{C}$ \cite{Kim:2001ai,Chu:2001cx} and in $\mathds{R}^2_{NC}\times \mathds{R}^2_{NC}$  
\cite{Kim:2000msa,Chu:2001cx,Furuuchi:2000vc,Sako:2010zza}.
For $k=N=1$, $B_1, B_2, I$ and $J$ are all complex numbers. 
As the noncommutative space (described by the coordinates $z$) has translational invariance, 
we can always choose the origin in such a way that $B_1$ and $B_2$  in  $\tau$ can be taken to be zero.
Thus (\ref{adhm_condition_r2}) or (\ref{adhm_condition_r4}) ensures that either  $I$ or $J$ is zero.
$I=0$ gives
\begin{eqnarray}
-\bar{z}_1\psi_1 -z_2\psi_2 + \bar{J} \xi = 0, \quad -\bar{z}_2\psi_1 + z_1 \psi_2 = 0.
\end{eqnarray}
while $J=0$ gives
\begin{eqnarray}
\bar{z}_2\psi_1 -z_1\psi_2 + I \xi = 0, \quad -\bar{z}_1\psi_1 - z_2 \psi_2 = 0.
\end{eqnarray}
for 
\begin{equation}
\Psi=
\left( \begin{array}{c}
\psi_1\\
\psi_2\\
\xi 
\end{array} \right)
\end{equation}
The two choices are related by a rotation in the plane of complex coordinates $z_1-z_2$, namely $z_1\rightarrow -z_2, z_2\rightarrow z_1$.
 We choose $J=0$ without the loss of any generality.
Now let us discuss the two cases seperately.

\subsubsection*{{\underline{$\mathds{R}^2_{NC}\times\mathds{R}^2_{C}$:}}}
Here we get $I= \sqrt{\theta}$ from the equation (\ref{adhm_condition_r2}).
The phase in $I$ does not effect the solution for the gauge field and hence has been taken to be zero.
The operator (\ref{d_asd}) for anti-self-dual instantons becomes 
\begin{equation}
\mathcal{D}^\dagger=
\left( \begin{array}{ccc}
-z_2 & -z_1 & \sqrt{\theta} \\
\bar{z}_1 & -\bar{z}_2 & 0
\end{array} \right)= \sqrt{\theta}\left( \begin{array}{ccc}
-a_2^\dagger & -a_1^\dagger & 1 \\
a_1 & -a_2 & 0
\end{array} \right)
\end{equation}
and its normalized zero mode solution is given by 
\begin{equation} \label{psi_form}
\psi_1= a_2 \frac{1}{\sqrt{\delta \Delta}}, \quad \psi_2= a_1 \frac{1}{\sqrt{\delta \Delta}},  
\quad \xi=\sqrt{\frac{\delta}{ \Delta}}
\end{equation}
with $ \delta= a_1^\dagger a_1+a_2^\dagger a_2$ and $ \Delta=\delta+1.$
Note that the inverse of the operator $\Delta$ is well--defined, but that of $\delta$ is not since 
 $|n_1=0$, $z_2=0\rangle$ is a zero--mode of $\delta$.

\subsubsection*{{\underline{$\mathds{R}^2_{NC}\times\mathds{R}^2_{NC}$:}}}
In this case, (\ref{adhm_condition_r4}) gives $I=\sqrt{2\theta}$ and the operator in (\ref{d_asd}) becomes
\begin{equation}
\mathcal{D}^\dagger=
\left( \begin{array}{ccc}
-z_2 & -z_1 & \sqrt{2\theta} \\
\bar{z}_1 & -\bar{z}_2 & 0
\end{array} \right)= \sqrt{\theta}\left( \begin{array}{ccc}
-a_2^\dagger & -a_1^\dagger & \sqrt{2} \\
a_1 & -a_2 & 0
\end{array} \right).
\end{equation}
The zero mode solution is again a 3-element column matrix $\Psi$  
where we write $\psi_1 = \sqrt{\theta}a_2v$ and $\psi_2 = \sqrt{\theta}a_1v$ . Then (\ref{zero_mode}) becomes 
\begin{equation} \label{eqn_v_xi_1}
\hat{\Delta}v = \sqrt{2\theta}\xi , \quad 
 v^\dagger\hat{\Delta}v + \xi^\dagger\xi = 1
\end{equation}
where $\hat{\Delta}=\theta \hat{N} = \theta (a_1^\dagger a_1 + a_2^\dagger a_2)$.
But this operator does not have an inverse since it has a zero--mode
\begin{equation}
\hat{\Delta}|0,0\rangle = 0 
\end{equation}
and hence finding $v$ and $\xi$ is a bit tricky.
We define a shift operator $S$ such that
 \begin{eqnarray}
 SS^\dagger = 1, \quad S^\dagger S = 1- P, \quad P = |0,0 \rangle \langle 0,0 |.
\end{eqnarray}
Note that although the inverse of $\hat{\Delta}$ is not defined otherwise, 
it is well--defined when sandwiched between $S$ and $S^\dagger$.
Now we can solve for $\xi$ and $v$:
\begin{equation}
 \xi = \hat{\Phi}^{-\frac{1}{2}} S^\dagger, \quad v = \sqrt{2\theta}\frac{1}{\hat{\Delta}}\hat{\Phi}^{-\frac{1}{2}} S^\dagger, \quad \quad \quad  
\mathrm{where} \quad \hat{\Phi}= 1+ \frac{2\theta}{\hat{\Delta}}=1+\frac{2}{\hat{N}}
\end{equation}
which satisfy (\ref{eqn_v_xi_1}).
Thus we get
\begin{equation} 
 \psi_1=\sqrt{\theta}a_2 \sqrt{2\theta} \frac{1}{\hat{\Delta}} \hat{\Phi}^{-\frac{1}{2}} S^\dagger, \quad 
\psi_2=\sqrt{\theta}a_1 \sqrt{2\theta} \frac{1}{\hat{\Delta}} \hat{\Phi}^{-\frac{1}{2}} S^\dagger, \quad 
\xi = \hat{\Phi}^{-\frac{1}{2}} S^\dagger.
\end{equation}
We can define  the components of the gauge field in terms of the complex coordinates as
\begin{eqnarray}
\hat{D}_1=\frac{1}{\sqrt{2}}\left(\hat{D}_{x_1}-i\hat{D}_{x_2}\right), \quad \hat{D}_2= \frac{1}{\sqrt{2}}\left(\hat{D}_{x_3}-i\hat{D}_{x_4}\right), \\ 
\hat{D}_{\bar{1}}=\frac{1}{\sqrt{2}}\left(\hat{D}_{x_1}+i\hat{D}_{x_2}\right), \quad \hat{D}_{\bar{2}}=\frac{1}{\sqrt{2}}\left(\hat{D}_{x_3}+i\hat{D}_{x_4}\right).
\end{eqnarray}
Then (\ref{instanton_field_connection}) translates to
\begin{equation}
\hat{D}_a = \frac{1}{\theta}\bar{z}_a + A_a, \quad \hat{D}_{\bar{a}} = -\frac{1}{\theta}z_a + A_{\bar{a}}, \quad \quad a= 1,2.
\end{equation}
with
\begin{eqnarray}
A_1=\frac{1}{\sqrt{2}}\left(A_{x_1}-iA_{x_2}\right), \quad A_2= \frac{1}{\sqrt{2}}\left(A_{x_3}-iA_{x_4}\right), \\ 
A_{\bar{1}}=\frac{1}{\sqrt{2}}\left(A_{x_1}+iA_{x_2}\right), \quad A_{\bar{2}}=\frac{1}{\sqrt{2}}\left(A_{x_3}+iA_{x_4}\right).
\end{eqnarray}
Also (\ref{instanton_field}) translates into 
\begin{equation} \label{A_a}
\hat{D}_a = \frac{1}{\theta}\Psi^\dagger \bar{z}_a\Psi, \quad  \hat{D}_{\bar{a}} = -\frac{1}{\theta}\Psi^\dagger z_a\Psi= -\hat{D}_a^\dagger .
\end{equation}
The solution for  $k=1$ $U(1)$ ASD instanton becomes
\begin{equation}\label{usual_instanton}
\hat{D}_{a} = \frac{1}{\sqrt{\theta}} S\hat{\Phi}^{-\frac{1}{2}} a_a\hat{\Phi}^{\frac{1}{2}} S^\dagger, \quad
\hat{D}_{\bar{a}} = -\frac{1}{\sqrt{\theta}} S\hat{\Phi}^{\frac{1}{2}} a^\dagger_a \hat{\Phi}^{-\frac{1}{2}} S^\dagger
\end{equation}
where the shift operator $S$,  written explicitly, is
\begin{equation} \left.
\begin{array}{c}
 S=  1- \displaystyle{\sum_{n'_1=0}^\infty} |n'_1 , 0\rangle \langle n'_1 , 0| \left(1- \frac{1}{\sqrt{N_1+1}}a_1\right) \\
 S^\dagger=  1-  \displaystyle{\sum_{n'_1=0}^\infty}\left(1- a_1^\dagger\frac{1}{\sqrt{N_1+1}}\right) |n'_1 , 0\rangle \langle n'_1 , 0|
\end{array} \right\rbrace.
\end{equation}
The field strengths are given by
\begin{equation}
F_{\alpha\beta} = -i\tilde{\theta}_{\alpha\gamma}\left[x^\gamma, A_{x^\beta}\right] +i\tilde{\theta}_{\beta\gamma}\left[x^\gamma, A_{x^\alpha}\right] + \left[A_{x^\alpha}, A_{x^\beta}\right] 
= -i\tilde{\theta}_{\alpha\beta} + \left[\hat{D}_{x^\alpha}, \hat{D}_{x^\beta}\right]
\end{equation}
Their dual are define as
\begin{equation}
\tilde{F}_{\alpha\beta} = \frac{1}{2}\epsilon_{\alpha\beta\gamma\delta} F_{\gamma\delta}
\end{equation}
The ASD condition
\begin{equation}
\tilde{F}_{\alpha\beta} + F_{\alpha\beta} = 0
\end{equation}
translates to
\begin{equation} \label{asd_r4}
F_{1\bar{1}} = - F_{2\bar{2}}, \quad F_{12} = F_{\bar{1}\bar{2}} = 0.
\end{equation}
where the field strengths, in our notation of complex coordinates, are given by 
\begin{eqnarray}
F_{a\bar{b}} =& \frac{1}{\sqrt{\theta}}\left[a_a,A_{\bar{b}}\right] + \frac{1}{\sqrt{\theta}}\left[a_b^\dagger,A_{a}\right] + \left[A_a,A_{\bar{b}}\right]
&=\frac{1}{\theta}\delta_{ab} + \left[ \hat{D}_a, \hat{D}_{\bar{b}} \right], \\
F_{ab} = & \frac{1}{\sqrt{\theta}}\left[a_a,A_{b}\right] - \frac{1}{\sqrt{\theta}}\left[a_b,A_{a}\right] + \left[A_a,A_{b}\right]
&= \left[ \hat{D}_a, \hat{D}_{b} \right], \\
F_{\bar{a}\bar{b}} =& - \frac{1}{\sqrt{\theta}}\left[a_a^\dagger,A_{\bar{b}}\right] + \frac{1}{\sqrt{\theta}}\left[a_b^\dagger,A_{\bar{a}}\right] + \left[A_{\bar{a}},A_{\bar{b}}\right]
&= \left[ \hat{D}_{\bar{a}}, \hat{D}_{\bar{b}} \right] \\
&F_{\bar{b}a} = -F_{a\bar{b}} &
\end{eqnarray}
The equations of motion in the $\mathds{R}^2_{NC}\times\mathds{R}^2_{NC}$ in terms of the fields $\hat{D}_{x^\alpha}$ are
\begin{equation} \label{eom_r4}
[\hat{D}_{x^\alpha},[\hat{D}_{x^\alpha},\hat{D}_{x^\beta}]]=0.
\end{equation}
The solution (\ref{usual_instanton}) satisfies both the ASD condition  and the equations of motion. 

\subsection{New Anti-Self-Dual $U(1)$ Instanton}
We can try to get new solutions for noncommutative instantons by using the GBOs.

\subsubsection*{{\underline{$\mathds{R}^2_{NC}\times\mathds{R}^2_C$:}}}
Let us first define two operators $b$ and $c$
\begin{eqnarray}\left.
\begin{array}{ccc}
&&b=\frac{1}{\sqrt{2}} a_1\frac{1}{\sqrt{N_1}}a_1 \Lambda_{1+} +\frac{1}{\sqrt{2}}a_1\frac{1}{\sqrt{N_1+1}}a_1 \Lambda_{1-}, \\
&&c=\frac{1}{\sqrt{2}} a_2\frac{1}{\sqrt{N_1}}a_1 \Lambda_{1+} +\frac{1}{\sqrt{2}}a_2\frac{1}{\sqrt{N_1+1}}a_1 \Lambda_{1-}
\end{array}\right.
\label{b_g_operator}
\end{eqnarray}
where $N_1$ is the number operator and $\Lambda_{1+}$ and $\Lambda_{1-}$ are the projection operators corresponding to $a_1$.
Here $b$ is the  generalized operator defined in (\ref{expression_b}).
We can easily show that 
\begin{equation}
 b^\dagger b= \frac{1}{2}(N_1-\Lambda_{1-}), \quad c^\dagger c= \frac{1}{2}N_2\frac{N_1-\Lambda_{1-}}{N_1-1  }, \quad {\rm where}\quad N_2=a_2^\dagger a_2.
\end{equation}
The zero-mode solution  $\Psi_0$ is given by
\begin{equation}
 \psi_1= \sqrt{2} c , \quad \psi_2=\sqrt{2} b ,  \quad \xi=\delta \left(\frac{1}{\sqrt{N_1}}a_1\Lambda_{1+} +\frac{1}{\sqrt{N_1+1}}a_1\Lambda_{1-}\right).
\end{equation}
But this solution is not normalized i.e. $\Psi_0^{\dagger}\Psi_0\neq1$.
Usually  the single instanton solution in $\mathds{R}_{NC}^2 \times \mathds{R}_{C}^2$ 
with  $U(1)$ gauge group is normalized as \cite{Chu:2001cx}
 \begin{equation}
 \Psi=\Psi_0 \frac{1}{\sqrt{\Psi_0^\dagger \Psi_0 }}.
 \end{equation}
But in our solution this technique cannot be used because in this case as
\begin{equation}
\Psi_0^\dagger \Psi_0=\frac{N_1-\Lambda_{1-}}{N_1-1}\delta (\delta-1)
\end{equation}
vanishes  when it operates on the state $|0\rangle$  and the inverse of $\Psi_0^\dagger \Psi_0$ 
does not exist.  
We fix this problem by defining 
\begin{equation}
  \Psi_{new}=\Psi_0 (1-p) \frac{1}{\sqrt{\Psi_0^\dagger \Psi_0 }}(1-p) u^\dagger, \quad \quad uu^\dagger = 1,\quad u^\dagger u =1-p
\end{equation}
where $p=|0\rangle\langle 0|$ is  a projection operator, and 
$u^\dagger$ is a shift operator projecting out the vacuum. The operator $u$ can be written as
\begin{equation}
u=\displaystyle{\sum_{n_1=0}^{\infty}}|n_1,z_2\rangle\langle n_1+1,z_2|, \quad 
u^\dagger=\displaystyle{\sum_{n_1=0}^{\infty}}|n_1+1,z_2\rangle\langle n_1,z_2|.
\end{equation}

It should be noted that the new solution in $\mathds{R}_{NC}^2 \times \mathds{R}_{C}^2$  is completely non-singular.

\subsubsection*{{\underline{$\mathds{R}^2_{NC}\times\mathds{R}^2_{NC}$:}}}
We claim the new solution to be
\begin{equation} \label{gauge_field_r4_new}
\hat{D}_{a}=\frac{1}{\sqrt{\theta}} S_{new}\hat{\Phi}_{new}^{-\frac{1}{2}} b_a\hat{\Phi}_{new}^{\frac{1}{2}} S_{new}^\dagger, 
\quad  
\hat{D}_{\bar{a}}= -\frac{1}{\sqrt{\theta}} S_{new}\hat{\Phi}_{new}^{\frac{1}{2}}b^\dagger_a \hat{\Phi}_{new}^{-\frac{1}{2}} S_{new}^\dagger
\end{equation}
where 
\begin{equation}
\hat{\Phi}_{new}=1+\frac{2}{M}
\end{equation} 
and 
\begin{equation}
M=M_1+M_2=\sum_{a=1}^2b_a^\dagger b_a.
\end{equation}
Here again the operator $\frac{1}{M}$ is not well-defined otherwise 
(as $M|0,0\rangle =M|0,1\rangle=M|1,0\rangle=M|1,1\rangle =0$), but is well-defined when sandwiched between $S_{new}$ and 
$S_{new}^\dagger$ (defined below) which projects out the states $|0,0\rangle$, $|1,0\rangle$, $|0,1\rangle$ and $|1,1\rangle$:
\begin{equation}
\begin{array}{c}
 S_{new}S_{new}^\dagger =1, \quad S_{new}^\dagger S_{new} = 1-P_{new} \\
P_{new} = |0,0 \rangle \langle 0,0 |+|1,0 \rangle \langle 1,0 |+|0,1 \rangle \langle 0,1 |+|1,1\rangle \langle 1,1 |.
\end{array}
\end{equation}
The explicit form for the new shift operator is as follows
\begin{equation}\left.
\begin{array}{l}
 S_{new} =  1- \displaystyle{\sum_{n'_1=0}^\infty} |n'_1 , 0\rangle \langle n'_1 , 0| \left(1- \frac{1}{\sqrt{M_1+1}}b_1\right)
- \displaystyle{\sum_{n'_1=0}^\infty} |n'_1 , 1\rangle \langle n'_1 , 1| \left(1- \frac{1}{\sqrt{M_1+1}}b_1\right) \\
 S_{new}^\dagger=  1-  \displaystyle{\sum_{n'_1=0}^\infty}\left(1- b_1^\dagger\frac{1}{\sqrt{M_1+1}}\right) |n'_1 , 0\rangle \langle n'_1 , 0|
-  \displaystyle{\sum_{n'_1=0}^\infty}\left(1- b_1^\dagger\frac{1}{\sqrt{M_1+1}}\right) |n'_1 , 1\rangle \langle n'_1 , 1|
\end{array}\right\rbrace
\end{equation}
The operator $b_a$ (corresponding to $a_a$ ) is defined as in (\ref{expression_b}).
We can check that this new solution  satisfies the ASD condition (\ref{asd_r4}) and the equations of motion (\ref{eom_r4}) 
 (for details see Appendices \ref{new_soln_eom} and \ref{new_soln_asd}).
The topological charge of this new solution can be shown to be 4 times the charge of the usual single ASD instanton:  $Q_{new} = -4$
 (see appendix \ref{topo_charge}).

This solution is different from the usual $k=-4$ instanton  for $U(1)$ gauge group despite the topological charge being the same.
We can understand this by observing  that the difference $q$ between the numbers of $a$'s and $a^\dagger$'s  in the two solutions is not same:
 $q=1$  for the usual ADHM solutions irrespective of its charge and the gauge group, whereas
 $q=2$ for the new solution (coming solely because of the operator $b$ in the expression of $\hat{D}_a$ 
given by (\ref{gauge_field_r4_new})).
(The number of $a$'s in $\Psi$ is equal to that of $a^\dagger$'s in $\Psi^\dagger$ and vice-versa.)
The new solution cannot be reduced to the usual instanton by a unitary transformation and hence represents gauge inequivalent configuration.

If we use the operator $b_a(z_{a+},z_{a-})$ defined in (\ref{translation_operator}), we can construct an instanton solution by 
repeating the steps we have outlined above. This instanton also has charge $-4$ as 
the trace of an operator is invariant under unitary transformation. Then the four complex parameters $z_{1+},z_{2+},z_{1-},z_{2-}$ 
can be thought as characterizing the ``locations'' of four instantons with charge $-1$. It is easy to see that in the coincident limit
 $z_{1+}=z_{2+}=z_{1-}=z_{2-}=0$, we recover (\ref{gauge_field_r4_new}).

We can use the above technique to find a new solution in terms of the GBO $b_a^{(p_a)}$:

\begin{eqnarray} \label{instanton_solution_b_pa}
 &&\hat{D}_{a}=\frac{1}{\sqrt{\theta}} S_{new}(\hat{\Phi}_{new})^{-\frac{1}{2}} b^{(p_a)}_a(\hat{\Phi}_{new})^
{\frac{1}{2}} S_{new}^{\dagger}, \nonumber \\
&& \hat{\Phi}_{new}=1+\frac{2}{M}, \quad M=M_1^{(p_1)}+M_2^{(p_2)}, \quad M_a^{(p_a)}=b^{(p_a)\dagger}_ab^{(p_a)}_a \nonumber \\
 &&S_{new} =1-\sum_{i=0}^{p_2-1}\sum_{n'_1=0}^{\infty} |n'_1,i\rangle\langle n'_1,i|\left(1-\frac{1}{\sqrt{M^{(p_1)}_1+1}}b^{(p_1)}_1\right),
\end{eqnarray}
with $S_{new}$ satisfying 
\begin{equation}
S_{new}S_{new}^{\dagger}=1, \quad S_{new}^{\dagger}S_{new}=1-\displaystyle\sum_{i=0,j=0}^{i=p_1-1,j=p_2-1}|i,j\rangle \langle i,j|.
\end{equation}
This solution represents an ASD instanton with charge $Q=-p_1p_2$. 
Again, it is gauge inequivalent to the $k=-p_1p_2$  instanton known in the literature.
We could as well have used a different shift operator given by
\begin{equation}\label{shift_operator_2}
S_{new}^\prime =1-\sum_{i=0}^{p_1-1}\sum_{n'_2=0}^{\infty} |i,n'_2\rangle\langle i,n'_2|\left(1-\frac{1}{\sqrt{M^{(p_2)}_2+1}}b^{(p_2)}_2\right).
\end{equation}
Their actions are given by
\begin{equation}
 S_{new}^\dagger |n_1,n_2\rangle = \left\{
\begin{array}{ll}
 |n_1+p_1,n_2\rangle & {\rm if } \,\, 0\leq n_2 \leq p_2-1 \\
|n_1,n_2\rangle & {\rm if } \,\, n_2 \geq p_2,
\end{array} \right.
\end{equation}
\begin{equation}
 S_{new}^{\prime \dagger} |n_1,n_2\rangle = \left\{
\begin{array}{ll}
 |n_1,n_2+p_2\rangle & {\rm if } \,\, 0\leq n_1 \leq p_1-1 \\
|n_1,n_2\rangle & {\rm if } \,\, n_1 \geq p_1.
\end{array} \right.
\end{equation}
Other multi--instanton solutions can be constructed using the reducible representations involving squeezed operators 
(\ref{reducible_squeeze}). 
Our construction of multi--instantonic solutions using reducible representations of the standard harmonic oscillator algebra
 may also be generalized for 4k-dimensional instantons as discussed in \cite{Ivanova:2005fh, Broedel:2007wc}. 
This exercise will be left as a future work.

In this section we found multi--instantons with charge $-p_1p_2$ ($p_1,p_2$ non-negative integers) which are 
not gauge equivalent to known solutions. The charge of the newly found multi--instantons has an explicit relation with the 
representation theory labels $p_1$ and $p_2$. Thus the instanton number does have the information about the reducibility of the space along with the topological nature of the solutions. 
Using the ``translated'' $b$ operators (\ref{translation_operator})
 we could construct multi--instantons that depend explicitly on $p_1p_2$ number of complex parameters. While the full moduli space of 
noncommutative multi--instantons is still not well understood, we hope that this identification contributes partially to this question.



\chapter{Quantum Mechanics with time-space noncommutativity 
} \label{qo}
\section{Introduction}
In the previous chapter we discussed the theories in noncommutative Moyal spacetime where time coordinate remained to be commutative.
In such situations where only the spatial coordinates do not commute with each other,
the quantum theory is conceptually straightforward (but nonetheless may display novel phenomena)
\cite{Dragovich:2003sj, Scholtz:2008zu, Ho:2001aa, Bolonek:2002cc, BenGeloun:2009zd, Rohwer:2010zq, Scholtz:2007ig}.
In this chapter we will concentrate on understanding some implications of quantum mechanics with time-space noncommutativity,
specifically we will work with the Moyal plane $\mathds{R}^{1,1}_\theta$.
We will use the formalism of unitary quantum mechanics on this space as developed by Balachandran et. al. \cite{Balachandran:2004rq} 
(see also \cite{Li:2001vc}).

When time and space do not commute with each other it is not unreasonable to expect that the dynamics of the time dependent 
processes get altered.
We will verify this explicitly in the context of a simple model of the forced harmonic oscillator (FHO) with the forcing term 
switched on only for a finite duration of time.
In the commutative case this is a much studied model.
We will compute deviations from the commutative case to leading order in $\theta$.
These deviations suggest that time-space noncommutativity can capture certain nonlinear effects seen in quantum optics.

This chapter is organized as follows: In section \ref{bala_paper} we will briefly review the formulation of unitary 
quantum mechanics on $\mathds{R}^{1,1}_\theta$ \cite{Balachandran:2004rq}. 
In section \ref{problem} we will solve the problem of the FHO perturbatively in $\theta$ and compute corrections to the 
transition probabilities between simple harmonic oscillator (SHO) states.
These corrections suggest the noncoherent nature of the time-evolved vacuum state 
and are the reminiscent of those seen in nonlinear quantum optics \cite{Walls_book}.
To flesh out this analogy better we study the time-evolution of uncertainties 
in position and momentum in section \ref{uncertainties}.
Encouraged by these results we, in section \ref{quantum_optics}, suggest a correspondence between the nonlinearity in 
quantum optics and the quantum mechanics on $\mathds{R}^{1,1}_\theta$.

\section{Unitary Quantum Mechanics on $\mathds{R}^{1,1}_\theta$}\label{bala_paper}
The noncommutative space $\mathds{R}^{1,1}_\theta$ is described by the coordinates $\hat{x}_\mu$'s satisfying
\begin{equation} \label{commutator}
 [\hat{x}_\mu,\hat{x}_\nu]=i\theta\varepsilon_{\mu\nu}\mbox{ with } 
\varepsilon_{\mu\nu}=-\varepsilon_{\nu\mu}\mbox{ and } \varepsilon_{01}=1,
\end{equation}
where $\mu$ and $\nu$ can take values 0,1.
Without loss of generality we can take $\theta>0$, as its sign can always be flipped by changing $\hat{x}_1$ to $-\hat{x}_1$.  
Let $\mathcal{A}_{\theta}(\mathds{R}^{1,1})$ be the unital algebra generated by $\hat{x}_0$ and $\hat{x}_1$.
We associate to each  $\hat{\alpha}\in\mathcal{A}_{\theta}\left(\mathds{R}^{1,1}\right)$,
 its left and right representations $\hat{\alpha}^{L}$ and $\hat{\alpha}^{R}$:
\begin{equation}
\hat{\alpha}^{L}\hat{\beta} = \hat{\alpha}\hat{\beta} \,\, , \,\,
\hat{\alpha}^{R}\hat{\beta} = \hat{\beta}\hat{\alpha} \,\, , \,\,
\hat{\beta}\in\mathcal{A}_{\theta}\left(\mathds{R}^{1,1}\right) \,\,.
\label{Left_Right_reps}
\end{equation}
Unless stated, we work with the left representation. \\
\indent For a quantum theory, what we need are:
(1) a suitable inner product on 
$\mathcal{A}_{\theta}\left(\mathds{R}^{1,1}\right)$;
(2) a Schr\"{o}dinger constraint on
$\mathcal{A}_{\theta}\left(\mathds{R}^{1,1}\right)$;
and (3) a self-adjoint (with respect to the inner product defined) Hamiltonian $\hat{H}$ and observables which act on the
constrained subspace of $\mathcal{A}_{\theta}\left(\mathds{R}^{1,1}\right)$.

{\it 1. The Inner Product:} \\
There are several suitable inner products and they are all equivalent to each other as discussed in \cite{Balachandran:2004rq}. Here we discuss only one such example. Using the commutator (\ref{commutator}) any $\hat{\alpha}\in\mathcal{A}_\theta(\mathds{R}^{1,1})$ can always be written as  $\hat{\alpha} = \int d^{2}k \, \tilde{\alpha}(k) e^{ik_{1}\hat{x}_{1}}e^{ik_{0}\hat{x}_{0}}$. We associate a symbol $\alpha_S$ corresponding to each such $\hat{\alpha}$ given by
\begin{equation}
\alpha_{S}(x_{0},x_{1}) = \int d^{2}k \,
\tilde{\alpha}(k) e^{ik_{1}x_{1}}e^{ik_{0}x_{0}} \,\, .
\label{alpha-s-symbol}
\end{equation}
Note that $x_0$ and $x_1$ and hence $\alpha_S$ are purely commutative.
The inner product is defined as
\begin{equation}
\left(\hat{\alpha},\hat{\beta}\right)_{t} =
\int dx_{1} \, \alpha_{S}^{*}(t,x_{1})\beta_{S}(t,x_{1}) \,\, .
\label{S-inner-product}
\end{equation}

{\it 2. The Schr\"{o}dinger Constraint and time evolution:} \\
The operators $\hat{p}_0$ and $\hat{p}_1$, given by 
\begin{equation}
i\frac{\partial}{\partial x_{0}}\equiv\hat{p}_{0} =
-\frac{1}{\theta}{\rm ad}_{\hat{x}_{1}}, \quad
-i\frac{\partial}{\partial x_{1}}\equiv\hat{p}_{1} =
-\frac{1}{\theta}{\rm ad}_{\hat{x}_{0}},
\end{equation}
generate time and space translations respectively. The adjoint action is defined as
\begin{equation}
{\rm ad}_{\hat{A}} \hat{\psi} = [\hat{A},\hat{\psi}].
\end{equation}
It can be easily checked that the canonical commutation relations are satisfied:
\begin{equation}
[\hat{x}_\mu,\hat{p}_\nu]= -i\eta_{\mu\nu} \mbox{ with } 
\eta_{\mu\nu}=\eta_{\nu\mu}\mbox{ and } \eta_{01}=0, \eta_{00}=1,\eta_{11}=-1
\end{equation}
The Hamiltonian $\hat{H}$, in general, may depend on $\hat{x}_1^L, \hat{x}_0^R$ and $\hat{p}_1$.
The possible dependence of $\hat{x}_1^R$ and $\hat{x}_0^L$ can be bypassed by
\begin{equation}
 \hat{x}_{1}^{R} = \theta\hat{p}_0+\hat{x}_{1}^{L}, \quad
\hat{x}_{0}^{L} = \theta \hat{p}_{1} + \hat{x}_{0}^{R}. \label{lr0_relation}
\end{equation}
Also, there is no dependence on $\hat{p}_0$ assumed in the line of the commutative case 
where there is never such dependence of H on $i\partial_{x_0}$.
Now note that the inner product (\ref{S-inner-product}) has an explicit dependence on the parameter $t$ and hence there exist more than one 
null vectors with respect to this inner product (actually any vector which vanishes at $\hat{x}_0 = t$ is a null vector).
But this fact need not bother us as we are only interested in those states that satisfy the Schr\"{o}dinger constraint
\begin{equation}
\left(\hat{p}_0-\hat{H} \right) \hat{\psi} = 0.
\end{equation}
The Hamiltonian $\hat{H}$ depends on $\hat{x}_1^L, \hat{x}_0^R$ and $\hat{p}_1$.
Since $\hat{x}_0^R$ commutes with $\hat{x}_1^L$ and $\hat{p}_1$ we will choose $\hat{x}_0^R$ as ``time''.

It is easy to write down the formal solution of the Schr\"{o}dinger constraint and find the time evolution.
The time evolution is given by $\hat{x}_0\rightarrow \hat{x}_0+\tau$ (or equivalently by $\hat{x}_0^R\rightarrow \hat{x}_0^R+\tau$).
Thus the amount of time-translation is always commutative, though the time-operator itself is noncommutative.
The time evolved wave functions satisfying the Schr\"{o}dinger constraint are of the form
$\hat{\psi}(\hat{x}_0,\hat{x}_1)=\hat{U}\left(\hat{x}_{0}^{R},\tau_{I}\right)\hat{\chi}(\hat{x}_1)$, where
\begin{equation}
\hat{U}\left(\hat{x}_{0}^{R},\tau_{I}\right) =
\left(\left. T \exp\left[-i\left(\int_{\tau_{I}}^{x_{0}}
d\tau\,\hat{H}\left(\tau,\hat{x}_{1}^{L},\hat{p}_{1}
\right) \right) \right]\right) \right|_{x_{0} =
\hat{x}_{0}^{R}} \,\,.
\label{formula_psi_2}
\end{equation}
As $\hat{x}_0^R$ commutes with $\hat{x}_1^L$ and $\hat{p}_1$, the time dependence of the time evolved wave function given above will mimic the same for the commutative case $\theta=0$. We know in the commutative case a wave function which vanishes at some finite time $t$, will vanish for all times. Hence the only null vector satisfying the Schr\"{o}dinger constraint for the case of nonzero $\theta$, is $\hat{\psi}=0$ and there are no other non-trivial null vectors!

{\it 3. The Spectral Map:} \\
Consider a time-independent Hamiltonian $\hat{H} = \frac{\hat{p}_{1}^{2}}{2m}+V(\hat{x}_{1})$.
The corresponding commutative Hamiltonian is $H = -\frac{1}{2m}\frac{\partial^{2}}{\partial x_{1}^{2}}+V(x_{1})$,
with eigenfunctions $\psi_{E}\left(x_{0},x_{1}\right) = \varphi_{E}(x_{1})e^{-iE x_{0}}$ and eigenvalues E. 
The spectrum of the corresponding noncommutative $\hat{H}$ will be given by
$\hat{\psi}_{E}=e^{-iE\hat{x}_{0}^R}\varphi_{E}(\hat{x}_{1})=\varphi_{E}(\hat{x}_{1})e^{-iE\hat{x}_{0}}$
with the same eigenvalues $E$ as
$\hat{H}\varphi_{E}(\hat{x}_{1})=E\varphi_{E}(\hat{x}_{1})$.
Here $\varphi_E(\hat{x}_1)$ has been obtained by replacing $x_1$ with $\hat{x}_1$ in $\varphi_E(x_1)$.

\section{QFHO in $\mathds{R}_\theta^{1,1}$  and their Transition Probabilities} \label{problem}
Let us recall the dynamics of a QFHO in ordinary spacetime. For a detailed discussion one can look into the section 14.6 of \cite{Merz}. The Hamiltonian of this system is given by
\begin{equation}
H(t)=\frac{p^2}{2m_0}+\frac{1}{2}m_0\omega^2 x^2+f(t)x+g(t)p,
\label{cfho}
\end{equation} 
where $m_0$ is the mass of the particle and $\omega$ is the angular frequency of the oscillator.
We are interested in real functions obeying 
\begin{equation}
f(t),g(t)=0\mbox{   for } t\rightarrow\pm\infty.
\label{bc}
\end{equation}
At $t\rightarrow -\infty$ the Hamiltonian is simple harmonic 
and we assume the system to be in one of the eigenstates of this SHO Hamiltonian.
At $t\rightarrow \infty$ the Hamiltonian again becomes simple harmonic
and we try to find the probability (the transition probability) for the system to be in any arbitrary eigenstate of 
the SHO Hamiltonian subjected to the fact that the system was in some already given eigenstate at $t\rightarrow -\infty$.
For this what we do is the following:
\begin{itemize}
\item First we assume our system to be in an eigenstate $\phi_n(x)$ at $t=t_i\rightarrow -\infty$.
\item The state $\phi_n(x)$ evolves under the SHO Hamiltonian from $t=t_i\rightarrow -\infty$ to $t=T_1$.
\item At $t=T_1$ the interaction gets switched on.
\item The system then evolves under the full Hamiltonian (\ref{cfho}) from $t=T_1$ to $t=T_2$.
\item At $t=T_2$ the interaction gets switched off.
\item The system again evolves under the SHO Hamiltonian from $t=T_2$ to $t=t_f\rightarrow \infty$.
\item We find the inner product of the final state we get at $t=t_f\rightarrow \infty$ with the eigenstate $\phi_m(x)$.
This gives the Transition Amplitude $A_{mn}$ while its absolute square gives the Transition Probability $P_{mn}$.
\end{itemize}
The generalization of the above Hamiltonian in $\mathds{R}_\theta^{1,1}$ is
\begin{equation}
\hat{H} = \frac{\hat{p}_1^2}{2m_0}+\frac{1}{2}m_0\omega^2\hat{x}_1^2
+\frac{1}{2}[f(\hat{x}_0)\hat{x}_1+\hat{x}_1f(\hat{x}_0)]+g(\hat{x}_0)\hat{p}_1
 = \hat{H}_0+\hat{H}_I,
\label{fho}
\end{equation}
with
\begin{equation}
\hat{H}_0 = \frac{\hat{p}_1^2}{2m_0}+\frac{1}{2}m_0\omega^2\hat{x}_1^2, \quad
\hat{H}_I = \frac{1}{2}[f(\hat{x}_0)\hat{x}_1+\hat{x}_1f(\hat{x}_0)]+g(\hat{x}_0)\hat{p}_1.
\label{h_0_i}
\end{equation}
As $\hat{x}_0$ and $\hat{p}_1$ commute with each other, the ordering does not matter in the last term.\\
\indent To define the transitions for the above Hamiltonian consider the time evolution by an amount $\tau$.
The functions $f(\hat{x}_0)$ and $g(\hat{x}_0)$ have the properties of vanishing in the far past and the far future, i.e.,
\begin{equation}
f,g(\hat{x}_0+\tau)\rightarrow 0 \mbox{\,\,\, \rm{as}\,\,\,} \tau\rightarrow \pm \infty.
\end{equation}
We shall find the transition probabilities ($P_{m,n}$) for an SHO state ``$n$'' 
at initial time ($\tau\rightarrow -\infty$) to go to some other SHO state ``$m$'' at final time ($\tau\rightarrow +\infty$) 
after evolving under the Hamiltonian (\ref{fho}).
The Spectral Map tells us that the energy spectrum of the SHO Hamiltonian in $\mathds{R}_\theta^{1,1}$ is same as 
that of the commutative one, i.e., 
\begin{equation}
E_n = \hbar\omega\left(n+\frac{1}{2}\right), \quad  
\psi_n(\hat{x}_0,\hat{x}_1) = \phi_n(\hat{x}_1)e^{-i\omega\left(n+\frac{1}{2}\right)\hat{x}_0},
\label{En_psin}
\end{equation}
where $\phi_n(x_1)$ is the eigenfunctions of the commutative SHO Hamiltonian.
The orthonarmality of the eigenfunctions 
$\psi_n(\hat{x}_0,\hat{x}_1)$ with respect to the inner product defined in section \ref{bala_paper} can easily be checked and is shown explicitly in Appendix \ref{orthonormality}.

The transition probabilities for our problem can be found by computing the same 
for the commutative Hamiltonian obtained after replacing
\begin{equation}
\quad \hat{x}_1 \rightarrow  x , \quad \hat{p}_1  \rightarrow  p, \quad 
\hat{x}_0 = \hat{x}_0^L =  -\frac{\theta}{\hbar} \hat{p}_1+ \hat{x}_0^R \rightarrow -\frac{\theta}{\hbar} p+t,
\label{replace}
\end{equation}
in the Hamiltonian (\ref{fho}).
Here $t$ has come in place of the ``time'' $\hat{x}_0^R$ which commutes with $\hat{x}_1$ and $\hat{p}_1$. Also, we have retained the fundamental constant $\hbar$ while it was taken to be $1$ in section \ref{bala_paper}.
To linear order in $\theta$ we obtain the following commutative Hamiltonian
\begin{equation}
H(t)= H_0 + H_I(t) = H_0 + H_{I0}(t)+\theta H_{I1}(t),
\label{h}
\end{equation}
with
\begin{equation}
H_0 = \frac{p^2}{2m_0}+\frac{1}{2}m_0\omega^2 x^2 
 = \hbar\omega\left(a^\dagger a+\frac{1}{2}\right),  \label{h0}
 \end{equation}
 \begin{equation}
 H_{I0} = f(t)x+g(t)p = z^*(t)a+z(t)a^\dagger,  \label{hi0}
 \end{equation}
 and
 \begin{eqnarray}
H_{I1} &=& \frac{1}{\hbar}\left(-g^{\prime}(t)p^2-\frac{1}{2}f^{\prime}(t)(xp+px)\right) \nonumber \\
&=& \frac{i}{\hbar}\sqrt{\frac{m_0\hbar\omega}{2}}\left(z^{*\prime}(t)a^2-z^{\prime}(t)a^{\dagger 2}+
i\sqrt{\frac{m_0\hbar\omega}{2}}g^{\prime}(t)(2a^{\dagger} a+1)\right)  \label{hi1}
\end{eqnarray}
The function $z(t)$ is related to $f(t)$ and $g(t)$ as
\begin{equation}
z(t)=\sqrt{\frac{\hbar}{2m_0\omega}}\left(f(t)+im_0\omega g(t)\right).
\label{z}
\end{equation}
Also, $a$ and $a^\dagger$ are the annihilation and creation operators respectively defined as
\begin{equation}
\begin{array}{rcl}
a=\sqrt{\frac{m_0\omega}{2\hbar}}\left(x+i\frac{p}{m_0\omega}\right) 
&& x=\sqrt{\frac{\hbar}{2m_0\omega}}\left(a^\dagger+a\right) \\
&\Rightarrow& \\
a^\dagger=\sqrt{\frac{m_0\omega}{2\hbar}}\left(x-i\frac{p}{m_0\omega}\right) 
&& p=i\sqrt{\frac{m_0\hbar\omega}{2}}\left(a^\dagger-a\right)
\end{array}
\end{equation}
The nonlinearity in the Hamiltonian (\ref{h}) is purely due to the noncommutativity.
This provokes us to model certain types of nonlinear phenomena in quantum optics by the noncommutativity between time and space coordinates.
This analogy will be further studied in section \ref{quantum_optics}. 
Let us now continue with calculating the transition amplitude which is given by
\begin{equation}
A_{m,n}(t_f,T_2;T_1,t_i)=\langle \phi_m|U_0^{\dagger}(t_f,t_i)U_0(t_f,T_2)U(T_2,T_1)U_0(T_1,t_i)|\phi_n\rangle,
\label{A}
\end{equation}  
where $U_0(t^{\prime},t)$ and $U(t^{\prime},t)$ are the time evolution operators from time $t$ to time $t^{\prime}$ for 
the Hamiltonians $H_0$ and $H(t)$ respectively, i.e.,
\begin{equation}
 U_0(t^{\prime},t) = e^{-\frac{i}{\hbar}H_0(t^{\prime}-t)}, \quad  
U(t^{\prime},t) = T\left[e^{-\frac{i}{\hbar}\int_{t}^{t^{\prime}}
d\tau H(\tau)}\right],
\end{equation}
the latter one being the time-ordered exponential.
This gives
\begin{equation}
A_{m,n}(t_f,T_2;T_1,t_i)=e^{\frac{i}{\hbar}[E_m(T_2-t_i)+E_n(t_i-T_1)]}\langle\phi_m|U(T_2,T_1)|\phi_n\rangle.
\label{A_}
\end{equation}
The state $|\psi(t)\rangle = U(t,T_1)|\phi_n\rangle$ evolves according to the Schr\"{o}dinger equation for the Hamiltonian (\ref{h}) 
\begin{equation}
\left( i\hbar \frac{d}{dt}-H_0\right)|\psi(t)\rangle = H_I(t)|\psi(t)\rangle,
\label{Schro}
\end{equation}
with the initial condition $|\psi(t=T_1)\rangle=|\phi_n\rangle$.
If we define the Green's operator function $G(t,t_0)$ as
\begin{equation}
\left( i\hbar \frac{\partial}{\partial t}-H_0\right)G(t,t_0)=\delta (t-t_0),
\label{G}
\end{equation}
then solution of the Schr\"{o}dinger  equation (\ref{Schro}) will be
\begin{equation}
|\psi(t)\rangle =|\phi(t)\rangle +\int_{-\infty}^{+\infty} dt_0 \,\,G(t,t_0)H_I(t_0)|\psi(t_0)\rangle,
\end{equation}
which in turn gives the Born series
\begin{eqnarray}
|\psi(t)\rangle &=& |\phi(t)\rangle +\int_{-\infty}^{+\infty} dt_0 G(t,t_0)H_I(t_0)|\phi(t_0)\rangle \nonumber \\
&&+\int_{-\infty}^{+\infty} dt_0 \int_{-\infty}^{+\infty} dt_1 G(t,t_0)H_I(t_0)G(t_0,t_1)H_I(t_1)|\phi(t_1)\rangle+.... \label{Born}
\end{eqnarray}
Here $|\phi(t)\rangle$ is the solution of the homogeneous equation 
\begin{equation} \label{phi_t}
\left( i\hbar \frac{d}{dt}-H_0\right)|\phi(t)\rangle=0
\end{equation}
which is nothing but the Schr\"{o}dinger  equation for SHO.
$G$ has been found in the Appendix \ref{greens} (see (\ref{G_sol})).
Note that the $\Theta$-function in the expression of the $G$  restricts the integration over $t_j$ in (\ref{Born}) 
within the limit of $-\infty$ to $t_{j-1}$ ($t_{-1}=t$).
Thus, at $t=T_1$ the integrations are only in the intervals when the interaction was switched off, i.e., $H_I=0$.
Hence, we get 
\begin{equation} \label{phi_T1}
|\psi(t=T_1)\rangle=|\phi(t=T_1)\rangle=|\phi_n\rangle.
\end{equation}
The solution of the homogeneous equation (\ref{phi_t}) with the initial condition (\ref{phi_T1}) is given by
\begin{equation}\label{hom_sol}
|\phi(t)\rangle=e^{-\frac{i}{\hbar}E_n(t-T_1)}|\phi_n\rangle.
\end{equation}
Putting this back in (\ref{Born}) we get
\begin{eqnarray}
|\psi(t)\rangle &=& e^{-\frac{i}{\hbar}E_n(t-T_1)}|\phi_n\rangle +\int_{-\infty}^{+\infty} dt_0 G(t,t_0)H_I(t_0)|\phi(t_0)\rangle \nonumber \\
&&+\int_{-\infty}^{+\infty} dt_0 \int_{-\infty}^{+\infty} dt_1 G(t,t_0)H_I(t_0)G(t_0,t_1)H_I(t_1)|\phi(t_1)\rangle+.... \label{Born1}
\end{eqnarray}
For $t=T_2$ we get
\begin{eqnarray}
U(T_2,T_1)|\phi_n\rangle = |\psi(T_2)\rangle &=& e^{-\frac{i}{\hbar}E_n(T_2-T_1)}|\phi_n\rangle +\int_{-\infty}^{+\infty} dt_0 G(T_2,t_0)H_I(t_0)|\phi(t_0)\rangle \nonumber \\
&&+\int_{-\infty}^{+\infty} dt_0 \int_{-\infty}^{+\infty} dt_1 G(T_2,t_0)H_I(t_0)G(t_0,t_1)H_I(t_1)|\phi(t_1)\rangle \nonumber \\
&& +.... \nonumber \\
\label{Born_T2}
\end{eqnarray}
We put this back in (\ref{A_}) to get
\begin{equation} \label{A__}
A_{m,n}(t_f,T_2;T_1,t_i)=\sum_{j=0}^{\infty}B_j(t_f,T_2;T_1,t_i),
\end{equation}
with
\begin{equation}
B_0(t_f,T_2;T_1,t_i) = \delta_{mn}
\end{equation}
and
\begin{equation}
B_j(t_f,T_2;T_1,t_i) = \int_{-\infty}^{+\infty}dt_0\int_{-\infty}^{+\infty}dt_1...\int_{-\infty}^{+\infty}dt_{j-1} F_{m,n}^{j}(t_f,T_2;t_0,t_1,...,t_{j-1};T_1,t_i)
\end{equation}
for $j=1,2,...$.
Here
\begin{eqnarray}
 F_{m,n}^{j}(t_f,T_2;t_0,t_1,...,t_{j-1};T_1,t_i) &=& e^{\frac{i}{\hbar}[E_m(T_2-t_i)+E_n(t_i-T_1)]} \langle \phi_m | G(T_2,t_0) H_I (t_0) \nonumber \\
 && G(t_0,t_1) H_I(t_1)... G(t_{j-2},t_{j-1}) H_I(t_{j-1}) | \phi(t_{j-1})\rangle \nonumber \\
\end{eqnarray}
We put the expressions for all $G(t-t')$'s as found in (\ref{G_sol}) to get
\begin{eqnarray}
 F_{m,n}^{j}(t_f,T_2;t_0,t_1,...,t_{j-1};T_1,t_i) &=&
\left(-\frac{i}{\hbar}\right)^j\Theta (T_2-t_0)\Theta (t_0-t_1)...\Theta (t_{j-2}-t_{j-1}) \nonumber \\
 && e^{\frac{i}{\hbar}\left[(E_n-E_m) t_i - E_n T_1\right]} \langle \phi_m|H_I^{int}(t_0)H_I^{int}(t_1)... \nonumber \\
 && H_I^{int}(t_{j-1}) e^{\frac{i}{\hbar}H_0 t_{j-1}}|\phi(t_{j-1})\rangle.
\end{eqnarray}
The $H^{int}$'s are defined as
\begin{equation} \label{int}
H_{(...)}^{int}(t) = e^{\frac{i}{\hbar}H_0t}H_{(...)}(t)e^{-\frac{i}{\hbar}H_0t}.
\end{equation}
We put $|\phi(t_{j-1})\rangle = e^{-\frac{i}{\hbar}E_n(t_{j-1}-T_1)}|\phi_n\rangle$ (see (\ref{hom_sol})) to get
\begin{eqnarray}
 F_{m,n}^{j}(t_f,T_2;t_0,t_1,...,t_{j-1};T_1,t_i) &=& 
\left(-\frac{i}{\hbar}\right)^j\Theta (T_2-t_0)\Theta (t_0-t_1)...\Theta (t_{j-2}-t_{j-1})\nonumber  \\
&& e^{\frac{i}{\hbar}(E_n-E_m)t_i}\langle \phi_m|H_I^{int}(t_0)H_I^{int}(t_1)...H_I^{int}(t_{j-1})|\phi_n\rangle. \nonumber \\
\end{eqnarray}
Writing $H_I(t) = H_{I0}(t) + \theta H_{I1}(t)$ (see (\ref{h})) and hence $H_I^{int}(t) = H_{I0}^{int}(t) + \theta H_{I1}^{int}(t)$ to separate out the $\theta$-dependent and independent parts up to linear order in $\theta$, gives 
\begin{equation}
A_{m,n}(t_f,T_2;T_1,t_i)=e^{\frac{i}{\hbar}(E_n-E_m)t_i}\langle \phi_m|[A^{(0)}(T_2,T_1)+\theta A^{(1)}(T_2,T_1)]|\phi_n\rangle,
\label{Amn}
\end{equation}
with
\begin{eqnarray}
A^{(0)}(T_2,T_1) &=& \mathbf{I}+\int_{-\infty}^{+\infty}dt_0\left( -\frac{i}{\hbar}\right) \Theta (T_2-t_0)H_{I0}^{int}(t_0) \nonumber \\
&&+\int_{-\infty}^{+\infty}dt_0\int_{-\infty}^{+\infty}dt_1\left( -\frac{i}{\hbar}\right)^2 \Theta (T_2-t_0)\Theta (t_0-t_1)H_{I0}^{int}(t_0)H_{I0}^{int}(t_1) \nonumber \\
&& +... \nonumber \\
&=& T\left[ e^{-\frac{i}{\hbar}\int_{-\infty}^{\infty}d\tau H_{I0}^{int}(\tau)}\right] \label{A0},
\end{eqnarray}
\begin{eqnarray}
A^{(1)}(T_2,T_1)]&=& -\frac{i}{\hbar}\int_{-\infty}^{+\infty}dt_0\Theta (T_2-t_0)H_{I1}^{int}(t_0) \nonumber \\
&& +\left(-\frac{i}{\hbar}\right)^2\int_{-\infty}^{+\infty}dt_0\int_{-\infty}^{+\infty}dt_1\Theta (T_2-t_0)\Theta (t_0-t_1) \nonumber \\
&& \hspace{2 cm}.[H_{I1}^{int}(t_0)H_{I0}^{int}(t_1) + H_{I0}^{int}(t_0) H_{I1}^{int}(t_1)] \nonumber \\
&& +\left(-\frac{i}{\hbar}\right)^3\int_{-\infty}^{+\infty}dt_0\int_{-\infty}^{+\infty}dt_1\int_{-\infty}^{+\infty}
dt_2\Theta (T_2-t_0)\Theta (t_0-t_1)\Theta (t_1-t_2) \nonumber \\
&& \,\,\,\,\, \quad \quad \quad .\left[H_{I1}^{int}(t_0)H_{I0}^{int}(t_1)H_{I0}^{int}(t_2) + H_{I0}^{int}(t_0)H_{I1}^{int}(t_1)H_{I0}^{int}(t_2) \right. \nonumber \\
&& \left. \quad \quad \quad \quad \quad \quad \quad \quad \quad \quad \quad \quad \quad \quad \quad + H_{I0}^{int}(t_0)H_{I0}^{int}(t_1)H_{I1}^{int}(t_2)\right] \nonumber \\
& &+....  \label{A1_series}
\end{eqnarray}
$H_{I0}(t)$ and $H_{I1}(t)$ are again defined in accordance to (\ref{int}). The arguments of $A^{(0)}(T_2,T_1)$ do not come in the formal expression as the integrands of all the integrals vanish for $t>T_2$.
The expression for $A^{(1)}(T_2,T_1)$ can be simplified to (see Appendix \ref{simplify_A1}) 
\begin{equation} \label{A1_A0}
A^{(1)}(T_2,T_1)=-\frac{i}{\hbar}A^{(0)}(T_2,T_1)\int_{-\infty}^{\infty}dt_0
\left[A^{(0)}(t_0,T_1)\right]^{-1}H_{I1}^{int}(t_0)A^{(0)}(t_0,T_1).
\end{equation}
where $A^{(0)}(t,t^{\prime})$ with arbitrary arguments is defined in (\ref{Gint_sol}).
Putting this in equation (\ref{Amn}) we get
\begin{eqnarray}
A_{m,n}(t_f,T_2;T_1,t_i) &=& 
e^{\frac{i}{\hbar}(E_n-E_m)t_i}\langle \phi_m|A^{(0)} \nonumber \\
&& \left[\mathbf{I}-\frac{i}{\hbar}\theta 
\int_{-\infty}^{\infty}dt_0[A^{(0)}(t_0,T_1)]^{-1}H_{I1}^{int}(t_0)A^{(0)}(t_0,T_1)\right]|\phi_n\rangle. \nonumber \\
\label{Amn_A0}
\end{eqnarray}
Here we have removed the arguments of $A^{(0)}(T_2,T_1)$ as they don't come in it's formal expression.
For the next few steps we are going to use the following identity extensively:
\begin{equation}
e^{\lambda A}Be^{-\lambda A}=B+\frac{\lambda}{1!}[A,B]+\frac{\lambda^2}{2!}\left[A,[A,B]\right]+\frac{\lambda^3}{3!}\left[A,\left[A,[A,B]\right]\right]+...
\label{identity}
\end{equation}
First of all to find $H_{I0}^{int}(t) $ and $H_{I1}^{int}(t) $ we calculate the following commutators:
\begin{eqnarray}
[H_0,H_{I0}(t)] &=& \hbar\omega \left(-z^*(t)a+z(t)a^{\dagger}\right) \\
\left[H_0,[H_0,H_{I0}(t)]\right] &=& \hbar^2\omega^2 H_{I0}(t) \\
&& ...\rm{and\,\,\,so\,\,\,on}. \nonumber \\
\left[H_0,H_{I1}(t)\right] &=& -2i\omega\sqrt{\frac{m_0\hbar\omega}{2}}\left(z^{*\prime}(t)a^2+z^{\prime}(t)a^{\dagger 2}\right) \\
\left[H_0,[H_0,H_{I1}(t)]\right] &=& 4\hbar\omega^2 \left(H_{I1}(t)+m_0g^{\prime}(t)H_0\right) \\
\left[H_0,\left[H_0,[H_0,H_{I1}(t)]\right]\right] &=& 4\hbar\omega^2[H_0,H_{I1}(t)] \\
&& ...\rm{and\,\,\,so\,\,\,on}. \nonumber
\end{eqnarray}
Now, using above commutators in the identity (\ref{identity}) gives
\begin{eqnarray}
H_{I0}^{int}(t) &=& e^{\frac{i}{\hbar}H_0t}H_{I0}(t)e^{-\frac{i}{\hbar}H_0t} \nonumber \\
&=& z^*(t)e^{-i\omega t}a+z(t)e^{i\omega t}a^{\dagger}
\end{eqnarray}
and
\begin{eqnarray}
H_{I1}^{int}(t) &=& e^{\frac{i}{\hbar}H_0t}H_{I1}(t)e^{-\frac{i}{\hbar}H_0t} \nonumber \\
&=& \frac{i}{\hbar}\sqrt{\frac{m_0\hbar\omega}{2}}z^{*\prime}(t)e^{-2i\omega t}a^2-\frac{i}{\hbar}\sqrt{\frac{m_0\hbar\omega}{2}}z^{\prime}(t)e^{2i\omega t}a^{\dagger 2} \nonumber \\
&& -\frac{m_0\omega}{2}g^{\prime}(t)\left( 2a^{\dagger}a+\mathbf{I}\right)
\end{eqnarray}
We put the expression of $H_{I0}^{int}(t)$ given above to find $A^{(0)}(t,t_0)$ (see (\ref{Gint_sol})) and follow the discussions given in pages 338-340 of \cite{Merz} to eliminate the time ordering. This gives
\begin{equation}
A^{(0)}(t,t_0) = e^{i\beta(t,t_0)}e^{-\xi^*(t,t_0)a+\xi(t,t_0)a^\dagger}
\end{equation}
with
\begin{equation}
\beta(t,t_0) = \frac{i}{2\hbar^2}\displaystyle{\int_{t_0}^t} d\tau_1\displaystyle{\int_{t_0}^{\tau_1}}d\tau_2 \left[z^*(\tau_1)z(\tau_2)e^{-i\omega(\tau_1-\tau_2)} - z(\tau_1)z^*(\tau_2)e^{i\omega(\tau_1-\tau_2)}\right]
\end{equation}
and
\begin{equation}
\xi(t,t_0) = -\frac{i}{\hbar}\displaystyle{\int_{t_0}^{t}}d\tau\, z(\tau) e^{i\omega\tau}
\end{equation}
The limit $t_0\rightarrow -\infty, t \rightarrow \infty$ gives the expression for $A^{(0)}$.
Now to simplify the integrand in (\ref{Amn_A0}) we find the following commutators
\begin{eqnarray*}
&[\xi^*(t_0,T_1)a-\xi(t_0,T_1)a^{\dagger},H_{I1}^{int}(t_0)]& \\
&=& \\
&\frac{2}{\hbar}\sqrt{\frac{m_0\hbar\omega}{2}}\left[ \left( -\sqrt{\frac{m_0\hbar\omega}{2}}g^{\prime}(t_0)\xi^*(t_0,T_1)+iz^{*\prime}(t_0)\xi(t_0,T_1)e^{-2i\omega t_0}\right)a\right.&\\
&\,\,\,\,\,\,\,\,\,\,\,\,\,\,\,\,\,\,\,\,\,\,\,\,\,\,\,\,\,\,\,\,\,\,\,\,
\left.+\left( -\sqrt{\frac{m_0\hbar\omega}{2}}g^{\prime}(t_0)\xi(t_0,T_1)-iz^{\prime}(t_0)\xi^*(t_0,T_1)e^{2i\omega t_0}\right)a^{\dagger}\right]& \\
\end{eqnarray*}
and
\begin{eqnarray*}
&\left[ \xi^*(t_0,T_1)a-\xi(t_0,T_1)a^{\dagger},[\xi^*(t_0,T_1)a-\xi(t_0,T_1)a^{\dagger},H_{I1}^{int}(t_0)] \right]& \\
&=& \\
&-2m_0\omega g^{\prime}(t_0)|\xi(t_0,T_1)|^2+\frac{i}{\hbar}\sqrt{2m_0\hbar\omega}\left\{ z^{*\prime}(t_0)\xi^2(t_0,T_1)e^{-2i\omega t_0}-z^{\prime}(t_0)\xi^{*2}(t_0,T_1)e^{2i\omega t_0}\right\}& \\
\end{eqnarray*}
All higher order commutators vanish! Thus we get
\begin{eqnarray}
[A^{(0)}(t_0,T_1)]^{-1}H_{I1}^{int}(t_0)A^{(0)}(t_0,T_1) &=&
\alpha_1(t_0,T_1)\mathbf{I} \nonumber \\
&& +\alpha_2(t_0,T_1)a+\alpha_2^*(t_0,T_1)a^{\dagger} \nonumber \\
&& +\alpha_3(t_0,T_1)a^2+\alpha_3^*(t_0,T_1)a^{\dagger 2} \nonumber \\
&& +\alpha_4(t_0,T_1)a^{\dagger}a \label{A0_H_A0}
\end{eqnarray}
with
\begin{eqnarray}
\alpha_1(t_0,T_1) &=& -\frac{m_0\omega}{2}g^{\prime}(t_0)-m_0\omega g^{\prime}(t_0)|\xi(t_0,T_1)|^2 \nonumber \\
&& \!\!\!\!\!\!\!\!\!\!\!\!\!\!\!\!\!\!\!\!\!\!\!\! +\frac{i}{\hbar}\sqrt{\frac{m_0\hbar\omega}{2}}\left[ z^{*\prime}(t_0)\xi^2(t_0,T_1)e^{-2i\omega t_0}-z^{\prime}(t_0)\xi^{*2}(t_0,T_1)e^{2i\omega t_0}\right] \label{alpha1} \\
\alpha_2(t_0,T_1) &=& -m_0\omega g^{\prime}(t_0)\xi^*(t_0,T_1)  \nonumber \\
&& +\frac{2i}{\hbar}\sqrt{\frac{m_0\hbar\omega}{2}}z^{*\prime}(t_0)\xi(t_0,T_1)e^{-2i\omega t_0}  \label{alpha2} \\
\alpha_3(t_0,T_1) &=& \frac{i}{\hbar}\sqrt{\frac{m_0\hbar\omega}{2}}z^{*\prime}(t_0)e^{-2i\omega t_0}  \label{alpha3} \\
\alpha_4(t_0,T_1) &=& -m_0\omega g^{\prime}(t_0) \label{alpha4}
\end{eqnarray}
Finally we get the expression of the transition amplitude as
\begin{eqnarray}
A_{m,n}(t_f,T_2;T_1,t_i) =& e^{i\beta}e^{\frac{i}{\hbar}(E_n-E_m)t_i}\left[D_{m,n}(\xi)
-\frac{i}{\hbar}\theta \{\beta_1 D_{m,n}(\xi)\right. +\beta_2\sqrt{n}D_{m,n-1}(\xi) \nonumber \\
 &+\beta_2^*\sqrt{n+1}D_{m,n+1}(\xi) 
+\beta_3\sqrt{n(n-1)}D_{m,n-2}(\xi) \nonumber \\
&\left.+\beta_3^* \sqrt{(n+1)(n+2)}D_{m,n+2}(\xi)\}\right],
\label{Amn_final}
\end{eqnarray}
with
\begin{eqnarray}\label{beta_xi}
\xi &=& -\frac{i}{\hbar}\int_{-\infty}^{\infty}d\tau e^{i\omega \tau}z(\tau) \\
\beta &=& \frac{i}{2\hbar^2}\int_{-\infty}^{\infty}d\tau_1\int_{-\infty}^{\infty}d\tau_2\left[ z^*(\tau_1)z(\tau_2)
e^{-i\omega (\tau_1-\tau_2)}-z(\tau_1)z^*(\tau_2)e^{i\omega (\tau_1-\tau_2)}\right] \nonumber \\
&=&  {\rm real} \\
\beta_1 &=& -m_0\omega \int_{-\infty}^{+\infty}d\tau g^{\prime}(\tau)|\xi(\tau)|^2 +\frac{i}{\hbar}\sqrt{\frac{m_0\hbar\omega}{2}} \nonumber \\
&& \quad \quad \quad \quad \quad \quad \int_{-\infty}^{+\infty}d\tau 
\left[ z^{*\prime}(\tau)\xi^2(\tau)e^{-2i\omega \tau}-z^{\prime}(\tau)\xi^{*2}(\tau)e^{2i\omega \tau}\right]   \label{beta1}\\
\beta_2 &=& -m_0\omega \int_{-\infty}^{+\infty}d\tau g^{\prime}(\tau)\xi^*(\tau)
+\frac{2i}{\hbar}\sqrt{\frac{m_0\hbar\omega}{2}}\int_{-\infty}^{+\infty}d\tau z^{*\prime}(\tau)\xi(\tau)e^{-2i\omega \tau}  \\
\beta_3 &=&  \frac{i}{\hbar}\sqrt{\frac{m_0\hbar\omega}{2}}\int_{-\infty}^{+\infty}d\tau z^{*\prime}(\tau)e^{-2i\omega \tau}
\end{eqnarray}
Here the function $\xi(t)$ is given as
\begin{equation} 
\xi(t) = -\frac{i}{\hbar}\int_{-\infty}^{t}d\tau e^{i\omega \tau}z(\tau).
\label{xi_t_0}
\end{equation}
$D_{m,n}(\xi)$'s are the matrix elements of the displacement operator $D(\xi)=e^{-\xi^*a+\xi a^\dagger}$ given by \cite{Perelomov}
\begin{equation}
D_{m,n}(\xi)=\sqrt{\frac{n!}{m!}}e^{-\frac{1}{2}|\xi|^2}\xi^{m-n}L_{n}^{m-n}(|\xi|^2),
\end{equation}
$L_{n}^{k}(x)$ are the associated Laguerre polynomials.
Also, the limits of the integrations have been extended to $-\infty$ and $\infty$ as the integrands are zero in the extended region. Note that the contribution from $\alpha_4$ given in (\ref{alpha4}) to the transition amplitude vanishes by virtue of (\ref{bc}).
The transition probability is given by
\begin{equation}\label{Pmn}
P_{m,n} =|A_{m,n}(t_f,T_2;T_1,t_i)|^2
\end{equation}
as usual. The arguments have been omitted as the formal expression of transition probability does not contain the times $t_f$,$T_2$;$T_1$,$t_i$.

\subsection{$n=0$} \label{n=0}
\begin{figure*}[h]
\begin{center}
\scalebox{0.5}{\includegraphics{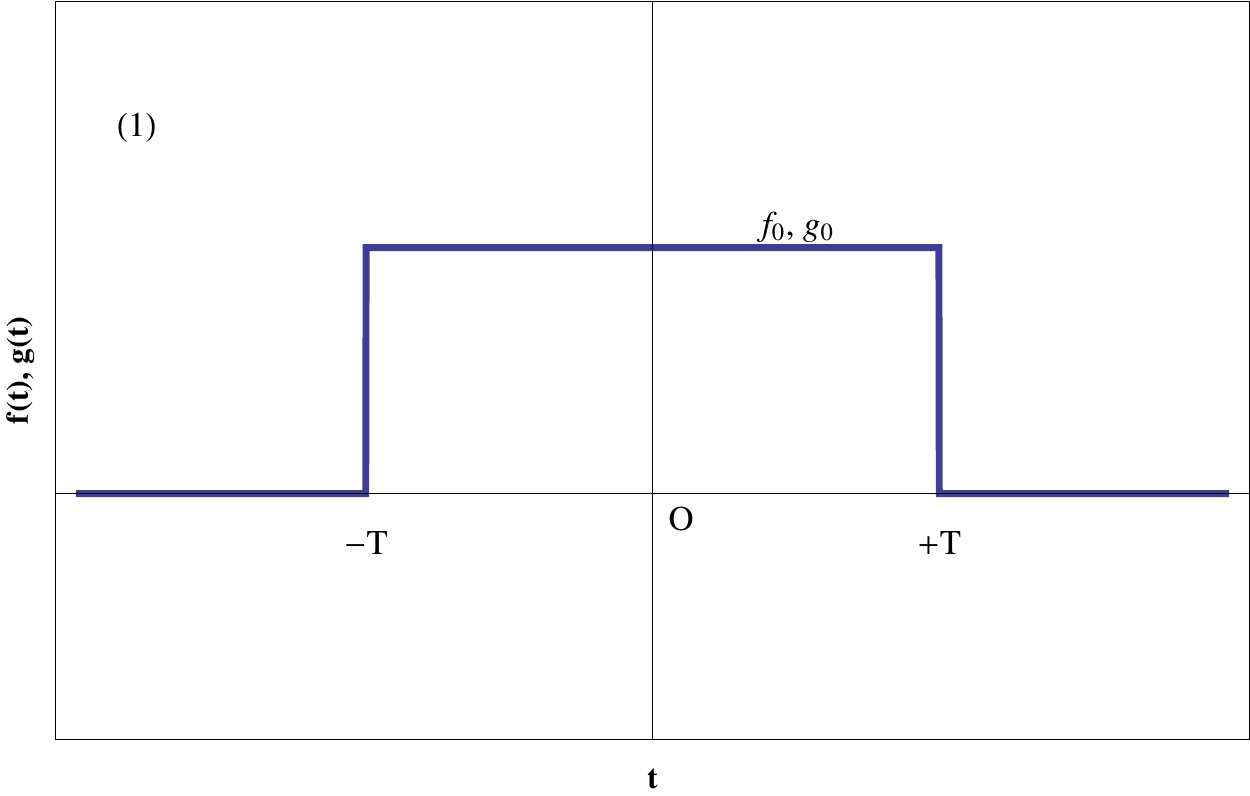}}
\scalebox{0.8}{\includegraphics{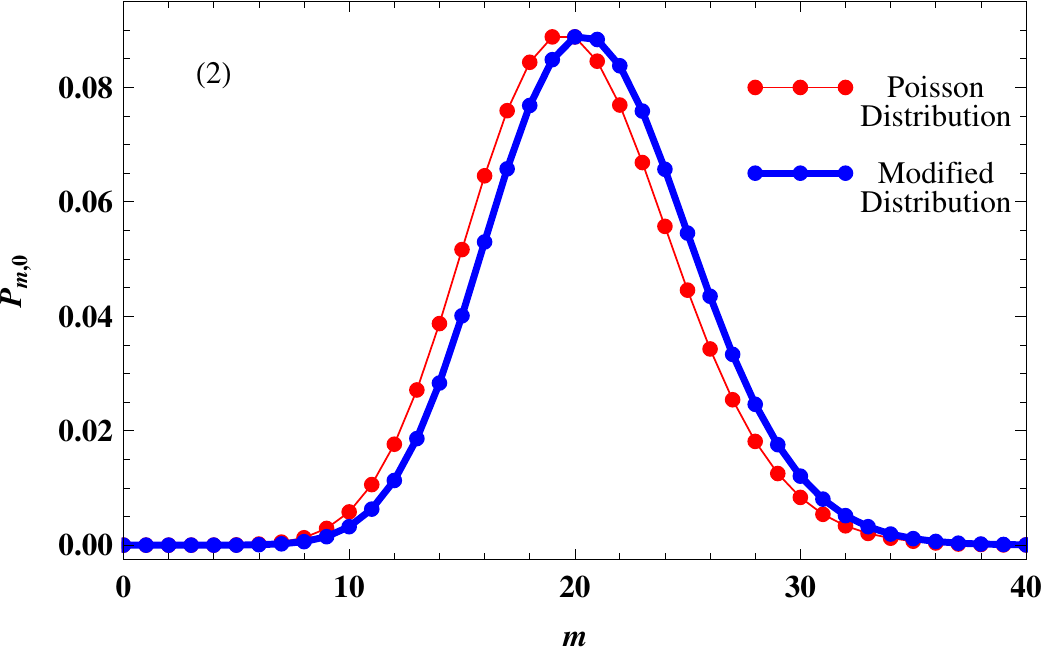}}
\end{center}
\caption{\label{fig_fg} The behaviour of functions $f(t)$ and $g(t)$ with $t$}
\caption{\label{fig_distribution} The modified distribution ($\theta=0.01$) along with the Poisson distribution ($\theta=0$) for
 $m_0=1$, $\omega=1$, $f_0=\sqrt{5}$, $g_0=\sqrt{5}$, $T=\frac{\pi}{2}$ and $\hbar=1$}
\end{figure*}

For vacuum $|\phi_0\rangle$ as the initial state, the transition amplitude is
\begin{eqnarray}
A_{m,0}(t_f,T_2;T_1,t_i) &=& e^{i\beta}e^{\frac{i}{\hbar}(E_0-E_m)t_i}e^{-\frac{|\xi|^2}{2}}\frac{\xi^m}{\sqrt{m!}}
\left[ 1-\frac{i}{\hbar}\theta \{ (\beta_1-\beta_2^*\xi^*+\beta_3^*\xi^{* 2}) \right. \nonumber \\
&&\left.  +m\frac{1}{|\xi|^2}(\beta_2^*\xi^*-2\beta_3^*\xi^{* 2}) 
+m(m-1)\frac{1}{|\xi|^4}\beta_3^*\xi^{* 2}\}\right].
\label{Am0}
\end{eqnarray}
We have used the explicit forms of the following associated Laguerre polynomials:
\begin{eqnarray}
L_0^m(x) &=& 1 \\
L_1^{m-1}(x) &=& m-x \\
L_2^{m-2}(x) &=& \frac{1}{2}\left[x^2-2mx+m(m-1)\right]
\end{eqnarray}
The transition probability becomes (upto linear order in $\theta$)
\begin{equation}
P_{m,0} = |A_{m,0}(t_f,T_2;T_1,t_i)|^2 = e^{-|\xi|^2}\frac{|\xi|^{2m}}{m!}\left[ 1
+\frac{2}{\hbar}\theta\{ A_1+mA_2+m(m-1)A_3\}\right],
\label{Pm0} 
\end{equation}
with
\begin{equation}
A_1 = Im(\beta_2\xi)-Im(\beta_3\xi^2), \quad A_2 = \frac{1}{|\xi|^2}\left(2Im(\beta_3\xi^2)-Im(\beta_2\xi)\right), \quad 
A_3 =-\frac{1}{|\xi|^4}Im(\beta_3\xi^2).
 \label{A3}
\end{equation}
Note that as $m\rightarrow \infty$, the $\theta$-correction starts dominating and in this case the expansion 
upto linear order in $\theta$ is no more meaningful!
Hence, the above result is valid only for those $m$-values which are far smaller than $1/\sqrt{m_0\omega\theta}$ (in the unit $\hbar=1$). 
For $\theta\rightarrow 0$, the transition probability becomes the well known Poisson distribution as expected. \\
\indent As a specific example let us work with the functions $f(t)$ and $g(t)$ of the form (see Figure \ref{fig_fg})
\begin{equation} \label{fg_expression}
\left.
\begin{array}{rcl}
 f(t) &=& f_0\left[\Theta (t+T)-\Theta (t-T)\right] \\
 g(t) &=& g_0\left[\Theta (t+T)-\Theta (t-T)\right]
\end{array}
\right\rbrace \quad\quad ;T>0.
\end{equation}
This gives
\begin{equation}
z(t) = z_0\left[\Theta (t+T)-\Theta (t-T)\right] , \quad z_0 = \sqrt{\frac{\hbar}{2m_0\omega}}\left[f_0+im_0\omega g_0\right]
\end{equation}
Their derivatives are
\begin{equation}
x'(t) = x_0 \left[\delta (t+T)-\delta (t-T)\right], \quad \mbox{{\rm for }} x=f,g,z.
\end{equation}
For these functions we calculate $\xi(t)$ to be
\begin{equation}
\xi(t) = \left\{
\begin{array}{cl}
0 & ;t<-T \\
-\frac{z_0\left(e^{i\omega t}-e^{-i\omega T}\right)}{\hbar\omega} & ;-T\leq t \leq T \\
-\frac{2iz_0\sin\omega T}{\hbar\omega} & ; t > T
\end{array} \right.
\end{equation}
We again calculate the following quantities:
\begin{equation}
\beta_2 = \frac{2}{\hbar} \sin \omega T \left[im_0 g_0 z_0^* - \sqrt{\frac{2m_0}{\hbar\omega}}|z_0|^2e^{-2i\omega T}\right]
\end{equation}
\begin{equation}
\beta_3 = -\sqrt{\frac{2m_0\omega}{\hbar}}\,z_0^*\sin 2\omega T
\end{equation}
\begin{equation}
\xi = -\frac{2iz_0\sin \omega T}{\hbar \omega}
\end{equation}
which further give
\begin{equation}
Im(\beta_2 \xi) = \frac{4}{\hbar^2\omega}\sqrt{\frac{2m_0}{\hbar\omega}}|z_0|^2\sin^2\omega T \left[Re(z_0) \cos 2\omega T + Im(z_0)\sin 2\omega T\right]
\end{equation}
and 
\begin{equation}
Im(\beta_3 \xi^2) = \frac{4}{\hbar^2\omega}\sqrt{\frac{2m_0}{\hbar\omega}}|z_0|^2\sin^2\omega T \sin 2\omega T \,Im(z_0)
\end{equation}
Thus we get
\begin{eqnarray}
 A_1 &=& \frac{2f_0}{m_0 \hbar \omega^3}(f_0^2+m_0^2\omega^2 g_0^2)\sin^2{\omega T}\cos{2\omega T} \\
A_2 &=& m_0\omega g_0 \sin{2\omega T}-f_0\cos{2\omega T} \\
A_3 &=& -\frac{m_0^2\hbar\omega^4 g_0}{f_0^2+m_0^2\omega^2g_0^2}\cot{\omega T}.
\end{eqnarray}
 The choices $m_0=1$, $\omega=1$, $f_0=\sqrt{5}$, $g_0=\sqrt{5}$ and $T=\frac{\pi}{2}$ (in natural units, i.e., $\hbar = 1$) in commutative case ($\theta=0$) give the 
following Poisson distribution: $P_{m,0}=e^{-20}\frac{20^m}{m!}$,
while for nonzero $\theta$ the probability distribution modifies to $P_{m,0}=e^{-20}\frac{20^m}{m!}[1+2\theta\sqrt{5}(m-20)]$.
The $\theta$-correction becomes of the order of the $\theta$-independent part when $m$ approaches the value 
$\tilde{m}(\theta)=\left(20+\frac{1}{2\sqrt{5}\theta}\right)$. 
Hence, our result is valid only in the region where $m<\tilde{m}(\theta)$.
Note that $\tilde{m}(\theta)\sim\frac{1}{\theta}$ rather than $\frac{1}{\sqrt{\theta}}$ 
because the $A_3$ is identically zero for the choices taken.
We choose $\theta=0.01$ ($\tilde{m}(\theta=0.01)\approx 42$) and get
 \begin{equation}
 P_{m,0}=e^{-20}\frac{20^m}{m!}[1+0.02\sqrt{5}(m-20)].
\end{equation}
This deformed distribution along with the Poisson distribution is shown in Figure \ref{fig_distribution}.
Such deformation of the Poisson distribution suggests that the vacuum does not evolve to be a coherent state anymore.
To explore this further let us look at the time-evolution of position and momentum uncertainties.

\section{The time evolution of $\Delta_x$ and $\Delta_p$}\label{uncertainties}
\begin{figure*}[h]
\begin{center}
\scalebox{0.7}{\includegraphics{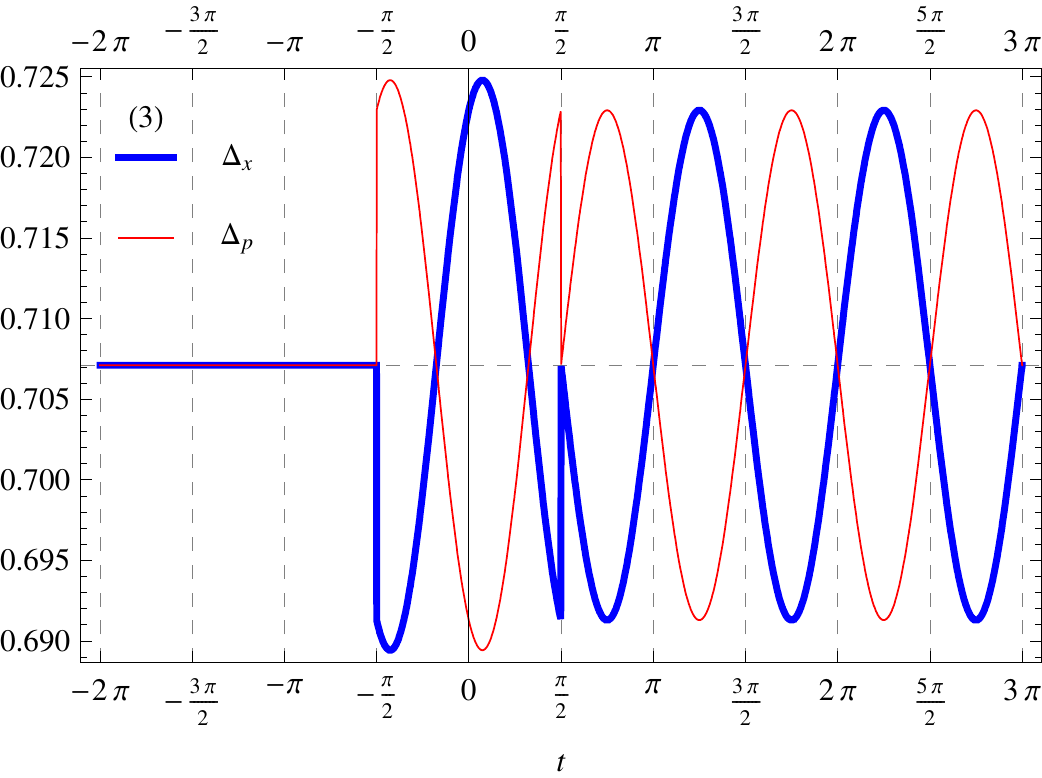}}
\scalebox{0.7}{\includegraphics{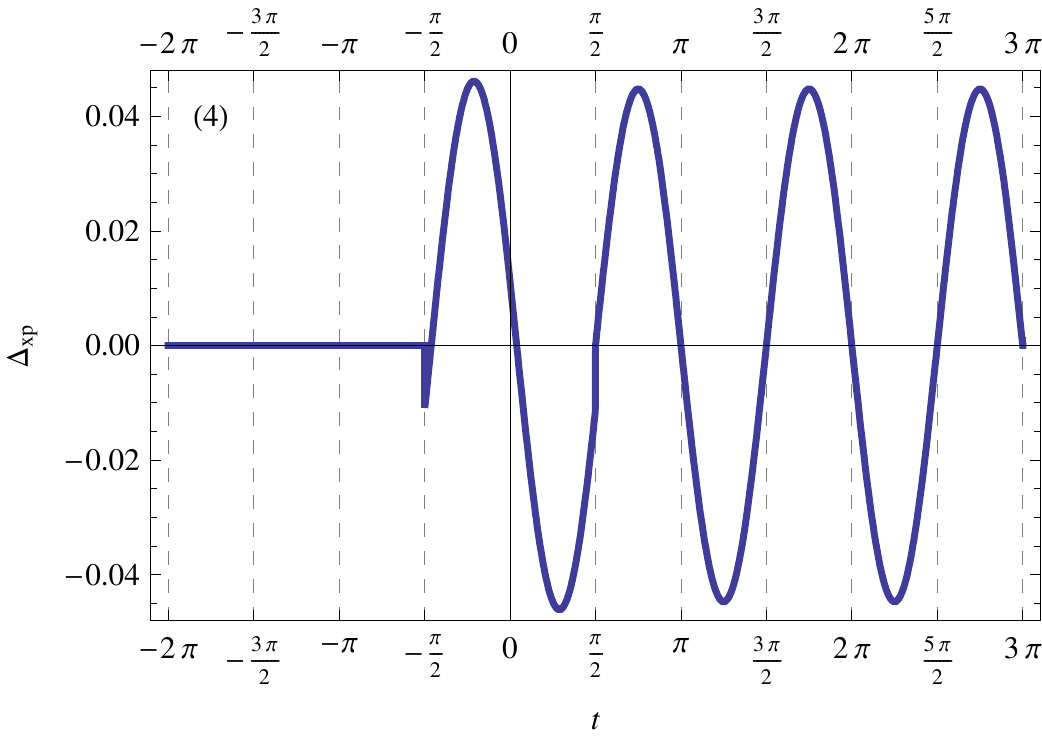}}
\end{center}
\caption{\label{fig_deltaxdeltap} The time-dependences of the uncertainties $\Delta_{x}$ and $\Delta_{p}$ 
for $m_0=1$, $\omega=1$, $f_0=\sqrt{5}$, $g_0=\frac{1}{2}\sqrt{5}$, $T=\frac{\pi}{2}$, $\theta=0.01$ and $\hbar = 1$}
\caption{\label{fig_deltaxp} The time-dependence of $\Delta_{xp}$ for the same choice of values}
\end{figure*}

The expectation value of any operator $\hat{\mathcal{O}}$ in a state $\hat{\psi}(\hat{x}_0,\hat{x}_1)$ at any time $t$ is defined to be
\begin{equation}
 \langle \hat{\mathcal{O}} \rangle_t =  \left(\hat{\psi},\hat{\mathcal{O}}\hat{\psi}\right)_t.
\end{equation}
Also,
\begin{equation}
 \langle \hat{\mathcal{O}} \rangle_{t+\tau} =  \left(\hat{\psi},\hat{\mathcal{O}}\hat{\psi}\right)_{t+\tau} =
\left(\hat{\psi}(\hat{x}_0+\tau,\hat{x}_1),\hat{\mathcal{O}}\hat{\psi}(\hat{x}_0+\tau,\hat{x}_1)\right)_t.
\end{equation}
Hence the time evolution of the expectation value of an operator is given by that of the state in which it is being calculated.
For the QFHO in $\mathds{R}_\theta^{1,1}$ the time evolution of any operator $\hat{\mathcal{O}}$ will be given by
\begin{equation}
\frac{d}{dt}\langle \hat{\mathcal{O}}\rangle=\frac{\partial}{\partial t}\langle \hat{\mathcal{O}}\rangle
+\frac{i}{\hbar}\left\langle\left[H(t),\hat{\mathcal{O}}\right]\right\rangle,
\end{equation} 
where $H(t)$ is the Hamiltonian (\ref{h}).
The uncertainty in any observable $\hat{\mathcal{O}}$ is given by
\begin{equation}
\Delta_{\mathcal{O}}=\sqrt{\langle\hat{\mathcal{O}}^2\rangle-\langle\hat{\mathcal{O}}\rangle^2}.
\end{equation}
For an observable not having an explicit dependence on time we will have
\begin{equation}
\frac{d}{dt}\Delta_{\mathcal{O}}^2 = \frac{i}{\hbar} \left(\langle [H(t),\hat{\mathcal{O}}^2]\rangle - 2\langle\hat{\mathcal{O}}\rangle\langle[H(t),\hat{\mathcal{O}}]\rangle\right)
\end{equation}
To find the evolution of $\Delta_x^2$ and $\Delta_p^2$ we calculate the following commutators
\begin{equation}
[H(t),x] = i\left(2\theta g'(t)-\frac{\hbar}{m_0}\right)p + i\theta f'(t)x - i\hbar g(t)
\end{equation}
\begin{equation}
[H(t),x^2] = i\left(2\theta g'(t)-\frac{\hbar}{m_0}\right)(xp+px) + 2i\theta f'(t) x^2 - 2i\hbar g(t)x
\end{equation}
\begin{equation}
[H(t),p] = i\hbar m_0\omega^2 x - i\theta f'(t) p + i\hbar f(t)
\end{equation}
\begin{equation}
[H(t),p^2] = i\hbar m_0\omega^2 (xp+px) - 2i\theta f'(t) p^2 + 2i\hbar f(t) p
\end{equation}
Thus the evolution of $\Delta_x^2$ and $\Delta_p^2$ is
\begin{equation}\left.
\begin{array}{rcl}
\frac{d}{dt}\Delta_x^2 &=& 2\left(\frac{1}{m_0}-\frac{2\theta}{\hbar} g^{\prime}(t)\right)\left(\frac{1}{2}\langle xp+px\rangle-\langle x\rangle\langle p\rangle\right)-\frac{2\theta}{\hbar} f^{\prime}(t)\Delta_x^2 \\
\frac{d}{dt}\Delta_p^2 &=& -2m_0\omega^2\left(\frac{1}{2}\langle xp+px\rangle-\langle x\rangle\langle p\rangle\right)+\frac{2\theta}{\hbar} f^{\prime}(t)\Delta_p^2 
\end{array}\right\rbrace.
\end{equation}
Let us define
\begin{equation}
\Delta_{xp}=\frac{1}{2}\langle xp+px\rangle-\langle x\rangle\langle p\rangle,
\end{equation}
Using
\begin{equation}
[H(t),xp+px] = 2i\left(2\theta g'(t)-\frac{\hbar}{m_0}\right)p^2 + 2i\hbar m_0 \omega^2 x^2 + 2i\hbar f(t) x - 2i\hbar g(t) p
\end{equation}
we find the following first order coupled equations
\begin{equation}\left.
\begin{array}{rcl}
\frac{d}{dt}\Delta_x^2 &=& 2\left(\frac{1}{m_0}-\frac{2\theta}{\hbar} g^{\prime}(t)\right)\Delta_{xp}-\frac{2\theta}{\hbar} f^{\prime}(t)\Delta_x^2 \\
\frac{d}{dt}\Delta_p^2 &=& -2m_0\omega^2\Delta_{xp}+\frac{2\theta}{\hbar} f^{\prime}(t)\Delta_p^2 \\
\frac{d}{dt}\Delta_{xp} &=& \left(\frac{1}{m_0}-\frac{2\theta}{\hbar} g^{\prime}(t)\right)\Delta_p^2 -m_0\omega^2\Delta_x^2 
\end{array}\right\rbrace.
\end{equation}
For vacuum as the initial state, the initial conditions for the above are
\begin{equation}
\Delta_x^2(t\rightarrow -\infty) = \frac{\hbar}{2m_0\omega}, \quad
\Delta_p^2(t\rightarrow -\infty) = \frac{m_0\hbar\omega}{2}, \quad
\Delta_{xp}(t\rightarrow-\infty) = 0 .
\end{equation} 
Our strategy for solving these equations is simple.
We do so perturbatively in $\theta$. We write the perturbative expansion of the solution up to linear order in $\theta$
\begin{equation}
\Delta_x^2 = \Delta_x^{2(0)}+\theta \Delta_x^{2(1)}, \quad 
\Delta_p^2 = \Delta_p^{2(0)}+\theta \Delta_p^{2(1)}, \quad 
\Delta_{xp} = \Delta_{xp}^{(0)}+\theta \Delta_{xp}^{(1)}.
\end{equation}
Splitting the equations in the part independent of $\theta$ and the part which is linear in $\theta$ we get
\begin{eqnarray}
\frac{d}{dt}\Delta_x^{2(0)} &=& \frac{2}{m_0}\Delta_{xp}^{(0)} \\
\frac{d}{dt}\Delta_p^{2(0)} &=& -2m_0\omega^2\Delta_{xp}^{(0)} \\
\frac{d}{dt}\Delta_{xp}^{(0)} &=&\frac{1}{m_0}\Delta_p^{2(0)} -m_0\omega^2\Delta_x^{2(0)} 
\end{eqnarray}
with initial conditions
\begin{equation}
\Delta_x^{2(0)}(t\rightarrow -\infty) = \frac{\hbar}{2m_0\omega}, \quad 
\Delta_p^{2(0)}(t\rightarrow -\infty) = \frac{m_0\hbar\omega}{2}, \quad 
\Delta_{xp}^{(0)}(t\rightarrow-\infty) = 0 
\end{equation} 
and
\begin{eqnarray}
\frac{d}{dt}\Delta_x^{2(1)} &=& \frac{2}{m_0}\Delta_{xp}^{(1)}-\frac{4}{\hbar} g^{\prime}(t)\Delta_{xp}^{(0)}-\frac{2}{\hbar} f^{\prime}(t)\Delta_x^{2(0)} \\
\frac{d}{dt}\Delta_p^{2(1)} &=& -2m_0\omega^2\Delta_{xp}^{(1)}+\frac{2}{\hbar} f^{\prime}(t)\Delta_p^{2(0)} \\
\frac{d}{dt}\Delta_{xp}^{(1)} &=&\frac{1}{m_0}\Delta_p^{2(1)}- \frac{2}{\hbar}g^{\prime}(t)\Delta_p^{2(0)} -m_0\omega^2\Delta_x^{2(1)} 
\end{eqnarray}
with initial conditions
\begin{equation}
\Delta_x^{2(1)}(t\rightarrow -\infty) = 0, \quad
\Delta_p^{2(1)}(t\rightarrow -\infty) = 0, \quad
\Delta_{xp}^{(1)}(t\rightarrow-\infty) = 0.
\end{equation}
The solutions of the $\theta$ independent part with the given initial conditions can be easily found to be
\begin{equation}
\Delta_x^{2(0)}(t) = \frac{\hbar}{2m_0\omega}, \quad
\Delta_p^{2(0)}(t) = \frac{m_0\hbar\omega}{2}, \quad
\Delta_{xp}^{(0)}(t) = 0
\end{equation} 
for all time $t$. The results are also obvious from the fact that for the commutative case ($\theta=0$) the time-evolved vacuum state is nothing but a coherent state in which the uncertainties are constants and are given by the above values. Putting above in the equations for the $\theta$-part, we get
\begin{eqnarray}
\frac{d}{dt}\Delta_x^{2(1)} &=& \frac{2}{m_0}\Delta_{xp}^{(1)}-\frac{1}{m_0\omega} f^{\prime}(t) \label{D_x}\\
\frac{d}{dt}\Delta_p^{2(1)} &=& -2m_0\omega^2\Delta_{xp}^{(1)}+m_0\omega f^{\prime}(t) \label{D_p}\\
\frac{d}{dt}\Delta_{xp}^{(1)} &=& \frac{1}{m_0}\Delta_p^{2(1)}- m_0\omega g^{\prime}(t) -m_0\omega^2\Delta_x^{2(1)} \label{D_xp}
\end{eqnarray}
The first two equations give
\begin{equation}
\frac{d}{dt}\left[\Delta_p^{2(1)}+m_0^2\omega^2\Delta_x^{2(1)}\right]=0
\end{equation}
whose solution satisfying the initial conditions is given by
\begin{equation}
\Delta_p^{2(1)}=-m_0^2\omega^2\Delta_x^{2(1)}\label{D_1}
\end{equation}
We put this in (\ref{D_xp}) to get
\begin{equation}
\frac{d}{dt}\Delta_{xp}^{(1)} =  -2m_0\omega^2\Delta_x^{2(1)} - m_0\omega g^{\prime}(t)
\end{equation}
Differentiating (\ref{D_x}) once and using the above equation we get the decoupled equation in $\Delta_x^{2(1)}$ as
\begin{equation}
\left(\frac{d^2}{dt^2}+4\omega^2\right)\Delta_x^{2(1)}= -2\omega g^{\prime}(t)-\frac{1}{m_0\omega}f^{\prime\prime}(t)
\end{equation}
with the initial conditions
\begin{equation}
\Delta_x^{2(1)}(t\rightarrow -\infty) = 0, \quad
\left.\frac{d}{dt}\Delta_x^{2(1)}(t)\right|_{t\rightarrow -\infty} = 0
\end{equation}
Here, the second condition comes from the equation (\ref{D_x}) and the initial conditions of $\Delta_{xp}^{(1)}(t)$ and $f^{\prime}(t)$. The solution is 
\begin{equation}
\Delta_x^{2(1)}(t)= -\frac{1}{m_0\omega} f(t) - \frac{2}{m_0}\displaystyle{\int_{-\infty}^t}d\tau \sin\{2\omega(\tau-t)\}f(\tau) - 2\omega \displaystyle{\int_{-\infty}^t}d\tau \cos\{2\omega(\tau-t)\}g(\tau)
\end{equation}
Using above in (\ref{D_1}) and (\ref{D_x}) respectively we get
\begin{eqnarray}
\Delta_p^{2(1)}(t) &=& m_0\omega \left[ f(t) + 2\omega \displaystyle{\int_{-\infty}^t}d\tau \sin\{2\omega(\tau-t)\}f(\tau)\right. \nonumber \\
&& \quad \quad \quad \left.+ 2m_0 \omega^2 \displaystyle{\int_{-\infty}^t}d\tau \cos\{2\omega(\tau-t)\}\right]g(\tau) \\
\Delta_{xp}^{(1)}(t) &=& m_0\omega \left[\frac{2}{m_0} \displaystyle{\int_{-\infty}^t}d\tau \cos\{2\omega(\tau-t)\}f(\tau)\right. \nonumber \\
&& \quad \quad \quad \left. - 2\omega\displaystyle{\int_{-\infty}^t}d\tau \sin\{2\omega(\tau-t)\}g(\tau) - g(t)\right]
\end{eqnarray}
Thus, the uncertainties upto linear order in $\theta$ are
\begin{eqnarray}
\Delta_x(t) &=& \sqrt{\frac{\hbar}{2m_0\omega}}-\frac{\theta}{2}\sqrt{\frac{m_0}{2\hbar\omega}}
\left[\frac{2}{m_0}f(t)+\frac{4\omega}{m_0}\displaystyle{\int_{-\infty}^t}d\tau \sin\{2\omega(\tau-t)\}f(\tau)\right. \nonumber \\
&& \hspace{2 cm} \left.+4\omega^2 \displaystyle{\int_{-\infty}^t}d\tau \cos\{2\omega(\tau-t)\}g(\tau)\right]
\label{uncertainties_x_final2}
\end{eqnarray}
\begin{eqnarray}
\Delta_p(t) &=& \sqrt{\frac{m_0\hbar\omega}{2}}+\frac{\theta m_0}{2}\sqrt{\frac{m_0\omega}{2\hbar}}
 \left[\frac{2}{m_0}f(t)+\frac{4\omega}{m_0}\displaystyle{\int_{-\infty}^t}d\tau \sin\{2\omega(\tau-t)\}f(\tau) \right.\nonumber \\
&& \left.+4\omega^2 \displaystyle{\int_{-\infty}^t}d\tau \cos\{2\omega(\tau-t)\}g(\tau)\right]
\label{uncertainties_p_final2}
\end{eqnarray}
\begin{equation}\label{uncertainties_xp_final2}
\Delta_{xp}(t) = \theta m_0\omega \left[- g(t)+\frac{2}{m_0}\displaystyle{\int_{-\infty}^t}d\tau \cos\{2\omega(\tau-t)\}f(\tau)
-2\omega \displaystyle{\int_{-\infty}^t}d\tau \sin\{2\omega(\tau-t)\}g(\tau)\right]
\end{equation}
The fundamental uncertainty product (to linear order in $\theta$) is
\begin{equation}
\Delta_x(t).\Delta_p(t)=\frac{\hbar}{2}.
\end{equation}
Thus the vacuum state evolves to a ``squeezed state'' rather than a coherent state as in the commutative case \cite{Walls}.  
The uncertainties in the commutative case depend only on the product $m_0\omega$.
But, their $\theta$-corrections change with $\omega$ even if $m_0\omega$ is kept constant.
Also, the squeezing effect is oscillatory in time as is obvious from the $\theta$-dependent terms in (\ref{uncertainties_x_final2}), (\ref{uncertainties_p_final2}) and (\ref{uncertainties_xp_final2}).
For the specific forms of $f(t)$ and $g(t)$ of (\ref{fg_expression}) we get
\begin{equation}
\Delta_{x}(t)=\left\{
 \begin{array}{ll}
  \sqrt{\frac{\hbar}{2m_0\omega}} &   ;\,\,t<-T \\
\sqrt{\frac{\hbar}{2m_0\omega}}-\frac{\theta}{2}\sqrt{\frac{1}{2\hbar m_0\omega}} 
\left[2f_0\cos\{2\omega(t+T)\} \right. \\
\left. \quad \quad \quad +2m_0g_0\omega \sin\{2\omega(t+T)\}\right] & ;\,\,-T<t<T \\
\sqrt{\frac{\hbar}{2m_0\omega}}-\frac{\theta}{2}\sqrt{\frac{1}{2\hbar m_0\omega}} 
\left[2f_0\left(\cos\{2\omega(t+T)\}-\cos\{2\omega(t-T)\}\right)\right. \\
\quad \quad \quad +\left. 2m_0g_0\omega \left(\sin\{2\omega(t+T)\}-\sin\{2\omega(t-T)\}\right)\right] 
& ;\,\,t>T
 \end{array}\right.
\end{equation}

\begin{equation}
\Delta_{p}(t)=\left\{
 \begin{array}{ll}
  \sqrt{\frac{m_0\hbar\omega}{2}}  & ;\,\, t<-T \\
\sqrt{\frac{m_0\hbar\omega}{2}}+\frac{\theta}{2}\sqrt{\frac{m_0\omega}{2\hbar}} 
\left[2f_0\cos\{2\omega(t+T)\} \right.\\
 \left. \quad \quad \quad +2m_0g_0\omega \sin\{2\omega(t+T)\}\right] \quad  & ;\,\, -T<t<T \\
\sqrt{\frac{m_0\hbar\omega}{2}}+\frac{\theta}{2}\sqrt{\frac{m_0\omega}{2\hbar}} 
\left[2f_0\left(\cos\{2\omega(t+T)\}-\cos\{2\omega(t-T)\}\right)\right. \\
\quad \quad \quad +\left. 2m_0g_0\omega \left(\sin\{2\omega(t+T)\}-\sin\{2\omega(t-T)\}\right)\right] \quad \quad  
& ;\,\, t >T
 \end{array}\right.
\end{equation}

and 
\begin{equation}
 \Delta_{xp}(t)=\left\{
 \begin{array}{ll}
  0 & ;\,\, t<-T \\
\frac{\theta}{2}\left[2f_0\sin\{2\omega(t+T)\} \right. \\
 \quad \quad \quad \left. -2m_0\omega g_0\cos\{2\omega(t+T)\}\right] & ;\,\, -T<t<T \\
\frac{\theta}{2} \left[2f_0\left(\sin\{2\omega(t+T)\}-\sin\{2\omega(t-T)\}\right)\right. \\
\quad \quad \quad \left. -2m_0\omega g_0\left(\cos\{2\omega(t+T)\}-\cos\{2\omega(t-T)\}\right)\right] & ;\,\, t >T.
\end{array} 
 \right.
\end{equation}
For the choice of parameters $m_0=1$, $\omega=1$, $f_0=\sqrt{5}$, $g_0=\frac{1}{2}\sqrt{5}$, $T=\frac{\pi}{2}$ and $\theta=0.01$ in natural units ($\hbar =1$) we get
\begin{equation}
\Delta_{x}(t)=\left\{
\begin{array}{l}
 \frac{1}{\sqrt{2}} \hspace*{5 cm} ;t<-\frac{\pi}{2} \\
\frac{1}{\sqrt{2}} + 0.01 \sqrt{\frac{5}{2}}\left(\cos{2t}+\frac{1}{2}\sin{2t}\right) \quad ;-\frac{\pi}{2}<t<\frac{\pi}{2} \\
\frac{1}{\sqrt{2}} + 0.01 \sqrt{\frac{5}{2}}\sin{2t} \quad ;t>\frac{\pi}{2},
\end{array}
\right.
\end{equation}

\begin{equation}
\Delta_{p}(t)=\left\{
\begin{array}{l}
 \frac{1}{\sqrt{2}} \hspace*{5 cm} ;t<-\frac{\pi}{2} \\
\frac{1}{\sqrt{2}} - 0.01 \sqrt{\frac{5}{2}}\left(\cos{2t}+\frac{1}{2}\sin{2t}\right) \quad ;-\frac{\pi}{2}<t<\frac{\pi}{2} \\
\frac{1}{\sqrt{2}} - 0.01 \sqrt{\frac{5}{2}}\sin{2t} \quad ;t>\frac{\pi}{2}
\end{array}
\right.
\end{equation}

\begin{equation}
{\rm and} \quad \quad \Delta_{xp}(t)=\left\{
\begin{array}{l}
 0 \hspace*{5 cm} ;t<-\frac{\pi}{2} \\
0.01 \sqrt{5}\left(\frac{1}{2}\cos{2t}-\sin{2t}\right) \quad ;-\frac{\pi}{2}<t<\frac{\pi}{2} \\
-0.02 \sqrt{5}\sin{2t} \quad ;t>\frac{\pi}{2}.
\end{array}
\right.
\end{equation}
Figures \ref{fig_deltaxdeltap} and \ref{fig_deltaxp} show the time-dependence of the different uncertainties.
The discontinuities at $t=\pm \frac{\pi}{2}$ is simply the manifestation of the fact that the functions $f(t)$ and $g(t)$ 
themselves are discontinuous at these times.
Before the interaction was switched on, the uncertainties were having values equal to those for the vacuum state.
During the time of nonvanishing interaction (and even after the interaction gets switched off!), 
they oscillate with frequency equal to twice that of the oscillator.

\section{Implications in Quantum Optics}\label{quantum_optics}
In Quantum Optics a monochromatic (single-mode) coherent light field is usually described by the harmonic oscillator coherent states \cite{Glauber:1963tx}.
It has also been shown that a coherent state (in particular the vacuum state)
remains to be coherent under the FHO Hamiltonian \cite{Carruthers}.
The annihilation and creation operators for photons are related to the field quadratures $X_1$ and $X_2$ by
\begin{equation}
a = X_1+iX_2, \quad  a^{\dagger} = X_1-iX_2,
\end{equation}
$X_1$ and $X_2$ being hermitian. 
The commutation $\left[a,a^{\dagger}\right]=1$ translates to $\left[X_1,X_2\right]=\frac{i}{2}$.
The coherent state has different uncertainties as $\Delta_{X_1} = \frac{1}{2}$, $\Delta_{X_2} = \frac{1}{2}$ and
 $\Delta_{X_1X_2} = 0$ $\Rightarrow$ $\Delta_{X_1}.\Delta_{X_2}=\frac{1}{4}$ which is the minimum. 
Also, the photon count (probability for having a certain number of photons) in the coherent state is given by 
the transition probabilities of the corresponding number eigenstate and the profile is Poissonian. \\
\indent The FHO Hamiltonian
\begin{eqnarray}
H(t) &=& \hbar\omega(X_1^2+X_2^2)+\sqrt{\frac{2\hbar}{m_0\omega}}f(t)X_1+\sqrt{2\hbar m_0\omega}g(t)X_2 \nonumber \\
&=& \hbar\omega\left(a^{\dagger}a+\frac{1}{2}\right)+z^*(t)a+z(t)a^{\dagger}
\end{eqnarray}
($z(t)$ is related with $f(t)$ and $g(t)$ by (\ref{z})) with the effective noncommutativity 
between time and the field quadrature $X_1$ of the form
\begin{equation}
\left[t,X_1\right]=i\sqrt{\frac{m_0\omega}{2\hbar}}\theta \label{QO_com}
\end{equation}
will allow us to use the calculation of the previous sections.
The photon count will be given by (\ref{Pm0}), while the uncertainties in the field quadratures $\Delta_{X_1} = \sqrt{\langle X_1^2 \rangle - \langle X_1 \rangle^2}$, $\Delta_{X_2} = \sqrt{\langle X_2^2 \rangle - \langle X_2 \rangle^2}$ and $\Delta_{X_1X_2} = \frac{1}{2}\langle X_1X_2 + X_2X_1\rangle - \langle X_1 \rangle \langle X_2 \rangle $ will get modified as
\begin{eqnarray}
\Delta_{X_1}(t) &=& \frac{1}{2}-\frac{\theta m_0}{4\hbar}
\left[\frac{2}{m_0}f(t)+\frac{4\omega}{m_0}\displaystyle{\int_{-\infty}^t}d\tau \sin\{2\omega(\tau-t)\}f(\tau)\right. \nonumber \\
&& \hspace{2 cm} \left.+4\omega^2 \displaystyle{\int_{-\infty}^t}d\tau \cos\{2\omega(\tau-t)\}g(\tau)\right]
\label{uncertainties_X1}
\end{eqnarray}
\begin{eqnarray}
\Delta_{X_2}(t) &=&  \frac{1}{2}+\frac{\theta m_0}{4\hbar}  \left[\frac{2}{m_0}f(t)+\frac{4\omega}{m_0}\displaystyle{\int_{-\infty}^t}d\tau \sin\{2\omega(\tau-t)\}f(\tau) \right.\nonumber \\
&& \left.\quad \quad \quad \quad \quad \quad +4\omega^2 \displaystyle{\int_{-\infty}^t}d\tau \cos\{2\omega(\tau-t)\}g(\tau)\right]
\label{uncertainties_X2}
\end{eqnarray}
\begin{eqnarray}
\Delta_{X_1X_2}(t) &=& \frac{\theta m_0\omega}{2\hbar} \left[- g(t)+\frac{2}{m_0}\displaystyle{\int_{-\infty}^t}d\tau \cos\{2\omega(\tau-t)\}f(\tau) \right. \nonumber \\
&& \left. \quad \quad \quad \quad \quad \quad -2\omega \displaystyle{\int_{-\infty}^t}d\tau \sin\{2\omega(\tau-t)\}g(\tau)\right]
\label{uncertainties_X1X2}
\end{eqnarray}
We further study the correlation among the photons. The time-evolved vacuum state
\begin{equation}
|i(t\rightarrow\infty)\rangle=\displaystyle{\sum_{m=0}^{\infty}}A_{m,0}|m\rangle
\end{equation}
will give
\begin{equation}
\bar{N}=\langle i(t\rightarrow\infty)|a^{\dagger}a|i(t\rightarrow\infty)\rangle=\displaystyle{\sum_{m=1}^{\infty}}mP_{m,0}
= |\xi|^2-\frac{2\theta}{\hbar^2}Im(\beta_2\xi),
\end{equation}
$\bar{N}$ being the average number of photons in state $|i(t\rightarrow\infty)\rangle$. Also
\begin{equation}
\langle i(t\rightarrow\infty)|a^{\dagger}a^{\dagger}aa|i(t\rightarrow\infty)\rangle=\displaystyle{\sum_{m=2}^{\infty}}m(m-1)P_{m,0}
= |\xi|^4-\frac{4\theta}{\hbar^2}\left(Im(\beta_3\xi^2)+|\xi|^2 Im(\beta_2\xi)\right).
\end{equation}
This, to linear order in $\theta$, gives the 2nd order correlation among photons with zero time delay to be equal to 
(see Appendix \ref{correlation})
\begin{equation}
g^{(2)}(0) = 1-\frac{\frac{4\theta}{\hbar^2} Im(\beta_3\xi^2)}{\left(|\xi|^4-\frac{4\theta}{\hbar^2}|\xi|^2 Im(\beta_2\xi)\right)}
= 1-\frac{\frac{4\theta}{\hbar^2} Im(\beta_3\xi^2)}{\bar{N}^2}.
\end{equation}
For the case $Im(\beta_3\xi^2)<0\Rightarrow g^{(2)}(0)>1$, the photons try to bunch together 
while for $Im(\beta_3\xi^2)>0\Rightarrow g^{(2)}(0)<1$, they anti-bunch \cite{Walls_book}.
For the functions (\ref{fg_expression}), we get
\begin{equation}
 Im(\beta_3\xi^2) = \frac{2(f_0^2+m_0^2\omega^2g_0^2)\sin^2 \omega T}{\omega^2}g_0\sin 2\omega T,
\end{equation}
which implies that the bunching or anti-bunching will depend only on the sign of the factor $g_0\sin 2\omega T$.
For the choices taken in figures \ref{fig_distribution}, \ref{fig_deltaxdeltap} $\&$ \ref{fig_deltaxp}, $\omega T = \frac{\pi}{2}$
and hence no bunching or anti-bunching occurs.

As a future work one can try to formulate the scattering process in higher dimensions and study its implications in quantum optics. 
The correspondence found between noncommutativity and quantum optics also encourages one to study such possibilities in other forms of time-sapce noncommutativity. 
As an example one can start with assuming the spacetime dependent noncommutative parameter $\theta$ \cite{Majid:1994cy, Sitarz:1994rh, Lukierski:1993wx}.



\chapter{Thermodynamics of Ideal Gas in Doubly Special Relativity} \label{DSR}
\section{Introduction}

 In DSR, apart from the constancy of speed-of-light scale, the Planck length $l_P$ or equivalently Planck energy $\kappa$ is also 
constant under coordinate transformation from one inertial frame to another. This leads to modification in  the dispersion relation. Consequences of the modified dispersion relations on the thermodynamics are being studied extensively to infer the effect of Planck scale physics \cite{AmelinoCamelia:2004xx, Camacho:2006qg, Gregg:2008jb, AmelinoCamelia:2009tv, Camacho:2007qy, Alexander:2001ck, Bertolami:2009wa}. 

The present chapter aims to study the thermodynamics of an ideal gas consisting of massive particles in DSR scenario. Both the modification in the dispersion relation of the constituent particles and the presence of a maximum energy scale are expected to contribute to new effects. DSR transformations can be of several type. 
We follow the formulation of \cite{Magueijo:2001cr, Magueijo:2002am} where the modified dispersion relation becomes
\begin{equation}
\e^2-p^2=m^2\l1-\frac{\e}{\kappa}\r^2.
\label{MS}
\end{equation}
As  $0\leq \l 1-\frac{\e}{\kappa}\r^{2} \leq 1$, the energy of a particle with a given momentum decreases in DSR with respect to that in SR. This has consequence on the thermodynamics as we will see in $\mathsection$ \ref{thermodynamics}.
The parameter $m$ can be called ``invariant mass'' as it remains invariant under a DSR transformation. 
Note that in contrary to the SR case, $m$ is no more the rest mass energy of the particle.
To get the rest mass energy $m_0$, we put $p=0$ in (\ref{MS}). We get two expressions 
for $m_0$, namely
\begin{equation}
m_0=\frac{\pm m}{1\pm \frac{m}{\kappa}}.
\end{equation}
The two solutions are connected by the redefinition of the parameter $m\rightarrow -m$.
Henceforth, without any loss of generality we use
\begin{equation} \label{rest_mass}
m_0=\frac{m}{1+\frac{m}{\kappa}}.
\end{equation}
The physical world is characterized by $E<\kappa$ \cite{Magueijo:2002am}.
In this sub-Planck regime ($E_{p=0}=m_0<\kappa$), the positivity of rest mass ($m_0\geq 0$) 
restricts the range of the invariant mass to $0\leq m<\infty$.
Thus, in (\ref{MS}), we have $0\leq p,E < \kappa$ and $0\leq m<\infty$.

We study the thermodynamics of an ideal gas in DSR 
setup. We obtain a series solution for the  partition function 
and compute the various thermodynamic quantities.
We show that our results go to the standard results in the 
SR limit ($\kappa \rightarrow \infty$) \cite{greinerbook} as well as in the 
massless DSR limit \cite{Das:2010gk}.

\section{The Partition Function}
We consider a gaseous system of non-interacting particles obeying Maxwell-Boltzmann
statistics whose macrostate is denoted by $(N,V,T)$ where $N$ is the number of particles
 in the system confined in volume $V$ at a temperature $T$. In the canonical
ensemble the thermodynamics of this system is derived from its partition function \cite{pathria}
\begin{equation}
 Z_N\l V,T\r=\sum_E\exp[-\beta E],
\end{equation}
where $\beta={1\over k_BT}$ and $\displaystyle{\sum_{E}}$ denotes sum over all the energy eigenvalues 
of the system. The total energy $E$ of the system can be written in terms of single particle energy $\e$
\beq
E=\sum_{\e}n_{\e}\e,
\eeq
where $n_{\e}$ is the number of particles in the single-particle energy state $\e$ and satisfy the following condition
\beqa
\sum_{\e}n_{\e}&=&N.
\label{cond}
\eeqa
We can rewrite $Z_N$ as
\begin{equation}
 Z_N\l V,T\r={\sum_{\{n_{\e}\}}}^\prime g\{n_{\e}\}\exp[-\beta\sum_{\e}n_{\e}\e],
\end{equation}
where $g\{n_{\e}\}$ is the statistical weight factor appropriate to the distribution set $\{n_{\e}\}$. 
The summation ${\sum}^{\prime}$ goes over all distribution sets
that conform to the above restrictive condition (\ref{cond}).
For Maxwell-Boltzmann statistics, it can be shown \cite{pathria}
\beq
Z_N\l V,T\r={1\over N!}[Z_1\l V,T\r]^N,
\eeq
where $Z_1$ is the single particle partition function given by
\beq
Z_1\l V,T\r=\sum_{\e}\exp[-\beta\e].
\label{Zsingle}
\eeq
While for ordinary spacetime, it is easy to show that in the large volume limit one can replace the sum by an integral \cite{pathria}
\beq
\sum_{\e} \rightarrow {V\over h^3}\int d^3p,
\label{replace}
\eeq
for more exotic spacetimes the measure of integration is expected to get modified
\beq
d^3p \rightarrow f(\vec{p}) d^3p.
\label{modmeas}
\eeq
Hence putting together (\ref{Zsingle}), (\ref{replace}), (\ref{modmeas}) and taking $\hbar = k_B = 1$ we get
\beqa
Z_1\l V,T\r &=& \frac{V}{\l2\pi\r^3}\int_{p=0}^{\kappa}d^3p\,\, f(\vec{p}) \exp[-\beta \l\e-m_0\r].
\label{Z1_modmeas}
\eeqa
Note that in accordance with standard practice, we have subtracted the rest mass $m_0$ from the
relativistic energy $\e$ of the particle.
Although there have been few attempts\cite{AmelinoCamelia:2009tv, AmelinoCamelia:1999pm, KowalskiGlikman:2001ct}, the form of $f(\vec{p})$ is far from settled.
Assuming isotropy of spacetime we may take $f(\vec{p})=f(p)$. For a possible deformation of the integration measure, $f(p)$ should be expandable in Taylor series in $\frac{\e}{\kappa}$
\begin{equation}
f(p) = \displaystyle{\sum_{n=0}^{\infty}} \frac{a_{n}}{n!}\left(\frac{\e}{\kappa}\right)^{n},
\end{equation}
with $a_{0}=1$ since in the limit $\kappa\rightarrow\infty$, $f(p)\rightarrow 1$. Hence $Z_1(V,T)$ becomes
\beqa
Z_1\l V,T\r &=& \frac{V}{\l2\pi\r^3}\int_{p=0}^{\kappa}d^3p\,\, \sum_{n=0}^{\infty} \frac{a_{n}}{n!}\left(\frac{\e}{\kappa}\right)^{n} \exp[-\beta \l\e-m_0\r]
\label{Z1_final1}\\
&=&\displaystyle{\sum_{n=0}^{\infty}} \frac{a_{n}}{n! \kappa^{n}} \l m_{0}-\frac{\partial}{\partial \beta}\r^{n} Z_1^0\l V,T\r,
\label{Z1_final2}
\eeqa
where $Z_1^0\l V,T\r$ is the single particle partition function with unmodified measure
\beq
Z_1^0\l V,T\r=\frac{V}{\l2\pi\r^3}\int_{p=0}^{\kappa}d^3p \exp[-\beta \l\e-m_0\r].
\label{z10}
\eeq
The derivation of (\ref{Z1_final2}) from (\ref{Z1_final1}) involves two steps. Firstly, the interchange of the summation
and integration which is allowed if (see theorem 1.38 of \cite{rudin})
\begin{equation}
\displaystyle{\sum_{n=0}^{\infty}} \frac{|a_{n}|}{n! \kappa^{n}} \int_{p=0}^{\kappa}d^3p\,\,  \e^{n} \exp[-\beta \l\e-m_0\r]
=\displaystyle{\sum_{n=0}^{\infty}} \frac{|a_{n}|}{n! \kappa^{n}} \l m_{0}-\frac{\partial}{\partial \beta}\r^{n} Z_1^0\l V,T\r < \infty.
\label{condition_modmeas_expansion}
\end{equation}
Secondly, writing 
\begin{eqnarray}
\int_{p=0}^{\kappa}d^3p\,\,  \e^{n} \exp[-\beta \l\e-m_0\r] &=&  4\pi \int_{0}^{\kappa}dp\,\, p^{2} \e^{n} \exp[-\beta \l\e-m_0\r] \nonumber \\
&=& 4\pi \l m_{0}-\frac{\partial}{\partial \beta}\r^{n}  \int_{0}^{\kappa}dp\,\, p^{2} \exp[-\beta \l\e-m_0\r] \nonumber \\
&=& \l m_{0}-\frac{\partial}{\partial \beta}\r^{n} \int_{p=0}^{\kappa}d^3p\,\,  \exp[-\beta \l\e-m_0\r]
\end{eqnarray}
since the integrand $p^{2} \e^{n} \exp[-\beta \l\e-m_0\r]$ remains to be continuous and bounded for $p\in[0,\kappa], \beta\in [0,\infty]$ (see $\S$ 5.12 of \cite{fleming}).
Hence our problem has boiled down to solving the integral in (\ref{z10})
where $\e$ and $p$ are related by the modified DSR dispersion relation given in (\ref{MS}). The solution of 
$Z_1^0$ in the massless case obtained in \cite{Das:2010gk} is
\beq
Z_{1ml}^0=\frac{2V}{(2\pi)^2\beta^3}\left(2-e^{-\beta\kappa}(\beta^2\kappa^2+2\beta\kappa+2)\right).
\label{DSRphoton}
\eeq
The term with $\displaystyle{e^{-\beta\kappa}}$ makes $Z_{1ml}^0$ non-analytic at $\frac{1}{\kappa}=0$. We anticipate
that even when $m_0\neq 0$, $Z_1^0$ continues to be non-analytic at $\frac{1}{\kappa}=0$ and hence does not
admit a straightforward Taylor series expansion in $\frac{1}{\kappa}$. Thus in order to find the leading order deviation of DSR thermodynamics 
from the SR case, one would require a non-trivial series expansion.

\subsection{Solution of $Z_1^0$}
The dispersion relation (\ref{MS}) gives
\begin{equation}
p = \left[\e^2-m^2\left(1-\frac{\e}{\kappa}\right)^2\right]^{1/2}
\end{equation}
\begin{equation}
pdp = \left[\e + \frac{m^2}{\kappa}\left(1-\frac{\e}{\kappa}\right)\right]d\e
\end{equation}
Hence changing the variable from $p$ to $\e$ in (\ref{z10}) we get
\begin{equation}
Z_1^0\l V, \beta\r = \frac{2V}{\l2\pi\r^2}\exp[\beta m_0]\int_{m_0}^{\kappa}\left[\e+
\frac{m^2}{\kappa}\l1-\frac{\e}{\kappa}\r\right]
\left[\e^2-m^2\l1-\frac{\e}{\kappa}\r^2\right]^{1/2}\exp[-\beta \e]d\e.
\label{Z_E}
\end{equation}
We now consider three different regions of values of $m$:

\subsubsection{Case I: $m=\kappa$}
This case is equivalent to the case when the rest mass energy
\begin{equation}
m_0 = \frac{\kappa}{2}
\end{equation}
In this case the dispersion relation (\ref{MS}) become
\begin{equation}
\e = \frac{\kappa}{2}\left(1+\frac{p^2}{\kappa^2}\right)
\end{equation}
Also the partition function (\ref{Z_E}) reduces to
\begin{eqnarray}
Z_1^0\l V,T\r &=& \frac{2V}{\l2\pi\r^2}\kappa^{3/2}\exp\left[{\beta\kappa\over2}\right]\int_{\kappa/2}^{\kappa}d\e(2\e-\kappa)^{1/2}\exp\left[-\beta \e\right]\nonumber\\
& = &\frac{2\sqrt 2V}{\l2\pi\r^2}\l\frac{\kappa}{\beta}\r^{3/2}\gamma\l\frac{3}{2},\frac{\beta\kappa}{2}\r,
\label{Zcase2}
\end{eqnarray}
where $\gamma(a,x)$ is the Incomplete Gamma Function (see (6.5.2) of 
\cite{abramowitz}). We get a very simple analytical form for $Z^0_1(V,T)$.

\subsubsection{Case II: $\kappa<m<\infty$}
This case is equivalent to the case when 
\begin{equation}
\frac{\kappa}{2}<m_0 < \kappa.
\end{equation}
We change the variable in (\ref{Z_E}) from $\e$ to $t=\frac{\e}{m}\left[\left(\frac{m}{\kappa}\right)^2-1\right]-\frac{m}{\kappa}$ which gives
\begin{eqnarray}
Z_1^0 &=& -\frac{2Vm^3}{\l2\pi\r^2\left[\l\frac{m}{\kappa}\r^2-1\right]^{3/2}}\exp\left[\beta m_0-\frac{\beta^{\prime}m^2}{\kappa}\right]
\int_{-1}^{-\kappa /m}dt \,\,\,\,\,\,t (1-t^2)^{1/2}\exp\left[-\beta^{\prime}mt\right]\nonumber\\
&=& -\frac{2Vm^3}{\l2\pi\r^2\left[\l\frac{m}{\kappa}\r^2-1\right]^{3/2}}\exp\left[\beta m_0-\frac{\beta^{\prime}m^2}{\kappa}\right]
\left[I^*\left(\beta^{\prime}m,1\right)-I^*\left(\beta^{\prime}m,\frac{\kappa}{m}\right)\right],
\label{Istar}
\end{eqnarray}
where 
\begin{equation}
\beta^{\prime}=\frac{\beta}{\l\frac{m}{\kappa}\r^2-1}
\end{equation} 
and
\begin{equation}
I^*(x,y) = \int_{-y}^{1}dt \,\,\,\,\,\,t (1-t^2)^{1/2}\exp\left[-xt\right].
\label{istar}
\end{equation}
We define Incomplete Modified Bessel function $I_{\nu}(z,y)$ of order $\nu$
\beq
I_{\nu}(z,y)=\frac{1}{\sqrt{\pi}\Gamma(\nu+\frac{1}{2})}\l\frac{z}{2}\r^{\nu}
\int_{-y}^1(1-t^2)^{\nu-\frac{1}{2}}\exp[-zt]dt\quad [Re\,\,\nu>0,|arg \,\,z|<\frac{\pi}{2}],
\eeq
such that 
\begin{equation}
I^*(x,y)=-\frac{\partial}{\partial x}\left[\frac{\pi}{x}I_1(x,y)\right].
\end{equation} 
In particular for $y=1$, 
using (3.387 (1)) of \cite{gradshteyn} and (9.6.26) of \cite{abramowitz} we get
\begin{equation}
I^*(x,1) = -\frac{\pi I_2(x)}{x},
\label{istar_i2}
\end{equation}
where $I_2(x)$ is the 2nd order Modified Bessel function. 

Let us consider {\bf a very interesting case} of $m\rightarrow\infty$ $(m_0\rightarrow \kappa)$\footnote[1]{We would like to thank Diptiman Sen for pointing out this interesting case.}. In this limit
one gets from (\ref{MS})
\beq
\frac{\e^2-p^2}{m^2}=\l1-\frac{\e}{\kappa}\r^2\Rightarrow \e\rightarrow\kappa \quad \forall p\in[0,\kappa].\nonumber
\eeq
Thus the total energy $E$ of the system becomes 
\begin{equation}
E=N\kappa
\end{equation} 
and the thermodynamics simplifies.
Entropy can be computed by counting the total number of microstates $\Omega_N$ available to the system
\beq
\Omega_N=\frac{\Omega_1^N}{N!}=\frac{1}{N!}\l V\int_{p=0}^{\kappa}\frac{d^3p}{h^3}\r^N=\frac{1}{N!}\l\frac{2V\kappa^3}{3\l2\pi\r^2}\r^N,
\eeq
where $\Omega_1$ is the total number of microstates available for a single particle. Thus the entropy $S$ of the system is
\beq
S=\ln\left[\frac{1}{N!}\l\frac{2V\kappa^3}{3\l2\pi\r^2}\r^N\right].
\label{s}
\eeq
The first law of thermodynamics in this case becomes 
\beq
dE=-PdV+\mu dN.
\eeq
Note that the usual term $TdS$ has been dropped as from (\ref{s}) it is evident that $S$ is a function of $N$ and $V$ alone.
The pressure of the system is zero as 
\begin{equation}
P=-\left.\frac{\partial E}{\partial V}\right|_{N}=0\end{equation} 
while the chemical potential is 
\begin{equation}
\mu=\left.\frac{\partial E}{\partial N}\right|_{V}=\kappa.
\end{equation}
Equation (\ref{z10}) can now be easily integrated to give
\beq
Z_1^0=\frac{2V}{\l2\pi\r^2}\frac{\kappa^3}{3},
\label{istarminf}
\eeq
which gives the limiting behaviour of $I^*\l\beta^{\prime}m,\frac{\kappa}{m}\r$ using (\ref{istar}) and (9.6.7) of \cite{abramowitz}
\beq
I^*\l\frac{\beta m}{\l\frac{m}{\kappa}\r^2-1},\frac{\kappa}{m}\r\stackrel{m\rightarrow\infty}{\longrightarrow}\frac{1}{3}.
\eeq

\subsubsection{Case III: $0<m<\kappa$}
This case is equivalent to the case when 
\begin{equation}
0<m_0 < \frac{\kappa}{2}.
\end{equation}
We change the variable in (\ref{Z_E}) from $\e$ to $t=\frac{\e}{m}\left[1-\l\frac{m}{\kappa}\r^2\right]+\frac{m}{\kappa}$
to get 
\begin{eqnarray}
Z_1^0 &=& \frac{2Vm^3}{\l2\pi\r^2\left[1-\l\frac{m}{\kappa}\r^2\right]^{3/2}}\exp\left[\beta m_0+\frac{\beta^{\prime \prime}m^2}{\kappa}\right]
\int_1^{\kappa /m}dt \,\,\,\,\,\,t (t^2-1)^{1/2}\exp\left[-\beta^{\prime \prime}mt\right]\nonumber\\
&=&\frac{2Vm^3}{\l2\pi\r^2\left[1-\l\frac{m}{\kappa}\r^2\right]^{3/2}}\exp\left[\beta m_0+\frac{\beta^{\prime \prime}m^2}{\kappa}\right]
\left[K^*\left(\beta^{\prime \prime}m,1\right)-K^*\left(\beta^{\prime \prime}m,\frac{\kappa}{m}\right)\right], \nonumber \\
\label{Z}
\end{eqnarray}
where 
\begin{equation}
\beta^{\prime \prime}=\frac{\beta}{1-\left(\frac{m}{\kappa}\right)^2}
\end{equation}
and
\begin{equation}
K^*(x,y) = \int_y^{\infty}dt \,\,\,\,\,\,t (t^2-1)^{1/2}\exp\left[-xt\right].
\label{kstar}
\end{equation}
As in Case II, we define Incomplete Modified Bessel function $K_{\nu}(z,y)$ of order $\nu$
\beq
K_{\nu}(z,y)=\frac{\sqrt{\pi}}{\Gamma(\nu+\frac{1}{2})}\l\frac{z}{2}\r^{\nu}\int_{y}^\infty(t^2-1)^{\nu-\frac{1}{2}}\exp[-zt]dt
\quad [Re\,\,\nu>-\frac{1}{2},|arg \,\,z|<\frac{\pi}{2}],
\eeq
such that 
\begin{equation}
K^*(x,y)=-\frac{\partial}{\partial x}\left[\frac{K_1(x,y)}{x}\right].
\end{equation}
In particular for $y=1$, using (9.6.23) and (9.6.26) of \cite{abramowitz} we get
\begin{equation}
K^*(x,1) = \frac{K_2(x)}{x},
\label{kstar_k2}
\end{equation}
where $K_2(x)$ is the 2nd order Modified Bessel function.

We shall now obtain the series solution of $K^*(x,y)$.
We rewrite (\ref{kstar}) as
\begin{equation}
K^*\l x,y\r = \int_{y}^{\infty}dt \,\,\,\,\,\,t^2 \l1-\frac{1}{t^2}\r^{1/2}e^{-xt}.
\end{equation}
Inside the integral $t\geq y$ and for $y>1$ (which is a valid assumption for the case of our interest) the factor $\left(1-\frac{1}{t^2}\right)^{1/2}$ can be expanded in series of $\frac{1}{t^2}$ to get
\begin{equation}
K^*\l x,y\r = \int_{y}^{\infty}d\mu_t \,\,\,\,\,\, \left[1+\sum_{r=1}^{\infty}f_r(t)\right]
\label{kstar_series}
\end{equation}
with
\begin{equation}
d\mu_t=t^2e^{-xt}dt 
\end{equation}
and
\begin{equation}
 f_r(t)=\frac{t_r}{t^{2r}},
\end{equation}
where
\begin{equation}
 t_r=\frac{(0-\frac{1}{2})(1-\frac{1}{2})...(r-1-\frac{1}{2})}{r!}=-\frac{(2r-2)!}{2^{2r-1}r!(r-1)!}.
\end{equation}
Now the integral and the summation in (\ref{kstar_series}) can be interchanged if (see theorem 1.38 of \cite{rudin})
\begin{equation}
\displaystyle{\sum_{r=1}^{\infty}} \displaystyle{\int_y^{\infty}}d\mu_t|f_r(t)|<\infty.
\end{equation}
Now as $t_r$ is $-ve$ for all $r\geq 1$ we have 
\begin{equation}
|f_r(t)|=-f_r(t).
\end{equation}
This allows us to interchange the summation and the integral if the final series is converging. Thus we get
\begin{equation}
 K^*\l x,y\r = M_0-\frac{1}{2}M_1+\sum_{r=2}^{\infty}t_rM_r,
\label{kstar_mr}
\end{equation}
if the above is a converging series (see Appendix \ref{kstar_convergence} for convergence of $K^*\l x,y\r$). Here
\begin{equation}
 M_r = \int_{y}^{\infty}dt \,\,\,\,\, t^{2(1-r)}e^{-xt}
\end{equation}
for $r=2, 3,...$. \\
$M_0$ and $M_1$ can be easily calculated to be
\begin{equation} \label{m0}
 M_0 = \frac{\exp\l-xy\r}{x^3}\l(xy)^2 + 2xy + 2\r,
\end{equation}
\begin{equation} \label{m1}
 M_1 = \frac{\exp\l-xy\r}{x}.
\end{equation}
Now, changing the variable to $t^\prime=xt$ in $M_r$ for $r\geq 2$ 
we get
\begin{equation}
 M_r = x^{2r-3}\int_{xy}^{\infty}dt^\prime \frac{e^{-t^\prime}}{(t^{\prime})^{2r-2}}.
\end{equation}
Taking $e^{-t^\prime}$ as first function, if we do the integration by parts again and again we finally get
\begin{equation}
 M_r=-\frac{x^{2r-3}}{(2r-3)!}E_1(xy)+e^{-xy}\sum_{k=1}^{2r-3}\frac{(-x)^{k-1}}{(2r-3)(2r-4)...(2r-2-k)}\left(\frac{1}{y}\right)^{2r-2-k}.
\label{mr_series}
\end{equation}
Here $E_1(x)$ is the Exponential Integral (see (5.1.1) of \cite{abramowitz}). A similar attempt to obtain the series 
solution of $I^*(x,y)$ fails!

Although we obtain the solutions of $Z_1^0$ in three different regions of 
values of $m$, $Z_1^0$ can be shown to be smooth in $m$ (see Appendix \ref{app1}) and hence we do not expect any phase transition like thermodynamic discontinuity as we vary $m$. We use the continuity of $Z_1^0$ to obtain the limiting behaviour of $I^*(\beta^{\prime}m,\frac{\kappa}{m})$ and $K^*(\beta^{\prime}m,\frac{\kappa}{m})$ as $m\rightarrow\kappa$. From (\ref{Zcase2}), (\ref{Istar}), (\ref{istar_i2}) and (9.7.1) of \cite{abramowitz} we obtain the leading order behaviour of $I^*\l\beta^{\prime}m,\frac{\kappa}{m}\r$ as $m \rightarrow \kappa^+$ to be
\beq
I^*\l\beta^{\prime}m,\frac{\kappa}{m}\r \stackrel{m\rightarrow \kappa^+}{\longrightarrow}{2\over (\beta \kappa)^{3/2}}\delta^{3/2}e^{{\beta \kappa\over 2\delta}}\left[-\pi^{1/2}+2e^{-{\beta \kappa\over 2}}\gamma\left({3\over 2}, {\beta \kappa\over 2}\right)\right],
\eeq
with $\delta=\frac{m}{\kappa}-1$.
For $m \rightarrow \kappa^-$, using (\ref{Zcase2}), (\ref{Z}), (\ref{kstar_k2}) and (9.7.2) of \cite{abramowitz} the leading order behaviour of $K^*\l\beta^{\prime\prime}m,{\kappa\over m}\r$ turns out to be
\begin{equation}
K^*(\beta^{\prime \prime}m,{\kappa\over m}) \stackrel{m \rightarrow \kappa^-}{\longrightarrow}{2\over (\beta \kappa)^{3/2}}\epsilon^{3/2}e^{-{\beta \kappa\over 2\epsilon}}\left[\pi^{1/2}-2e^{-{\beta \kappa\over 2}}\gamma\left({3\over 2}, {\beta \kappa\over 2}\right)\right],
\end{equation}
where $\epsilon = 1-{m\over\kappa}$.
\subsection{Leading order deviations}
Having obtained the series solution of $Z_1^0$ in Case III, we shall now obtain the leading order
corrections from the massless and the SR cases.

\subsubsection{Leading order deviation from the massless case}
\begin{figure*}[h]
 \begin{center}
  \scalebox{0.74}{\includegraphics{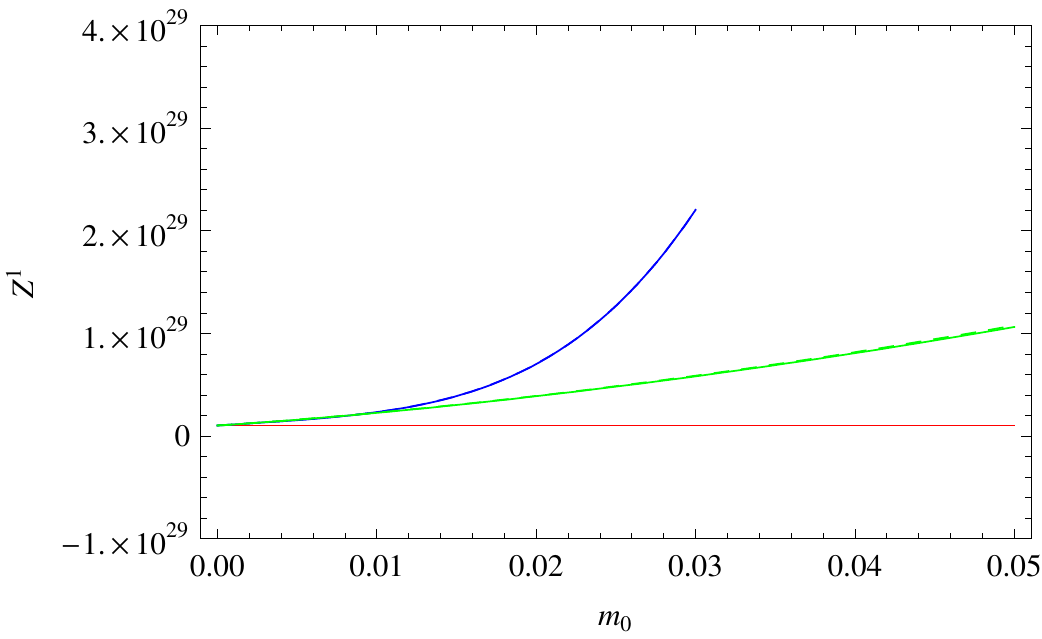}}
  \scalebox{0.74}{\includegraphics{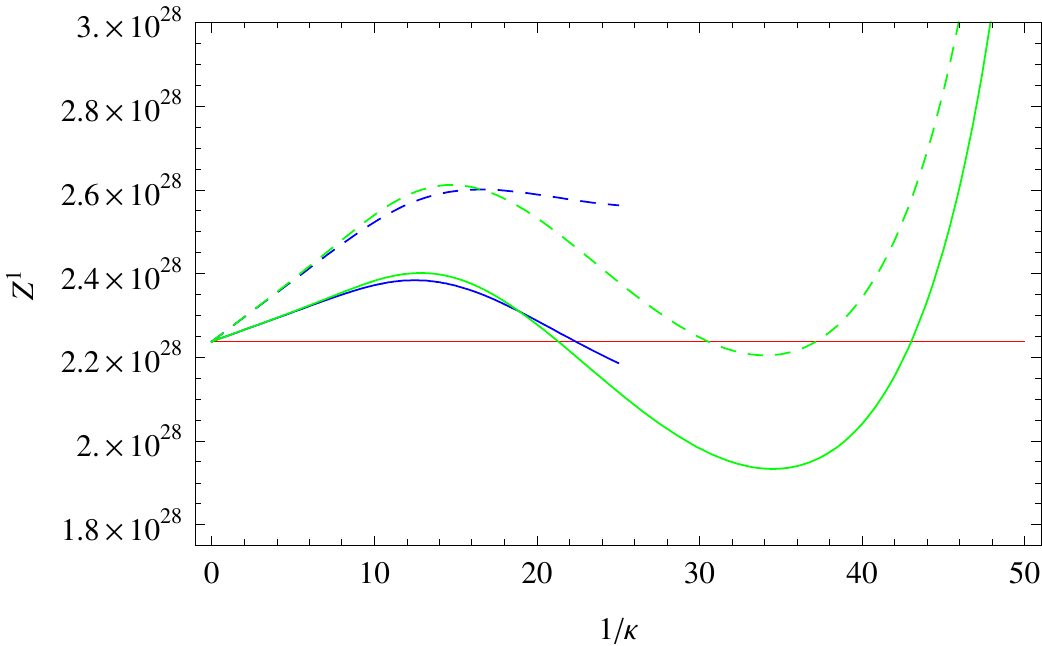}}
\end{center}
\caption{The single particle partition function is plotted vs $m_0$ (left) and ${1\over\kappa}$ (right). The plots for the unmodified measure are in solid lines while those for modified measure are in dashed lines.
The plots in massless (left) and SR (right) cases are shown in red.
The leading order behaviours are plotted in blue. The numerical plots for (\ref{Z}) and (\ref{Z1_final2}) are shown in green (solid and dashed respectively) for comparison.
Different values for the parameters in natural units are as follows: $V=10^{35}, N=10^{25}, T=0.01, a_{0}=1, a_{1}=0.2, a_{2}=a_{3}=...=0$ and $\kappa=1$ (left), $m_0=0.01$ (right).
}
\label{fg.Z}
\end{figure*}

Thermodynamics of a photon gas in DSR with dispersion relation (\ref{MS}) and unmodified measure has been worked out in \cite{Das:2010gk}. Here we calculate the deviation of single particle partition function from that of a photon gas. On expanding $Z_1^0$ in $\eta = \frac{m_0}{\kappa}$ with $m_0 \rightarrow 0$ (assuming $\kappa$ to be finite) and using (9.6.10) and (9.6.11) of \cite{abramowitz}, we get
\beq
Z_1^0=Z_{1 ml}^0+Z_{1 ml corr}^0,
\eeq
where $Z_{1 ml}^0$ is the single particle partition function of photon gas in DSR scenario with unmodified measure\cite{Das:2010gk} and $Z_{1 ml corr}^0$ is $\mathcal{O}\l\eta\r$:
\beqa
Z_{1ml}^0&=&\frac{2V}{(2\pi)^2\beta^3}\left(2-e^{-\beta\kappa}(\beta^2\kappa^2+2\beta\kappa+2)\right),\nonumber\\
Z_{1mlcorr}^0&=& - \frac{2V}{(2\pi)^2\beta^3}\frac{(\beta\kappa)^4}{8}\ln(\eta)\l\eta^4+\mathcal{O}(\eta^5)\r+\l\beta\kappa Z_{1ml}^0\r \eta + \mathcal{O}(\eta^2).
\label{zm0}
\eeqa
Note that the correction due to mass of the constituent particle is non-perturbative in nature as the first term in $Z_{1mlcorr}^0$ which contains $\ln(\eta)$ is the non-analytic piece and does not allow a trivial Taylor series expansion at $\eta=0$.
We can rewrite (\ref{Z1_final2}) as
\beq
Z_1=Z_{1 ml}+Z_{1 ml corr},
\eeq
where
\begin{equation}
Z_{1 ml}=\displaystyle{\sum_{n=0}^{\infty}} (-1)^{n}\frac{a_{n}}{n!\kappa^{n}}\frac{\partial^{n}Z_{1 ml}^0}{\partial \beta^{n}},
\end{equation}
\begin{equation}
Z_{1 ml corr}=Z^0_{1 ml corr} + \displaystyle{\sum_{n=1}^{\infty}}(-1)^{n}\frac{a_{n}}{n!\kappa^{n}}\l \frac{\partial^{n}Z^0_{1 ml corr}}{\partial\beta^{n}}-\eta n \kappa \frac{\partial^{n-1}Z_{1 ml}^0}{\partial\beta^{n-1}} \r.
\end{equation}
The above leading order behaviours have been plotted in Fig \ref{fg.Z}.
For our choice of parameters they match with the numerical plots up to $\frac{m_0}{\kappa}\sim 0.012$. 

\subsubsection{Leading order deviation from the SR case}
On expanding $Z_1^0$ in $\eta = \frac{m_0}{\kappa}$ with $\kappa\rightarrow\infty$ (assuming $m_{0}$ to be finite), we get
\beq
Z_1^0=Z_{1 SR}^0+Z_{1 SR corr}^0,
\eeq
where $Z_{1 SR}^0$ is the single particle partition function in SR and $Z_{1 SR corr}^0$ is $\mathcal{O}\l\eta\r$.
\begin{equation}
Z_{1 SR}^0=\frac{2Vm_0^3}{(2\pi)^2}e^{\beta m_0} \frac{K_2(\beta m_0)}{\beta m_0}
\end{equation}
\begin{equation}
Z_{1 SR corr}^0 = - \frac{2Vm_0^3}{(2\pi)^2}\frac{e^{\beta m_0}}{\beta m_0}\frac{e^{-\beta m_0/\eta}}{\eta^2} \l1+\mathcal{O}(\eta)\r 
+\left(1 -\frac{K_1(\beta m_0)}{K_2(\beta m_0)} \right)\beta m_0Z^0_{1SR}\eta
+ \mathcal{O}\left({\eta^2}\right).
\label{sr_corr}
\end{equation}
Note that the DSR correction is non-perturbative in nature as the first term in $Z_{1 SR corr}^0$ which contains $e^{-\beta m_0/\eta}$ is the non-analytic piece and does not allow a Taylor series expansion at $\eta=0$. 
This is a novel feature in DSR as we know that SR thermodynamics is perturbative
in the non-relativistic limit:
\beqa
Z_{1SR}^0&=&\frac{4\pi V\l k_BTm_0\r^{3/2}}{h^3}u^{1/2}e^{u}K_2(u)\nonumber\\
&&\stackrel{u\rightarrow \infty}{\longrightarrow}V\l\frac{2\pi m_0k_BT}{h^2}\r^{3/2}\l 1+\frac{15}{8u}+\mathcal{O}\l\l\frac{1}{u}\r^2\r\r\nonumber\\
&&=Z_{1NR}\l 1+\frac{15}{8u}+\mathcal{O}\l\l\frac{1}{u}\r^2\r\r,
\eeqa
where $u=\frac{m_0c^2}{k_BT}$ and $Z_{1NR}$ is the single particle partition function in the non-relativistic case.
We can rewrite (\ref{Z1_final2}) as
\beq
Z_1=Z_{1 SR}+Z_{1 SR corr},
\eeq
where
\beqa
Z_{1 SR}&=&Z_{1 SR}^0,\nonumber\\
Z_{1 SR corr}&=&Z^0_{1 SR corr} + \frac{a_{1}m_0}{\kappa}Z_{1 SR}^0-\frac{a_{1}}{\kappa}\frac{\partial Z^0_{1 SR}}{\partial \beta}.
\eeqa
The above leading order behaviours have been plotted in Fig \ref{fg.Z}.
For our choice of parameters they match with the numerical plots up to $\frac{m_0}{\kappa}\sim 0.08$.
Having obtained the leading order correction to $Z_1$ due to DSR, we shall now compute its
effect on the various thermodynamic quantities. 

\section{Thermodynamic quantities} \label{thermodynamics}
The free energy $F$, pressure $P$, entropy $S$, internal energy $U$, internal energy density $\rho$ and heat capacity $C_{V}$ are defined as
\beqa
F &=& -{1\over \beta}\ln\left(Z_N(V,\beta,m_0)\right) = -\frac{1}{\beta} N\left\{\ln\left({Z_1\over N}\right)+1\right\} \\
P &=& -\left({\partial F\over \partial V}\right)_{N,T} \\
S &=& -\left({\partial F\over \partial T}\right)_{V,N} \\
U &=& F+TS \\
\rho &=& \frac{U}{V} \\
C_V &=& \left({\partial U\over \partial T}\right)_{N,V}
\eeqa 
\begin{figure*}[h]
\vspace{-1.25 cm}
 \begin{center}
  \scalebox{0.74}{\includegraphics{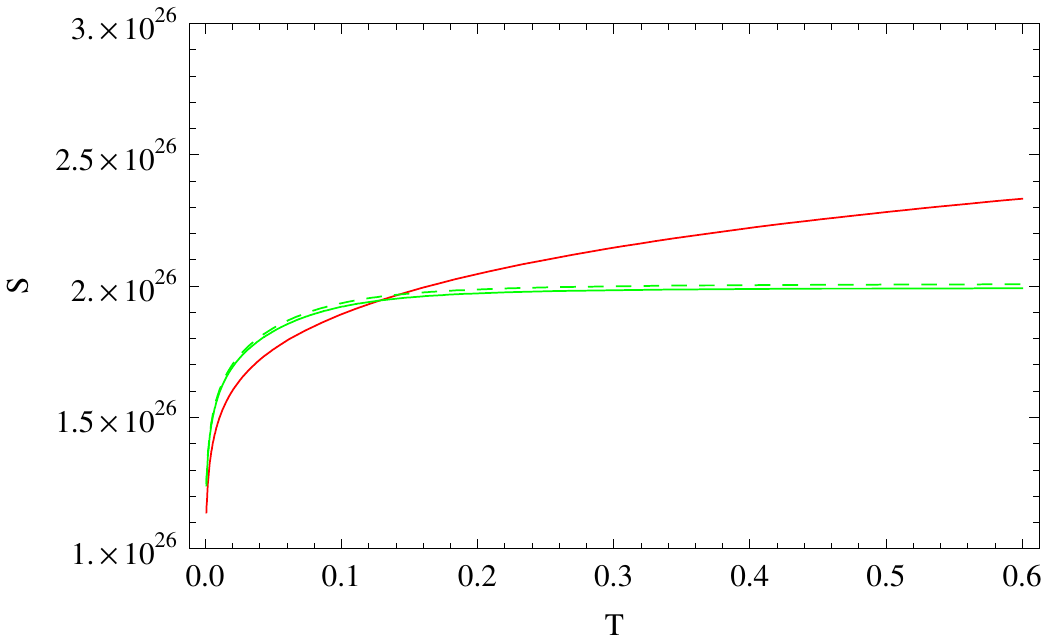}}
  \scalebox{0.74}{\includegraphics{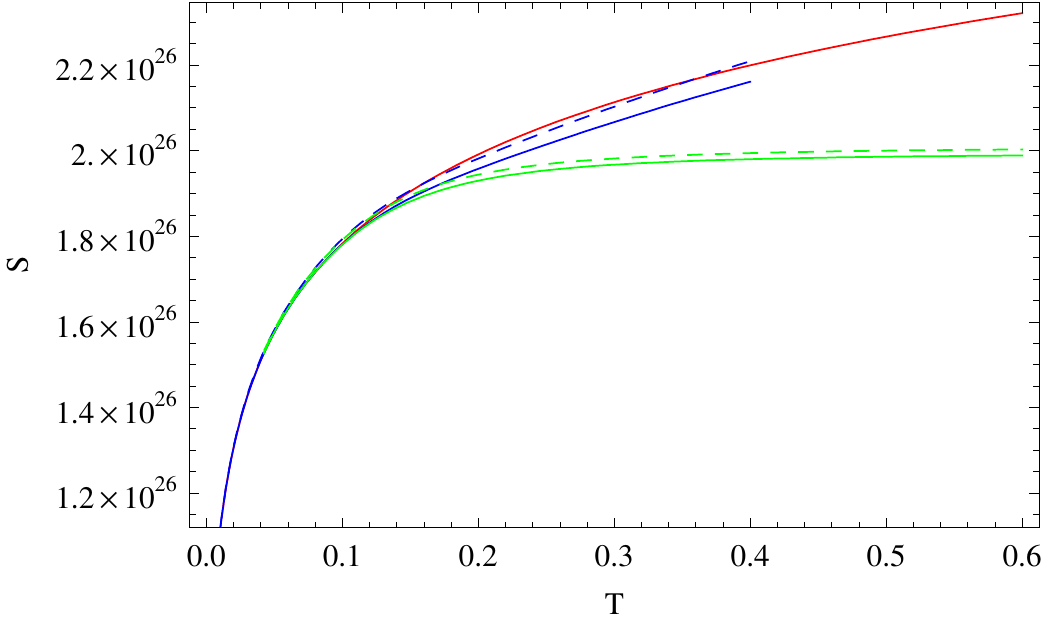}} \\
  \scalebox{0.74}{\includegraphics{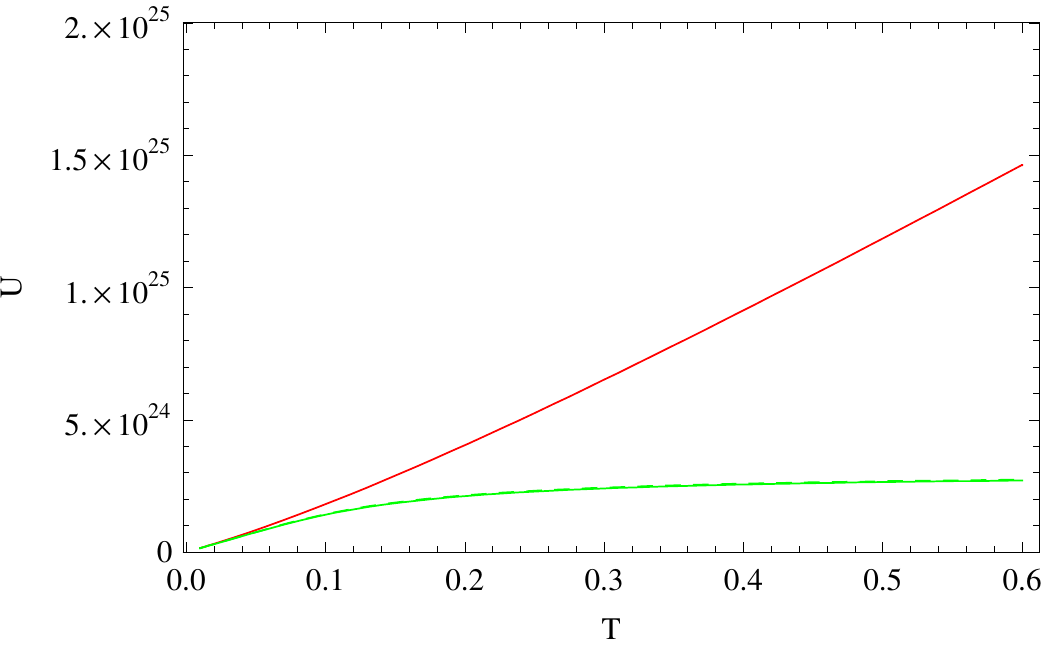}}
  \scalebox{0.74}{\includegraphics{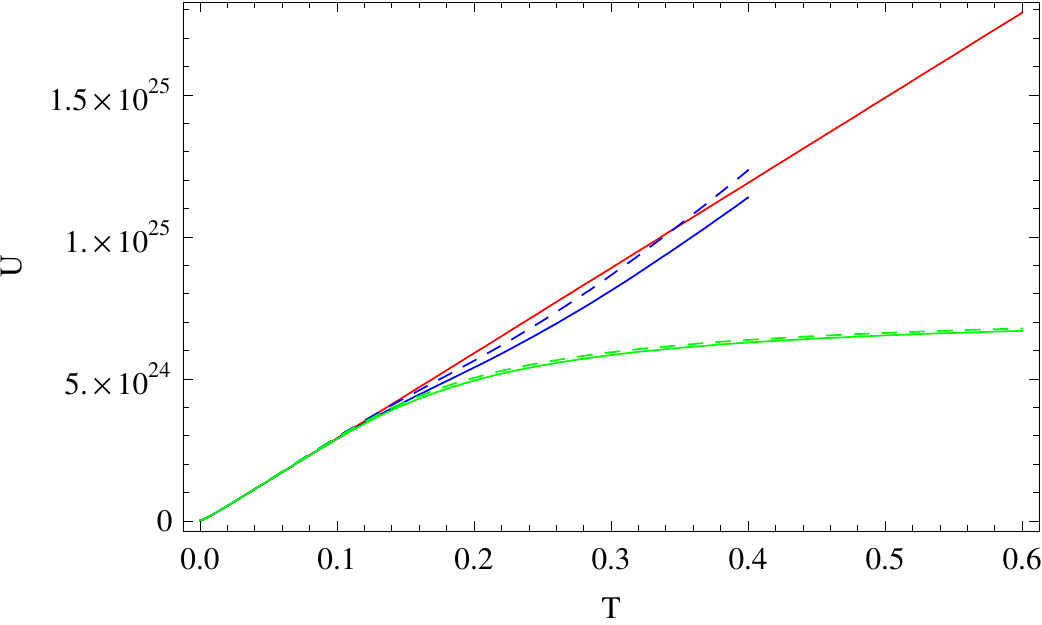}}\\
  \scalebox{0.74}{\includegraphics{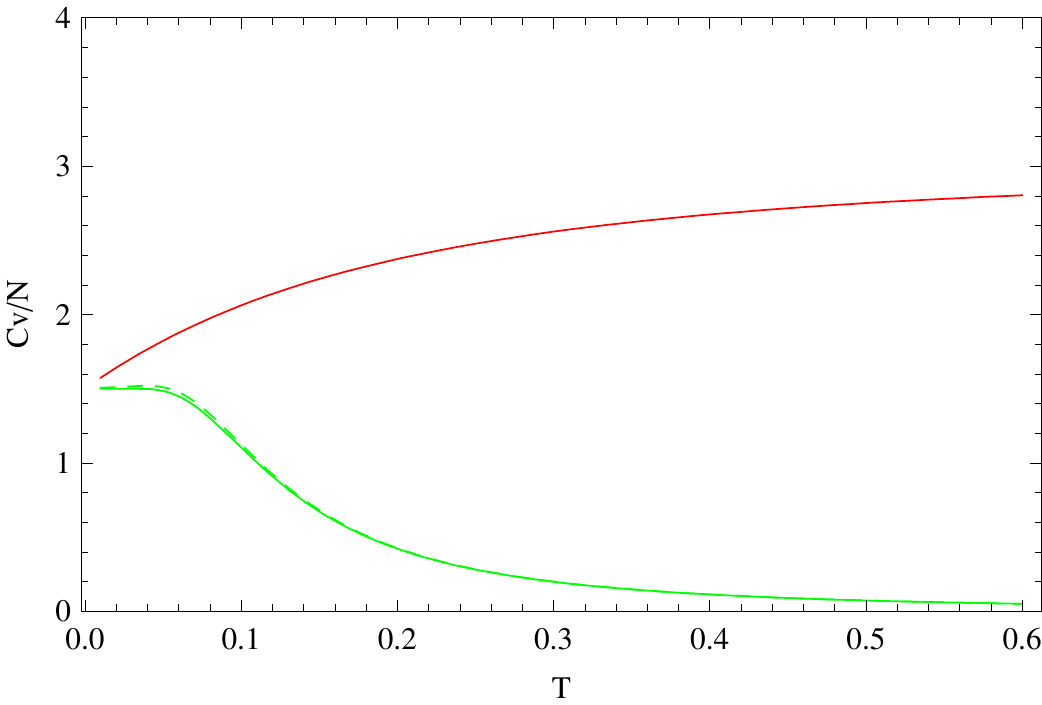}}
  \scalebox{0.74}{\includegraphics{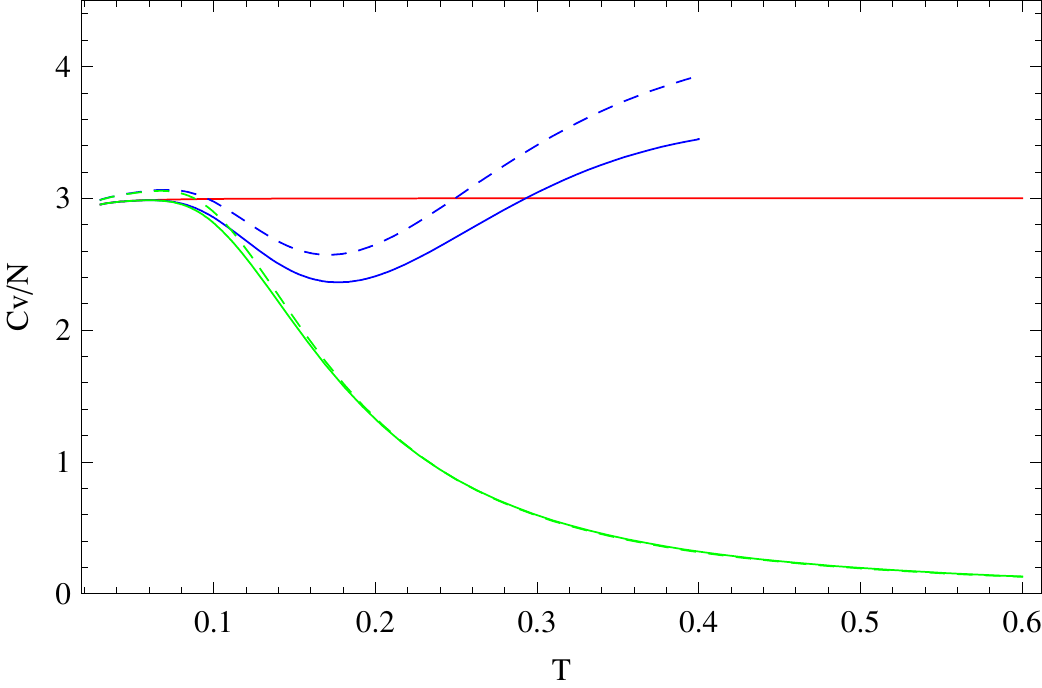}}\\
  \scalebox{0.74}{\includegraphics{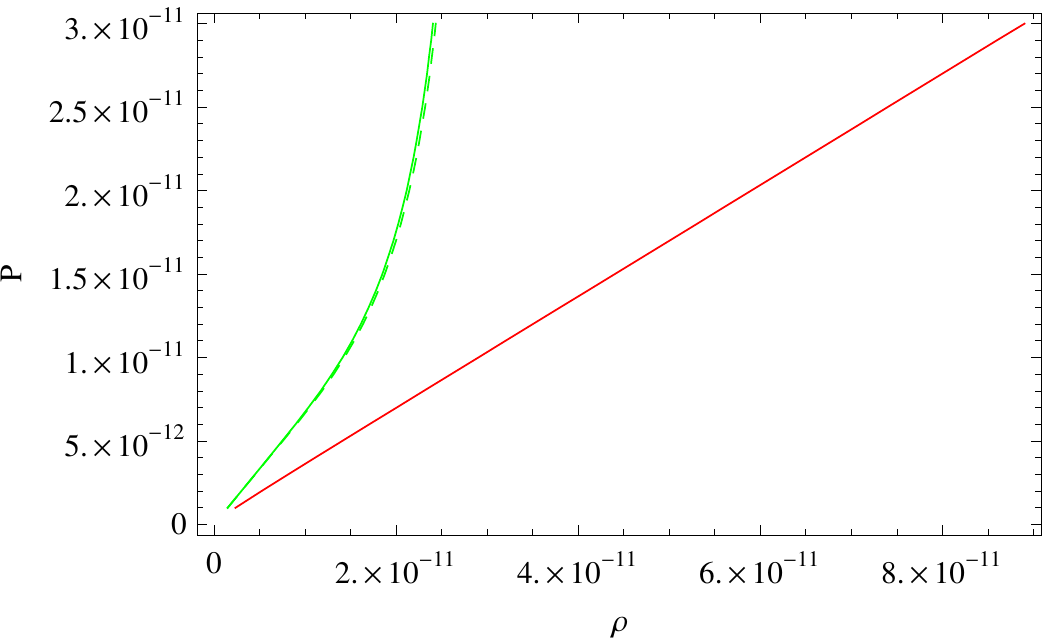}}
  \scalebox{0.74}{\includegraphics{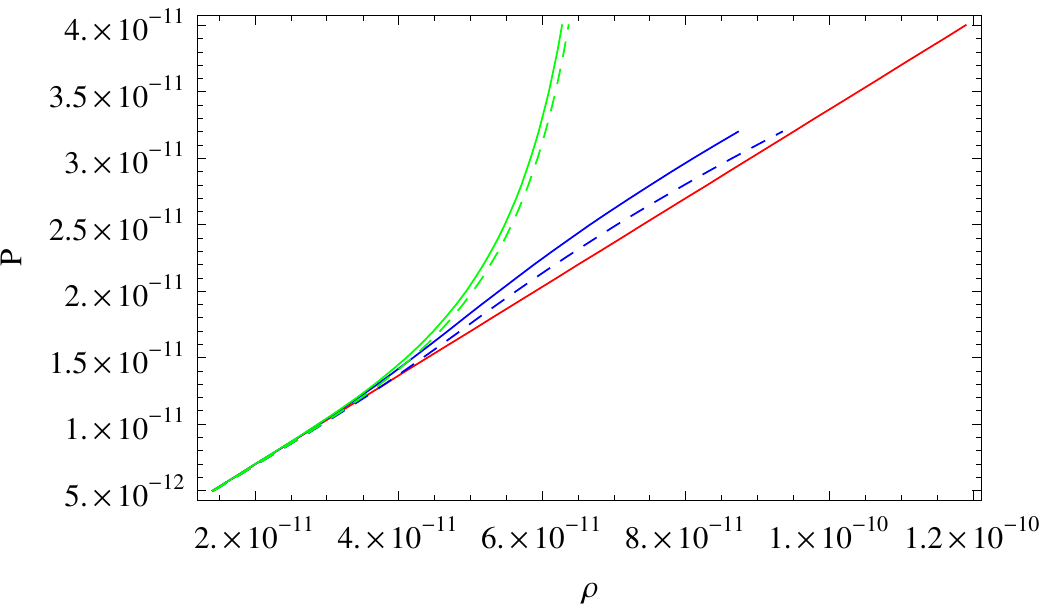}}\\
\end{center}
\caption{The plots for the unmodified measure are in solid lines while those for modified measure are in dashed lines.
The SR plots are shown in red. The figures in the left column are for Case I while those in the right column are for Case III. The DSR plots for Case I are plotted in green. For Case III, the leading order behaviours are plotted in blue. The numerical plots for Case III are also shown in green for comparison. Different values for the parameters in natural units are as follows: $V=10^{35}, N=10^{25},\kappa=1, m_{0} = 1 \, {\rm (Case \,\, I)}, 0.01 \, {\rm (Case\,\, III)}, a_{0}=1, a_{1}=0.2, a_{2}=a_{3}=...=0$.
}
\label{fg.TQ}
\end{figure*}
%
%
The above quantities for Case I can be found by using (\ref{Zcase2}) and they have been plotted in Fig \ref{fg.TQ}. Now we shall obtain the leading order thermodynamics for Case III.
If we denote  the free energy, pressure, entropy, internal energy, internal energy density and heat capacity obtained in the SR or massless cases by $F_0, P_0, S_0, U_0$, $\rho_0$ and $C_{V0}$ respectively, and write $Z_1=Z_{10}+Z_{1corr}$, where $\l Z_{10},Z_{1corr}\r=\l Z_{1SR},Z_{1SRcorr}\r$ or $\l Z_{1ml},Z_{1mlcorr}\r$, we have
\begin{eqnarray}
F&=&F_{0} - {N\over \beta}ln \left(1+{Z_{1corr}\over Z_{10}}\right)=F_{0}-{N\over \beta} {Z_{1corr}\over Z_{10}} + \mathcal{O}\l\left({Z_{1corr}\over Z_{10}}\right)^2\r \\
S&=&N\left[\ln\left({Z_1\over N}\right)+1\right]-{\beta N\over Z_1}{\partial Z_1\over \partial \beta}\nonumber\\
&=&S_{0}+{NZ_{1corr}\over Z_{10}}-{\beta N\over Z_{10}}
\left({\partial Z_{1corr}\over \partial \beta}-{Z_{1corr}\over Z_{10}}{\partial Z_{10}\over \partial \beta}\right)+ \mathcal{O}\l\left({Z_{1corr}\over Z_{10}}\right)^2\r \\
U&=&U_{0}\l1-{Z_{1corr}\over Z_{10}}\r-{N\over Z_{10}}{\partial Z_{1corr}\over \partial \beta}+ \mathcal{O}\l\left({Z_{1corr}\over Z_{10}}\right)^2\r \\
C_V&=&C_{V0}\l1-\frac{Z_{1corr}}{Z_{10}}\r \nonumber \\
&& -\beta^2\left[-
{U_{0}\over Z_{10}}{\partial Z_{1corr}\over \partial \beta}+{Z_{1corr}\over Z_{10}^2}U_{0}{\partial Z_{10}\over \partial \beta}+
{N\over Z_{10}^2}{\partial Z_{10}\over \partial \beta}{\partial Z_{1corr}\over \partial \beta}-{N\over Z_{10}}{\partial^2 Z_{1corr}\over \partial^2 \beta}\right]\nonumber\\
&&+ \mathcal{O}\l\left({Z_{1corr}\over Z_{10}}\right)^2\r \\
\rho&=&\rho_{0}\left(1-{Z_{1corr}\over Z_{10}}\right)-{n\over Z_{10}}{\partial Z_{1corr}\over \partial \beta}+ \mathcal{O}\l\left({Z_{1corr}\over Z_{10}}\right)^2\r
\end{eqnarray}
where $n = \frac{N}{V}$ is the number density.
The correction to $F$ depends on the ratio of $Z_{1corr}$ and $Z_{10}$ and is independent of the volume $V$ of the system. Note that this is true to all orders.
Hence, the pressure $P$ of the system which is defined as $ P=-\left({\partial F\over \partial V}\right)_{N,T}$ gets no correction:
\begin{equation}
  P=P_0.
\end{equation}
The equation of state in SR is (see $(8.128)$ and $(8.134)$ of \cite{greinerbook}) 
\beq
P_{SR}={\rho_{SR}\over3-\beta m+\frac{K_1(\beta m)}{K_2(\beta m)}\beta m},
\eeq 
which gives the following DSR equation of state
\begin{eqnarray}
 P&=&{\left(\rho+{Z_{1SRcorr}\over Z_{1SR}}\rho+{n\over Z_{1SR}}{\partial Z_{1SRcorr}\over \partial \beta}\right)\over3-\beta m+\frac{K_1(\beta m)}{K_2(\beta m)}\beta m}+ \mathcal{O}\l\left({Z_{1SRcorr}\over Z_{1SR}}\right)^2\r.
\end{eqnarray}
In Fig \ref{fg.TQ}, we have plotted the various thermodynamic quantities for Cases I and III as a function of $T$ and compared them with the SR case. The $P$ vs $\rho$ plots have been obtained by varying $T$ keeping all other parameters fixed. The qualitative natures of the plots for different cases are same. In case of $S$ there are two competing effects: while the  cutoff tries to reduce $S$ by limiting the number of accessible states, the modified dispersion tries to increase $S$ by enhancing the Boltzmann weight $\exp(-\e/T)$ (note that $\e_{DSR}(p)< \e_{SR}(p)$ for a given momentum state $p$, the change being more for greater value of the parameter $m$). At low temperatures, the latter is dominant and $S_{DSR}> S_{SR}$. For our choice of parameters this is clearly visible for the plot of $S$ in Case I. In the high $T$ regime, the cutoff effect comes into play and  $S_{DSR}< S_{SR}$. The cutoff also saturates $U$ as $T$ increases, and $C_V \rightarrow 0$, resulting in a steeper equation of state. Here we make an interesting observation. There have been attempts to define velocity in DSR \cite{Kosinski:2002gu}. If we adopt the usual definition for the speed of sound $c_s = \sqrt{\frac{\partial P}{\partial \rho}}$, then we observe that $c_s$ grows without any bound. Possibility of such scenarios has been discussed in \cite{Kim:2004uj}.

For given choice of parameters in case of quantities like $S$ and $U$, 
the leading order behaviours for Case III match with the numerical plots up to $T \sim 0.15$, 
while in case of $C_V$ which contains second order derivatives of the partition function with respect to $T$, 
the leading order behaviours match with the numerical plots up to $T \sim 0.09$. 
Note that the leading order behaviours have been obtained assuming $\frac{1}{\beta m_0}$ to be finite and $\frac{m_0}{\kappa}\rightarrow 0$ 
which in turn implies $\frac{1}{\beta\kappa}=\frac{T}{\kappa}\rightarrow 0$. Hence as $T$ increases, the leading order plots 
depart from their numerical counterparts.


\chapter{Conclusions}
 
In chapter \ref{instanton} we described static classical solutions of noncommutative gauge theories in various spacetime dimensions 
and showed that the GBOs are significant in constructing solutions with higher topological numbers.
We started with describing the static magnetic flux tube solutions of higher moment found by Polychronakos in terms of GBOs. In doing so we get a better understanding of the winding numbers $n_{0i}$'s. They correspond to the radii of the flux tube in different irreducible parts of the full Fock space. The increase in the degrees of freedom (higher number of required $n_{0i}$'s) due to use of reducible representation of the oscillator algebra cause to expand the solution space of the flux tube solitons.
On the other hand, the vortices with higher winding numbers correspond to known solutions.

The case of 
multi--instantons is different. The multi--instantons with charge $-p_1p_2$ ($p_1,p_2$ non-negative integers) are 
not gauge equivalent to known solutions. Another significant result of this study is an explicit relation between the instanton number and the 
representation theory labels $p_1$ and $p_2$. The charge of a multi--instanton does have the information about the reducibility of the space along with the topological nature of the solutions. 
Using the ``translated'' $b$ operators (\ref{translation_operator})
 we have constructed multi--instantons that depend explicitly on $p_1p_2$ complex parameters. While the full moduli space of 
noncommutative multi--instantons is still not well understood, we hope that this identification contributes partially to this question.
 
Though we have only considered a few cases, there is actually a large variety of situations in noncommutative 
gauge theories where GBOs may be used. 
In particular we expect that this procedure may shade new light on merons, monopoles,
dyons, skyrmions etc. We plan to revisit some of these questions in future.

In chapter \ref{qo} we studied the effect of noncommutative time coordinate. To do so we took a simple quantum mechanical system with time-space noncommutativity in 1+1 dimension. We developed a formalism to compute the transitions between asymptotic  states 
of the quantum mechanical system with noncommutative time.
We found that for a free Hamiltonian in $\mathds{R}_\theta^{1,1}$ which is independent of time, the transitions are equal to
the same for a different Hamiltonian in $\mathds{R}^{1,1}$ found after the replacements (\ref{replace}).
The time evolution of an operator and its expectation value (and hence also its uncertainty) can also be found in a similar manner.
Specifically, for FHO the transition probabilities get modified and is given by (\ref{Amn_final}) and (\ref{Pmn}).
The Poissonian distribution for the ``vacuum to any state transition'' also gets modified and is given by (\ref{Pm0}).
The study of uncertainties in position and momentum says that the time-evolved state is no more coherent.
It gets some squeezing effect due to the noncommutativity, keeping the product of the uncertainties minimum.
These uncertainties are explicitly found and is given in (\ref{uncertainties_x_final2}), (\ref{uncertainties_p_final2}) and (\ref{uncertainties_xp_final2}).
The leading order corrections in these uncertainties are oscillatory in time and they depend independently on the mass of the
particle $m_0$ and the frequency of the oscillator $\omega$ (note that the commutative uncertainties depend only on the product $m_0\omega$). 
These results suggest a possible modelling of the noncommutativity for the nonlinear phenomena in Quantum Optics.
The noncommutativity results in the following nonlinear effects:
\begin{enumerate}
\item{The photon-count gets modified from the usual Poisson distribution.}
\item{The uncertainties in the field quadratures change keeping the product minimum (the squeezing effect).}
\item{The second order correlation function $g^{(2)}(0)$ gets modified producing new effects 
like bunching or anti-bunching of photons depending on the value of $Im(\beta_3\xi^2)$.}
\end{enumerate}
All these observations suggest that the noncommutativity produces incoherency in the otherwise coherent field.

In chapter \ref{DSR}
we used the dispersion relation (\ref{MS}) and have considered the modified phase space measure (the modification being isotropic and expandable in Taylor series).
We considered three cases separately ($m=\kappa, m>\kappa, m<\kappa$). 
The single particle partition function has been shown to be smooth in $m_0 \in (0,\kappa)$ (see Appendix \ref{app1}).
For the case $m=\kappa$, a simple analytical form for the partition function has been obtained (see (\ref{Zcase2})) while a series solution for the partition function has been obtained for $m<\kappa$ (see (\ref{Z}) and (\ref{kstar_mr})).
In doing so, new type of special functions (Incomplete Modified Bessel functions) emerged.
We observed that DSR thermodynamics is non-perturbative in the SR and massless limits.
Using the leading order solutions, we derived thermodynamic quantities like the 
free energy, pressure, entropy, internal energy and heat capacity (see Fig.\ref{fg.TQ}).
A stiffer equation of state has been found.


\appendix

\chapter{Appendices for Chapter \ref{instanton}}

\section{Computing the flux tube solutions} \label{flux_tube_calculation}
We start with the ansatz
\begin{equation}
\mathcal{D} = af(N), \quad \quad \bar{\mathcal{D}} = - \mathcal{D}^\dagger = -f^*(N)a^\dagger.
\end{equation}
The equation of motion (\ref{eom_flux}) reduces to 
\begin{equation}
af(N)\left\{(N+1)|f(N+1)|^2 - 2N|f(N)|^2 + (N-1)|f(N-1)|^2 \right\} = 0.
\end{equation}
We write $f(N)$ in the basis of number states
\begin{equation}
f(N) = \sum_{n=0}^\infty f(n) |n\rangle \langle n|
\end{equation} 
to get
\begin{equation}
f(n) = 0,
\quad \mbox{{\rm or }} \quad (n+1)|f(n+1)|^2 - 2n|f(n)|^2 + (n-1)|f(n-1)|^2 = 0.
\end{equation}
We try to find a solution for which $f(n)$ satisfies the first of the above equations, i.e., $f(n) = 0$ for $0\leq n\leq n_0$ and the second of the above equations for $n> n_0$. This choice leads to the localized flux tube solutions as discussed in the section \ref{sec:flux_sec}. The most general solution of this type is given by
\begin{eqnarray}
f(n) &=& \left\{
\begin{array}{cc}
0 & \mbox{{\rm for}} \, \, 0\leq n\leq n_0 \\
e^{i\varphi_n}\left| f(n_0+1)\right| \sqrt{\frac{(n_0+1)(n-n_0)}{n}} & \mbox{{\rm for}} \, \, n> n_0 
\end{array}
\right. \\
&=& \left\{
\begin{array}{cc}
0 & \mbox{{\rm for}} \, \, 0\leq n< n_0 \\
e^{i\varphi_n}\left| f(n_0+1)\right| \sqrt{\frac{(n_0+1)(n-n_0)}{n}} & \mbox{{\rm for}} \, \, n\geq n_0 
\end{array}
\right.
\end{eqnarray}
For the choice $\varphi_n = 0, \left| f(n_0+1)\right| = \frac{1}{\sqrt{(n_0+1)\theta}}$ we get
\begin{equation}
f(N) = \frac{1}{\sqrt{\theta}}\sqrt{\frac{N-n_0}{N}}\sum_{n=n_0}^\infty |n\rangle \langle n|
\end{equation}
which corresponds to the solution (\ref{polychronakos_sol}).

Let us now start with the ansatz
\begin{equation}
\mathcal{D} = b \, G(N).
\end{equation}
The equation of motion reduces to
\begin{equation}
b \, G(N) \left\{ (M+1) \left|G(N+2)\right|^2 - 2M\left|G(N)\right|^2 + (M-1)\left|G(N-2)\right|^2 \right\} = 0.
\end{equation}
We expand $G(N)$ in the number state basis, i.e.,
\begin{equation}
G(N) = \sum_{n=0}^\infty G(n) |n\rangle \langle n|
\end{equation}
to get
\begin{equation}
G(n) = 0, \quad
\mbox{{\rm or }} \quad \left(\frac{n}{2}+1\right)|G(n+2)|^2 - n|G(n)|^2 + \left(\frac{n}{2}-1\right)|G(n-2)|^2 = 0.
\end{equation}
for $n=even$ and
\begin{equation}
G(n) = 0, \quad
\mbox{{\rm or }} \quad \left(\frac{n-1}{2}+1\right)|G(n+2)|^2 - (n-1)|G(n)|^2 + \left(\frac{n-1}{2}-1\right)|G(n-2)|^2 = 0.
\end{equation}
for $n=odd$.
There exists a solution of the form
\begin{eqnarray}
G(n) &=& \left\{
\begin{array}{cc}
0 & \mbox{{\rm for}} \, \, 0\leq n\leq n_{0+} \\
\frac{1}{2}e^{i\varphi_{+n}}\left| G(n_{0+}+2)\right| \sqrt{\frac{(n_{0+}+2)(n-n_{0+})}{n}} & \mbox{{\rm for}} \, \, n> n_{0+} 
\end{array}
\right. \\
&=& \left\{
\begin{array}{cc}
0 & \mbox{{\rm for}} \, \, 0\leq n< n_{0+} \\
\frac{1}{2}e^{i\varphi_{+n}}\left| G(n_{0+}+2)\right| \sqrt{\frac{(n_{0+}+2)(n-n_{0+})}{n}} & \mbox{{\rm for}} \, \, n\geq n_{0+} 
\end{array}
\right.
\end{eqnarray}
for $n=even$ and
\begin{eqnarray}
G(n) &=& \left\{
\begin{array}{cc}
0 & \mbox{{\rm for}} \, \, 0\leq n\leq n_{0-} \\
\frac{1}{2}e^{i\varphi_{-n}}\left| G(n_{0-}+2)\right| \sqrt{\frac{(n_{0-}+1)(n-n_{0-})}{n-1}} & \mbox{{\rm for}} \, \, n> n_{0-} 
\end{array}
\right. \\
&=& \left\{
\begin{array}{cc}
0 & \mbox{{\rm for}} \, \, 0\leq n< n_{0-} \\
\frac{1}{2}e^{i\varphi_{-n}}\left| G(n_{0-}+2)\right| \sqrt{\frac{(n_{0-}+1)(n-n_{0-})}{n-1}} & \mbox{{\rm for}} \, \, n\geq n_{0-} 
\end{array}
\right.
\end{eqnarray}
for $n=odd$. Here $n_{0+}$ and $n_{0-}$ are even and odd integers respectively. For the choice $G(n_{0+}+2)=\frac{2}{\sqrt{(n_{0+}+2)\theta}}, G(n_{0-}+2)=\frac{2}{\sqrt{(n_{0-}+1)\theta}}, \,\,\varphi_{+n} = \varphi_{-n} = 0$ we get
\begin{equation}
G(N) = \frac{1}{\sqrt{\theta}}\sqrt{\frac{N-n_{0+}\Lambda_+ -n_{0-} \Lambda_-}{2M}}
\left(\sum_{n=n_{0+}}^\infty |n\rangle \langle n|\Lambda_+ +\sum_{n=n_{0-}}^\infty |n\rangle \langle n|\Lambda_-\right)
\end{equation}
which corresponds to the solution (\ref{flux_tube_sol_b}).

\section{Determination Of $A_\infty$ and $\bar{A}_\infty$ \label{det_gauge_field}}
We multiply $\bar{\Phi}_{\infty}$ to the first equation of (\ref{eom_sublead}) from right and $\Phi_\infty$ to the second equation of (\ref{eom_sublead}) from left  and use (\ref{eom_lead}) to get
\begin{equation}
 \bar{A}_\infty=-i\left(a- \Phi_{\infty}a\bar{\Phi}_{\infty}\right), \quad A_\infty = i \left(a^\dagger - \Phi_\infty a^\dagger \bar{\Phi}_\infty\right).
\label{expression_gauge_field}
\end{equation}
For the solutions (\ref{witten_vortex}) we get
\begin{eqnarray}
 \bar{A}_\infty&=&-i\left(a- \Phi_{\infty}a\bar{\Phi}_{\infty}\right)\\
&=& -i\left(a-a\frac{\sqrt{(N+n)(N+n-1).....(N+1)} }{\sqrt{(N+n-1)(N+n-2).....N}}\right)\\
&=& -ia \frac{\sqrt{N}-\sqrt{N+n}}{\sqrt{N}}\\
&=& -i \frac{1}{\sqrt{N+1}} a\left(\sqrt{N}-\sqrt{N+n}\right).
\end{eqnarray}
Here we have used the identity 
\begin{eqnarray}
 a^n a^{\dagger n} &=& 
(N+n)(N+n-1).....(N+1).
\end{eqnarray}
In a similar fashion we get the expression for $\bar{A}_\infty$. 
The same technique can be used to compute $A^{new}_\infty$ and $\bar{A}^{new}_\infty$ as in (\ref{new_gauge_field}).

\section{Large Distance Behavior \label{l_d_b}}
Let $|\omega\rangle $ be the coherent states of the operator $a$, where $\omega$ is a complex number labeling the states. In the large 
$\omega $ limit, the expectation value $\langle\omega|\Phi_{\infty}| \omega 
\rangle$ and $\langle\omega|\bar{A}_\infty| \omega \rangle$ gives the large distance behavior of the solution:
\begin{eqnarray}
 \langle\omega|\Phi_{\infty}| \omega \rangle &=&\langle\omega| \frac{1}{\sqrt{a^n a^{\dagger n}}}a^n| \omega \rangle  \\
&=& \omega^n \langle\omega|  \frac{1}{\sqrt{(N+n)(N+n-1)....(N+1)}}| \omega \rangle  \\
&\approx& \omega^n \langle\omega|  \frac{1}{(\sqrt{N})^n}| \omega \rangle  \\
&\approx& \omega^n (\langle\omega| \frac{1}{N}| \omega \rangle)^{\frac{n}{2}}\\
 &\approx& \omega^n \frac{1}{(\bar{\omega}\omega)^{\frac{n}{2}}} \\
&\approx& e^{i n\varphi}.
\end{eqnarray}
For the gauge field,
\begin{eqnarray}
 \langle\omega|\bar{A}_\infty| \omega \rangle &=& \langle\omega|-i \frac{1}{\sqrt{N+1}} a(\sqrt{N}-\sqrt{N+n})| \omega \rangle \\
&\approx&-i \langle\omega| \frac{N+1}{\sqrt{N+1}}a (1-(1+\frac{n}{2N})| \omega \rangle\\
&\approx& i \frac{n}{2}\langle\omega|a\frac{1}{N}|\omega \rangle\\
&\approx&  i \frac{n}{2}\langle\omega|\frac{1}{N}a|\omega \rangle\\
&\approx&  i\frac{n}{2\bar{\omega}}.
\end{eqnarray}
Similar thing can done for  $\Phi^{new}_{\infty}$ and $\bar{A}^{new}_\infty$ to get (\ref{large_distance2}).

\section{Computation of the leading order magnetic field for the Nielsen-Olesen Vortices} \label{appendix_B}
Using (\ref{expression_gauge_field}) in the expression of $B_\infty$ (\ref{B_infinity}) we get
\begin{eqnarray}
B_\infty &=& -1 + \left[\Phi_\infty a \bar{\Phi}_\infty, \Phi_\infty a^\dagger \bar{\Phi}_\infty\right] \\
&=& \Phi_\infty \left(a\bar{\Phi}_\infty\Phi_\infty a^\dagger - a^\dagger \bar{\Phi}_\infty\Phi_\infty a\right) \bar{\Phi}_\infty -1
\end{eqnarray}
Using (\ref{Phibar_Phi}) one can calculate
\begin{equation}
a \bar{\Phi}_\infty \Phi_\infty a^\dagger - a^\dagger \bar{\Phi}_\infty \Phi_\infty a =
 1+ n |n\rangle \langle n| 
+ (n-1) |n-1\rangle \langle n-1|
- \sum_{m=0}^{n-2} |m \rangle \langle m|.
\end{equation}
Finally for the solution (\ref{witten_vortex}) we get the magnetic field as given in (\ref{B_witten}). Similar technique can be used to calculate $B_\infty^{new}$.

\section{The New Solution Satisfies The Equation Of Motion \label{new_soln_eom}}
The equations of motion are
\begin{equation}
 [\hat{D}_{x^\beta}, [\hat{D}_{x^\beta},\hat{D}_{x^\alpha}]]=0, \quad \alpha,\beta=1,2,3,4.
\end{equation}
There are four equations for different values of $\alpha$. 
We will show only for $\alpha=1$ and other follow similarly.
For $\alpha=1$ the equation becomes
\begin{equation}
 [\hat{D}_{x^2}, [\hat{D}_{x^2},\hat{D}_{x^1}]]+[\hat{D}_{x^3}, [\hat{D}_{x^3},\hat{D}_{x^1}]]+[\hat{D}_{x^4}, [\hat{D}_{x^4},\hat{D}_{x^1}]]=0.
\end{equation}
In terms of the complex coordinates, 
\begin{equation}
 \left([\hat{D}_1,[\hat{D}_{\bar{1}},\hat{D}_1]]+[\hat{D}_2,[\hat{D}_{\bar{2}},\hat{D}_1]]\right)+\left([\hat{D}_{\bar{1}},[\hat{D}_{\bar{1}},\hat{D}_1]]-
[\hat{D}_{\bar{2}},[\hat{D}_{2},\hat{D}_{\bar{1}}]]\right)=0.
\end{equation}
Now using the explicit expression for the solutions (\ref{gauge_field_r4_new}) we get
\begin{eqnarray}
[\hat{D}_{\bar{1}}, [\hat{D}_{\bar{1}},\hat{D}_1]]&=& S_{new} b_1^\dagger \sqrt{\frac{M+3}{M+1}} \sqrt{\frac{M}{M+2}}
\left( \frac{M+2}{M} \frac{M-1}{M+1} M_1 -  
\frac{M}{M+2} \frac{M+3}{M+1}( M_1+1)\right.\nonumber \\  
&& \left. - \frac{M}{M+2} \frac{M+3}{M+1}( M_1+1)+ \frac{M+1}{M+3} \frac{M+4}{M+2}( M_1+2)\right) S_{new}^\dagger,
\end{eqnarray}
\begin{eqnarray}
 [\hat{D}_{\bar{2}}, [\hat{D}_{2},\hat{D}_{\bar{1}}]]&=& -S_{new} b_1^\dagger \sqrt{\frac{M+3}{M+1}} \sqrt{\frac{M}{M+2}}
\left( \frac{M+2}{M} \frac{M-1}{M+1} M_2 -  
\frac{M}{M+2} \frac{M+3}{M+1}M_2\right. \nonumber \\  
&&\left. - \frac{M}{M+2} \frac{M+3}{M+1}( M_2+1)+ \frac{M+1}{M+3} \frac{M+4}{M+2}( M_2+2)\right) S_{new}^\dagger.
\end{eqnarray}
Adding above two we get
\begin{equation}
 [\hat{D}_{\bar{1}}, [\hat{D}_{\bar{1}},\hat{D}_1]]+[\hat{D}_{\bar{2}}, [\hat{D}_{2},\hat{D}_{\bar{1}}]]=0.
\end{equation}
Similarly we can show $[\hat{D}_{\bar{1}},[\hat{D}_{\bar{1}},\hat{D}_1]]-[\hat{D}_{\bar{2}},[\hat{D}_{2},\hat{D}_{\bar{1}}]]=0$
and hence the equation of motion is satisfied.

\section{The New Solution Satisfies The ASD Condition \label{new_soln_asd}}
The commutator between $\hat{D}_a$ and $\hat{D}_{\bar{a}}$ is
\begin{equation}
[\hat{D}_a , \hat{D}_{\bar{a}}] = -\frac{1}{\theta}S_{new}\left(\frac{M}{M+2}\frac{M+3}{M+1} 
(M_{a}+1)-\frac{M+2}{M}\frac{M-1}{M+1}M_{a}\right)S_{new}^\dagger.
\end{equation}
Summing over $a=1,2$ we get
\begin{equation}
 [\hat{D}_1 , \hat{D}_{\bar{1}}]+[\hat{D}_2 , \hat{D}_{\bar{2}}]= -\frac{1}{\theta}S_{new}\left(
M\frac{M+3}{M+1}-(M+2)\frac{M-1}{M+1}\right)S_{new}^\dagger 
=-\frac{2}{\theta}.
\end{equation}
Therefore, $ F_{1\bar{1}}+F_{2\bar{2}} =\frac{2}{\theta}+ [\hat{D}_1 , \hat{D}_{\bar{1}}]+[\hat{D}_2 , \hat{D}_{\bar{2}}]=0$.
Similarly one can show $F_{12}= F_{\bar{1}\bar{2}}=0$.
\section{The Topological Charge Of Instantons \label{topo_charge}}

The topological charge of the instanton is defined as
\begin{equation}
 Q \propto \theta^2 \,\,Tr\left(F_{1\bar{1}}F_{2\bar{2}}-F_{1\bar{2}}F_{2\bar{1}}\right) \implies Q 
= c\theta^2 \displaystyle{\sum_{n_1,n_2=0}^\infty}\langle n_1,n_2|\left(F_{1\bar{1}}F_{2\bar{2}}-F_{1\bar{2}}F_{2\bar{1}}\right)|n_1,n_2\rangle
\end{equation}
where we denote the constant of proportionality by $c$.

Further, the ASD condition gives
\begin{equation}
 F_{1\bar{1}}F_{2\bar{2}} = -F_{1\bar{1}}^2 = -\left( \frac{1}{\theta^2}+\frac{2}{\theta}\left[\hat{D}_1, \hat{D}_{\bar{1}}\right]
+\left[\hat{D}_1, \hat{D}_{\bar{1}}\right]^2\right).
\end{equation}
\subsection{Usual Single ASD solution}
The solutions (\ref{usual_instanton}) can be simplified to
\begin{equation}
 \hat{D}_{a} = \frac{1}{\sqrt{\theta}}S \sqrt{\frac{N(N+3)}{(N+1)(N+2)}}a_a S^{\dagger}, \quad
 \hat{D}_{\bar{a}} = -\frac{1}{\sqrt{\theta}}S a_a^\dagger \sqrt{\frac{N(N+3)}{(N+1)(N+2)}} S^{\dagger}.
\end{equation}
An arbitrary commutator becomes
\begin{equation} \label{d_comm_alpha_beta}
\left[\hat{D}_{a}, \hat{D}_{\bar{b}}\right] = 
-\frac{4}{\theta}S a_b^\dagger a_a\frac{1}{N(N+1)(N+2)} S^\dagger 
-\frac{\delta_{ab}}{\theta}S \frac{N(N+3)}{(N+1)(N+2)} S^\dagger
\end{equation}
and the components of the field strength (and their products) can be calculated to be
\begin{equation}\label{f11_f22}
 F_{1\bar{1}}F_{2\bar{2}} = -\frac{4}{\theta^2}S\frac{(N_2-N_1)^2}{N^2(N+1)^2(N+2)^2}S^\dagger.
\end{equation}
Again using (\ref{d_comm_alpha_beta}) we get
\begin{equation}
 F_{1\bar{2}} =  -\frac{4}{\theta}Sa_2^\dagger a_1\frac{1}{N(N+1)(N+2)}S^\dagger, \quad
 F_{2\bar{1}} =  -\frac{4}{\theta}Sa_1^\dagger a_2\frac{1}{N(N+1)(N+2)}S^\dagger.
\end{equation}
Multiplying the above two and simplifying gives
\begin{equation}
F_{1\bar{2}} F_{2\bar{1}} = \frac{16}{\theta^2}S\frac{(N_1+1)N_2}{N^2(N+1)^2(N+2)^2} S^\dagger \label{f12_f21}.
\end{equation}
Thus we get
\begin{equation}
 Q = -4c \displaystyle{\sum_{n_1=0}^\infty}\displaystyle{\sum_{n_2=1}^\infty}\frac{(n_1+n_2)^2+4n_2}{(n_1+n_2)^2(n_1+n_2+1)^2(n_1+n_2+2)^2}
-4c \displaystyle{\sum_{n_1=0}^\infty}\frac{1}{(n_1+2)^2(n_1+3)^2}
\end{equation}
\subsection{The New Solution}
Again the solutions (\ref{gauge_field_r4_new}) can be written as
\begin{equation}
 \hat{D}_a = \frac{1}{\sqrt{\theta}}S_{new} \sqrt{\frac{M(M+3)}{(M+1)(M+2)}}b_a S_{new}^{\dagger},
\quad
 \hat{D}_{\bar{a}} = -\frac{1}{\sqrt{\theta}}S_{new} b_a^\dagger \sqrt{\frac{M(M+3)}{(M+1)(M+2)}} S_{new}^{\dagger}.
\end{equation}
Then the commutators evaluate to
\begin{equation}
\left[\hat{D}_a, \hat{D}_{\bar{b}}\right] = 
-\frac{4}{\theta}S_{new} b_b^\dagger b_a\frac{1}{M(M+1)(M+2)} S_{new}^\dagger 
-\frac{\delta_{ab}}{\theta}S_{new} \frac{M(M+3)}{(M+1)(M+2)} S_{new}^\dagger.
\end{equation}
The product of the field strength becomes
\begin{eqnarray}
 F_{1\bar{1}}F_{2\bar{2}} = -\frac{4}{\theta^2}S_{new}\frac{(M_2-M_1)^2}{M^2(M+1)^2(M+2)^2}S_{new}^\dagger \\
F_{1\bar{2}} F_{2\bar{1}} = \frac{16}{\theta^2}S_{new}\frac{(M_1+1)M_2}{M^2(M+1)^2(M+2)^2} S_{new}^\dagger.
\end{eqnarray}
Hence the charge is
\begin{eqnarray}
 Q_{new}= -4c \displaystyle{\sum_{n_1=0}^\infty}\left(\displaystyle{\sum_{n_2=2}^\infty}
\frac{(m_{n_1}+m_{n_2})^2+4m_{n_2}}{(m_{n_1}+m_{n_2})^2(m_{n_1}+m_{n_2}+1)^2(m_{n_1}+m_{n_2}+2)^2}\right.\nonumber\\ 
 \left. + \frac{2}{(m_{n_1}+2)^2(m_{n_1}+3)^2}\right)
\end{eqnarray}
with
\begin{eqnarray}
 m_n = \frac{1}{2}(n-\lambda^-_n),\quad \lambda^-_n = \left\{\begin{array}{lll} 0 \,\,\,\,\,\,;n={\rm even} \\
                                                     \,1 \,\,\,\,\,\,;n={\rm odd} \end{array}\right.
\end{eqnarray}

Now for all even $n$'s we have $m_n=m_{n+1}$. Hence any absolutely convergent series over 
$n_1$ (or $n_2$) whose terms depend only on $m_{n_1}$ and $m_{n_2}$ can be broken into equal sums to give
\begin{equation}
 \displaystyle{\sum_{n_1=0}^\infty} G\left(m_{n_1},m_{n_2}\right)= 2\displaystyle{\sum_{n_1=0,2,4,...}} G\left(m_{n_1},m_{n_2}\right).
\end{equation}
Hence
\begin{eqnarray}
 \displaystyle{\sum_{n_1=0}^\infty}\displaystyle{\sum_{n_2=2}^\infty} F\left(m_{n_1},m_{n_2}\right)
&=& 2\displaystyle{\sum_{n_1=0,2,4,...}}2\displaystyle{\sum_{n_2=2,4,6,...}} F\left(m_{n_1},m_{n_2}\right) \nonumber \\
&=& 4\displaystyle{\sum_{n_1=0,2,4,...}}\displaystyle{\sum_{n_2=2,4,6,...}}F\left(m_{n_1},m_{n_2}\right).
\end{eqnarray}
Again $m_n=\frac{n}{2}$ for all even $n$'s. Thus the charge becomes
\begin{eqnarray}
 Q_{new} &=& 4\left[ -4c \displaystyle{\sum_{n_1=0,2,4,...}}\displaystyle{\sum_{n_2=2,4,6,...}}\frac{\left(\frac{n_1}{2}+\frac{n_2}{2}\right)^2+4\left(\frac{n_2}{2}\right)}{\left(\frac{n_1}{2}+\frac{n_2}{2}\right)^2\left(\frac{n_1}{2}+\frac{n_2}{2}+1\right)^2\left(\frac{n_1}{2}+\frac{n_2}{2}+2\right)^2} \right. \nonumber\\
&& \left. \,\,\,\,\,\,\,\, -4c \displaystyle{\sum_{n_1=0,2,4,...}}\frac{1}{\left(\frac{n_1}{2}+2\right)^2\left(\frac{n_1}{2}+3\right)^2}\right].
\end{eqnarray}
Redefining $\frac{n_1}{2}\rightarrow n_1$ and $\frac{n_2}{2}\rightarrow n_2$, we get
\begin{equation}
 Q_{new} = 4Q.
\end{equation}

\chapter{Appendices for Chapter \ref{qo}}
\section{Orthonormality of the eigenfunctions of the noncommutative SHO} \label{orthonormality}
The eigenfunctions of the noncommutative SHO Hamiltonian $\hat{H}_0$ given by (\ref{h_0_i}) are given in (\ref{En_psin}). If we define $\tilde{\phi}_n(k_1)$ such that 
\begin{equation}
\phi_n(\hat{x}_1) = \int dk_1 \, \tilde{\phi}_n(k_1)e^{ik_1\hat{x}_1}
\end{equation}
we can write
\begin{equation}
\psi_n(\hat{x}_0,\hat{x}_1) = \int dk_odk_1 \, \tilde{\phi}_n(k_1)\delta\left[k_0+\omega\left(n+\frac{1}{2}\right)\right] e^{ik_1\hat{x}_1}e^{ik_0\hat{x}_0}
\end{equation}
Hence
\begin{eqnarray}
\psi_{nS}(x_0,x_1) &=& \int dk_odk_1 \, \tilde{\phi}_n(k_1)\delta\left[k_0+\omega\left(n+\frac{1}{2}\right)\right] e^{ik_1x_1}e^{ik_0x_0} \nonumber \\
&=& \phi_n (x_1) e^{-i\omega\left(n+\frac{1}{2}\right)x_0},
\end{eqnarray}
$\phi_n(x_1)$ being the orthonormal eigenfuntions in the commutative case $\theta=0$. 
The inner product of two eigenfunctions become
\begin{eqnarray}
\left(\psi_n(\hat{x}_0,\hat{x}_1),\psi_m(\hat{x}_0,\hat{x}_1\right)_t &=& \int dx_1 \, \psi^*_{nS}(t,x_1)\psi_{mS}(t,x_1) \nonumber \\
&=&  \int dx_1 \, \phi^*_{n}(x_1)e^{i\omega\left(n+\frac{1}{2}\right)t} \phi_{m}(x_1) e^{-i\omega\left(m+\frac{1}{2}\right)t}\nonumber \\
&=& e^{-i\omega\left(m-n\right)t} \int dx_1 \, \phi^*_{n}(x_1)\phi_{m}(x_1) \nonumber \\
&=& \delta_{nm}
\end{eqnarray}

\section{The Green's operator function} \label{greens}
Expanding $G(t,t_0)$ as a Fourier integral we get
\begin{equation}
G(t,t_0) = G(t-t_0)
= \frac{1}{2\pi} \int_{-\infty}^{+\infty}d\omega^{\prime}g(\omega^{\prime})e^{-i\omega^{\prime}(t-t_0)},
\end{equation}
with
\begin{equation}
g(\omega^{\prime}) = (\hbar\omega^{\prime}-H_0)^{-1} = \sum_{j=0}^{\infty}\frac{1}{\hbar(\omega^{\prime}-\frac{E_j}{\hbar})}|\phi_j\rangle\langle\phi_j|.
\end{equation}
Thus
\begin{equation}
G(t,t_0)=\sum_{j=0}^{\infty}\left(\lim_{\epsilon\rightarrow 0^+} \frac{1}{2\pi}\int_{-\infty}^{+\infty}
e^{-i\omega^{\prime}(t-t_0)}\frac{1}{\hbar(\omega^{\prime}-\frac{E_j}{\hbar}+i\epsilon)}\right) |\phi_j\rangle\langle\phi_j|.
\end{equation}
Here, $i\epsilon$ has been introduced to avoid the pole on the contour (real axis). 
After finding the integral inside the summation using the complex analysis we get the Green's function to be
\begin{equation}
G(t,t_0) = -\frac{i}{\hbar}\Theta (t-t_0) e^{-i\frac{H_0}{\hbar}(t-t_0)} \label{G_sol},
\end{equation}
where $\Theta(x)$ is the Heaviside Step Function.
\section{Simplifying $A^{(1)}(T_2,T_1)$} \label{simplify_A1}
To simplify (\ref{A1_series}) what we do is to find the differential equation for $A^{(1)}(t,T_1)$ with independent variable $t$ and solve it with proper initial conditions.
The differential equation has been found to be
\begin{equation}
\left( i\hbar\frac{\partial}{\partial t}-H_{I0}^{int}(t)\right)A^{(1)}(t,T_1)=H_{I1}^{int}(t)A^{(0)}(t,T_1),
\end{equation}
which is a first order equation and hence $A^{(1)}(t,T_1)$ is unique if an initial condition is given.
The initial condition comes from the fact that  $A^{(0)}(t,T_1)$ and $A^{(1)}(t,T_1)$ must become identity and zero respectively for $t=T_1$, i.e., no interaction.
Thus, $A^{(1)}(T_1,T_1)=0$.
To solve the equation we define the Green's operator function $G_{int}(t,t_0)$ as
\begin{equation}
\left( i\hbar\frac{\partial}{\partial t}-H_{I0}^{int}(t)\right)G_{int}(t,t_0)=\delta(t-t_0).
\label{Gint}
\end{equation}
Now generalizing solution (\ref{G_sol}) for the time-dependent case we get
\begin{equation}
G_{int}(t,t_0) = -\frac{i}{\hbar}\Theta (t-t_0)T\left[ e^{-\frac{i}{\hbar}\int_{t_0}^{t}d\tau H_{I0}^{int}(\tau)}\right]
= -\frac{i}{\hbar}\Theta (t-t_0)A^{(0)}(t,t_0).
\label{Gint_sol}
\end{equation}
It can be easily checked that the above expression for the $G_{int}(t,t_0)$ satisfies the corresponding differential equation. 
The solution for $A^{(1)}(t,T_1)$ is then given by
\begin{equation}
A^{(1)}(t,T_1)=A_{hom}^{(1)}(t,T_1)+\int_{-\infty}^{+\infty}dt_0\,\,G_{int}(t,t_0)H_{I1}^{int}(t_0)A^{(0)}(t_0,T_1),
\end{equation}
where $A_{hom}^{(1)}(t,T_1)$ is the solution of the homogeneous equation
\begin{equation}
\left( i\hbar\frac{\partial}{\partial t}-H_{I0}^{int}(t)\right)A_{hom}^{(1)}(t,T_1)=0.
\end{equation}
The $\Theta$-function in the expression of $G_{int}(t,t_0)$
and the fact that interaction was off before $T_1$, with the initial condition for $A^{(1)}(T_1,T_1)=0$,
gives the initial condition for $A_{hom}^{(1)}(t,T_1)$, i.e., $A_{hom}^{(1)}(T_1,T_1)=0$.
The only solution of the homogeneous equation with this initial condition is $A_{hom}^{(1)}(t,T_1)=0$.
Thus we get
\begin{equation}
A^{(1)}(t,T_1)=\int_{-\infty}^{+\infty}dt_0 \,\,G_{int}(t,t_0) H_{I1}^{int}(t_0)A^{(0)}(t_0,T_1).
\end{equation}
Now, putting the expression of $G_{int}(t,t_0)$ above and 
introducing $A^{(0)}(t_0,T_1)\left[A^{(0)}(t_0,T_1)\right]^{-1}$ before $H_{I1}^{int}(t_0)$, we get (\ref{A1_A0}).
Here we have also used the following property of $A^{(0)}(t,T_1)$:
\begin{equation}
 A^{(0)}(t,t_0)A^{(0)}(t_0,T_1)=A^{(0)}(t,T_1).
\end{equation}

\section{The correlation function} \label{correlation}
An operator corresponding to the detection of a photon by a detector 
should be proportional to the annihilation operator $a$  (say $ka$) \cite{Walls_book, Glauber:1963fi}.
Hence if $|i\rangle$ is the initial state of the radiation field, the state after the detection of one photon is $ka|i\rangle$.
The amplitude for going to the final state $|f\rangle$ is given by $k\langle f|a|i\rangle$.
The corresponding probability is $|k|^2|\langle f|a|i\rangle|^2$.
Thus the probability of detection of one photon in the state $|i\rangle$
\begin{equation}
P_1 = \displaystyle{\sum_f}|k|^2|\langle f|a|i\rangle|^2
= |k|^2\displaystyle{\sum_f}\langle i|a^{\dagger}|f\rangle\langle f|a|i\rangle
= |k|^2\langle i|a^{\dagger}a|i\rangle.
\end{equation}
Similarly, probability of detection of two photons with a time delay of $\tau$ is
\begin{equation}
\begin{array}{r}
P_2 = \displaystyle{\sum_f}|k|^4|\langle f|a(t+\tau)a(t)|i\rangle|^2
= |k|^4\displaystyle{\sum_f}\langle i|a^{\dagger}(t)a^{\dagger}(t+\tau)|f\rangle\langle f|a(t+\tau)a(t)|i\rangle \\
= |k|^4\langle i|a^{\dagger}(t)a^{\dagger}(t+\tau)a(t+\tau)a(t)|i\rangle.
\end{array}
\end{equation}
The 2nd order correlation function with a time delay $\tau$ is defined as 
\begin{equation}
g^{(2)}(\tau) = \frac{P_2(\tau)}{P_1^2}
= \frac{\langle i|a^{\dagger}(t)a^{\dagger}(t+\tau)a(t+\tau)a(t)|i\rangle}{\langle i|a^{\dagger}(t)a(t)|i\rangle^2}.
\end{equation}
For $\tau=0$
\begin{equation}
g^{(2)}(0)= \frac{\langle i|a^{\dagger}(t)a^{\dagger}(t)a(t)a(t)|i\rangle}{\langle i|a^{\dagger}(t)a(t)|i\rangle^2} 
=  \frac{\langle i(t)|a^{\dagger}a^{\dagger}aa|i(t)\rangle}{\langle i(t)|a^{\dagger}a|i(t)\rangle^2},
\label{g20}
\end{equation}
For a coherent state it can be calculated to be equal to 1.

\chapter{Appendices for Chapter \ref{DSR}}
\section{Convergence of $K^*\l x,y\r$}\label{kstar_convergence}
$K^*\l x,y\r$ given by (\ref{kstar_mr},\ref{m0},\ref{m1},\ref{mr_series}) is convergent if the following two series are convergent:
\begin{equation}
 S=\sum_{r=2}^{\infty}t_r\frac{(-x)^{2r-3}}{(2r-3)!}
\end{equation}
and
\begin{equation}
 S^\prime=\sum_{r=2}^{\infty}t_r\sum_{k=1}^{2r-3}\frac{(-x)^{k-1}}{(2r-3)(2r-4)...(2r-2-k)}\left(\frac{1}{y}\right)^{2r-2-k}.
\end{equation}
$S$ can be easily proved to be absolutely convergent using Ratio test.
For $S^\prime$ first consider the following double series:
\begin{equation}
 S^{\prime\prime}=\sum_{r=2}^{\infty}t_r\sum_{k=1}^{\infty}\frac{(-x)^{k-1}}{(2r-3)(2r-4)...(2r-2-k)}\left(\frac{1}{y}\right)^{2r-2-k}
=\sum_{r=2}^{\infty}\sum_{k=1}^{\infty}a_{r,k}.
\end{equation}
Let us first test the convergence of $S^{\prime\prime}$ (See theorem (2.7) of \cite{balmohan}).
The row series $S_r$ (for a fixed $r$) and the column series $S_k$ (for a fixed $k$) are defined as
\begin{equation}
 S_{r}=\sum_{k=1}^{\infty}a_{r,k},
\end{equation}
\begin{equation}
 S_{k}=\sum_{r=2}^{\infty}a_{r,k}.
\end{equation}
The ratio tests for $S_r$ and $S_k$ show that they are absolutely convergent (for $y>1$).
Also $\displaystyle{\lim_{r,k\rightarrow \infty}}\left|\frac{a_{r,k+1}}{a_{r,k}}\right|=0<1$. Hence, $S^{\prime\prime}$ is absolutely convergent.
Now
\begin{equation}
 |S^{\prime\prime}|=\sum_{r=2}^{\infty}\sum_{k=1}^{\infty}|a_{r,k}|=\sum_{r=2}^{\infty}\sum_{k=1}^{2r-3}|a_{r,k}|+L \quad \quad \quad ;L\geq 0.
\end{equation}
As $|S^{\prime\prime}|$ is convergent (or in other words $S^{\prime \prime}$ is absolutely convergent) 
we must have $\left|S^{\prime}\right|=\displaystyle{\sum_{r=2}^{\infty}\sum_{k=1}^{2r-3}}|a_{r,k}|$ to be convergent (or in other words $S^\prime$ to be absolutely convergent).
Thus the series expansion of $K^*\l x,y\r$ is absolutely convergent.
\section{Continuity and differentiability of the partition function in $m_0$}
\label{app1}
We shall show that $Z_1^0$ is continuous in $m_0$ for $m_0\in [0,\kappa]$.
After integrating over the angular coordinates (\ref{z10}) gives
\begin{equation}
Z_1^0\l m_0 \r = \frac{2V}{\l2\pi\r^2}\int_{0}^{\kappa}dp \,\,\, p^2 e^{-\beta[\e(p,m_0)-m_0]} 
=\frac{2V}{\l2\pi\r^2}\int_{0}^{\kappa}dp \,\,\, f(p,m_0).
\label{Z10_m0}
\end{equation}
The integrand $f(p,m_0) = p^2 e^{-\beta[\e(p,m_0)-m_0]}$ is a continuous bounded function of 
$p$ and $m_0$ in the range $m_0\in [0,\kappa], p\in [0,\kappa]$.
Thus $Z_1^0\l m_0 \r$ is a continuous function of $m_0$ as the function $g(p)=C$,
where $C$ is the upper bound of $|f(p,m_0)|$ in the range $m_0\in [0,\kappa], p\in [0,\kappa]$, satisfies
$g(p)\geq |f(p,m_0)|$ for all $m_0\in [0,\kappa], p\in [0,\kappa]$ and is integrable as
 $\displaystyle{\int_{0}^{\kappa}}dp\,\,\, g(p) = C\kappa<\infty$ (see Lemma 1 in $\S$ 5.12 of \cite{fleming}).\\
\indent The derivative of the integrand with respect to $m_0$ is given by
\begin{equation}
 \frac{\partial f(p,m_0)}{\partial m_0} = \beta f(p,m_0) 
\l1-\frac{\e^2(p,m_0)-p^2}{m_0 \l1-\frac{2m_0}{\kappa}\r^{1/2}\l p^2-\frac{m_0^2}{\frac{2m_0}{\kappa}-1}\r^{1/2}}\r.
\end{equation}
It has 2 poles (and also branch points) in the complex $p$-plane at $p=\pm \frac{m_0}{\l\frac{2m_0}{\kappa}-1\r^{1/2}}$.
We note that the poles and the branch points remain to be at the same positions for all higher order derivatives of $f(p,m_0)$
with respect to $m_0$. 
For $m_0=0$ both the poles are at $p=0$ and as $m_0$ increases the poles separate towards the imaginary axis.
They keep on moving on the imaginary axis till they reach $\pm i \infty$ at $m_0=\frac{\kappa}{2}$.
After that they start to come closer to each other on the real line and finally at $m_0=\kappa$ they stop at $p=\pm \kappa$.
Note that for all $m_0\in(0,\kappa)$ the poles are never on the contour of integration (the real line from $p=0$ to $p=\kappa$)
and the functions $\l\frac{\partial f(p,m_0)}{\partial m_0},\frac{\partial^2 f(p,m_0)}{\partial m_0^2}, etc.\r$ 
remain to be bounded.
This (by the same argument as given in the case of the continuity of $Z_1^0\l m_0 \r$) ensures the infinite-order differentiability
of $Z_1^0\l m_0 \r$ in $m_0\in(0,\kappa)$ and the derivatives can be found by using the Leibniz rule 
(see Lemma 2 in $\S$ 5.12 of \cite{fleming}). Note that the fact that we are not being able to say about the differentiability 
of $Z_1^0$ at $m_0=0$ could be a relic of the non-analytic part in (\ref{zm0}).



\bibtitle{References}

\end{document}